\documentclass[aps,pre,twocolumn]{revtex4}
\usepackage{amsmath,amsfonts}
\usepackage{amssymb,amsthm,newlfont,setspace,rotating}
\usepackage{color}
\usepackage[normalem]{ulem}
\usepackage{graphicx}
\usepackage{subfigure}
\usepackage{epstopdf}

\newcommand{\be}{\begin{equation}}
\newcommand{\ee}{\end{equation}}

\begin{document}

\title{Opinion formation and distribution in a bounded confidence model\\ on various networks}
\author{X. Flora Meng\( ^{1,2} \), Robert A. Van Gorder\( ^1 \), and Mason A. Porter\( ^{1,3,4,*} \)}
\affiliation{\mbox{\( ^1 \)Mathematical Institute, University of Oxford, Oxford OX2 6GG, UK}\\
\mbox{\( ^2 \)Department of Electrical Engineering and Computer Science,}\\\mbox{Massachusetts Institute of Technology, Cambridge, MA 02139, USA}\\
\mbox{\( ^3 \)CABDyN Complexity Centre, University of Oxford, Oxford OX1 1HP, UK}\\
\mbox{\( ^4 \)Department of Mathematics, University of California, Los Angeles, CA 90095, USA}\\
\( ^* \)mason@math.ucla.edu}

\begin{abstract}

In the social, behavioral, and economic sciences, it is an important problem to predict which individual opinions will eventually dominate in a large population, if there will be a consensus, and how long it takes a consensus to form. This idea has been studied heavily both in physics and in other disciplines, and the answer depends strongly on both the model for opinions and for the network structure on which the opinions evolve. One model that was created to study consensus formation quantitatively is the Deffuant model, in which the opinion distribution of a population evolves via sequential random pairwise encounters. To consider the heterogeneity of interactions in a population due to social influence, we study the Deffuant model on various network structures (deterministic synthetic networks, random synthetic networks, and social networks constructed from Facebook data) using several interaction mechanisms. We numerically simulate the Deffuant model and conduct regression analyses to investigate the dependence of the convergence time to equilibrium on parameters, including a confidence bound for opinion updates, the number of participating entities, and their willingness to compromise. We find that network structure and parameter values both have an effect on the convergence time, and for some network topologies, the convergence time undergoes a transition at a critical value of the confidence bound. We discuss the number of opinion groups that form at equilibrium in terms of a confidence-bound threshold for a transition from consensus to multiple-opinion equilibria. 
\end{abstract}

\maketitle



\section{Introduction}

Social interactions play a central role in the process of decision-making and opinion formation in populations of humans and animals \cite{jackson08,couzin05}. Discussions among acquaintances, coworkers, friends, and family members often lead interlocutors to adjust their viewpoints on politics, participation in a social movement, adoption of technological innovations, or other things \cite{degroot,oliver1985,siegel09,yariv2011,friedkin2015}; and the prediction of collective opinion formation in a population from attributes of individuals is one of the most important problems in the social sciences \cite{castellano09,friedkin2016}. Consensus dynamics is also a key problem in areas such as control theory \cite{ali-flock,ali2003} and collective dynamics more generally \cite{vicsek-review2012}. From a physical and mathematical standpoint, the study of opinion dynamics is one of the key motivating examples for studying the effects of network structure on dynamical processes on networks \cite{porter2016}.

There are various methods for studying opinion formation in social networks, such as through Bayesian learning or generative social-interaction mechanisms \cite{acemoglu11b}. Bayesian updating requires some unrealistic assumptions about individuals' knowledge and reasoning ability, and it becomes computationally infeasible in complex settings \cite{acemoglu11b,jackson08}. Even in opinion models that do not suffer from these issues, there remains significant arbitrariness in the choice of specific models and parameters to use, and different choices can lead to markedly (and qualitatively) different results \cite{acemoglu11b,sobkowicz09}. A substantial amount of work on non-Bayesian approaches to opinion formation employs models and tools from dynamical systems, probability theory, and statistical physics \cite{castellano09}. Moreover, a major theme in statistical physics is how global properties can emerge from local rules, which is similar to the question in social sciences of how the collective opinion of a population evolves as the result of individual attitudes and the mutual influence of individuals on each other \cite{kozma08a}. Some notable generative models of opinion formation include voter models \cite{clifford73,holley75,holme06,durrett12}, majority-rule models \cite{galam02}, models based on social-impact theory \cite{latane81,nowak90}, the Sznajd model \cite{sznajdweron00,sznajdweron05}, and bounded confidence models \cite{deffuant00,hegselmann02,krause00,weisbuch00}.

Bounded-confidence models, first introduced (to our knowledge) by Deffuant \textit{et al.} \cite{deffuant00,weisbuch02} and Hegselmann and Krause \cite{hegselmann02,krause00}, capture the notion of a tolerance threshold based on experimental social psychology \cite{psych2003,weisbuch05}. Bounded confidence reflects the psychological concept of \emph{selective exposure}, which refers to an individual's tendency to favor information that supports their views while neglecting conflicting arguments \cite{lorenz07b,sullivan09}. The Deffuant model and the Hegselmann--Krause (HK) model both consider a set of agents who hold continuous opinions that can vary. Agents are connected to each other by an interaction network, and neighboring agents adjust their opinions at discrete time steps whenever their opinions are sufficiently close to each other. The two models differ primarily in their communication regime. In the HK model, agents interact with all of their compatible neighbors simultaneously at each time step, and they update their opinions to agree with the mean opinion of these neighbors. In contrast, the Deffuant model adopts a sequential updating rule and can be viewed as a discrete-time repeated game that is played in pairwise fashion among a set of agents until the agents' opinions converge to either a single opinion or multiple opinions \cite{fudenberg95,jackson08,jackson14}. One can also tune the speed at which opinions converge in the Deffuant model through an additional parameter, sometimes called a \textit{cautiousness parameter}, that describes openness to compromises. The Deffuant model was developed to study opinion-formation processes in large populations in which people interact in small groups (such as pairwise interactions in a network), whereas the HK model is suitable for contexts such as meetings with many participants. Two questions have drawn considerable interest: (1) how does the parameter space influence the number of opinion groups in an equilibrium state; and (2) how long does it take for a system to reach an equilibrium state \cite{deffuant00,laguna04,lorenz05a,weisbuch02,weisbuch04}?

Despite its seeming simplicity, the Deffuant model is not analytically solvable in general, and most results about it have been obtained from Monte Carlo simulations. It has been shown numerically, for a few values of the cautiousness parameter, that consensus occurs for large confidence bound values on complete graphs with probability close to \( 1 \) in the large-population limit, whereas multiple opinion groups persist at equilibrium for low confidence bounds \cite{fortunato04b,laguna04,weisbuch02,weisbuch04}. However, different confidence-bound thresholds have been proposed for transition from consensus to multiple opinion groups at equilibrium. In the latter case, one can approximate the number of groups by a function of the confidence-bound value \cite{weisbuch02,weisbuch04}. Numerical simulations have suggested that the time to opinion equilibrium is proportional to the number of agents in the network \cite{laguna04}. Moreover, a higher value of the cautiousness parameter increases not only the convergence speed but also the number of agents that hold extreme opinions at equilibrium \cite{laguna04}. Numerical simulations also have illustrated the possibility of forcing or preventing a consensus within a population by manipulating the initial opinion distribution \cite{bennaim03,carro12}.

There has been some research that compares results for the Deffuant model on complete graphs with those on other networks. Results for complete graphs and square lattices are similar for large confidence-bound values, except that a few extreme opinions remain on square lattices at equilibrium \cite{weisbuch02}. The Deffuant model has also been simulated on random graphs generated by Barab\'{a}si--Albert (BA), Erd\H{o}s--R\'{e}nyi (ER), and Watts--Strogatz (WS) mechanisms \cite{fortunato04b,gandica,jalili13,stauffer03a}. However, different assumptions and update rules are often used, and this poses a major barrier for comparing results across different networks.

There have also been efforts to study the Deffuant model from an analytical perspective using a density function that determines the agents' density in opinion space \cite{bennaim03, lorenz05a}. Such an approach adopts a classical strategy in statistical physics by deriving a rate equation (also called a ``master equation'') and can be interpreted as taking the infinite limit of the number of agents \cite{lorenz07a}. These derivations have not led to analytical solutions of the Deffuant model, but they require numerical integration only of the master equation, which is faster than running Monte Carlo simulations of the original model. Unfortunately, however, such density-based method requires fairly restrictive assumptions, such as homogeneous mixing and averaging agents' opinions as the means of compromise.

The Deffuant model itself also has limitations, and numerous efforts have been made to extend it in order to better reflect reality. For instance, the confidence bound imposes a boundary on interacting agents' decision whether or not to adjust their opinions. A small change in the difference between their opinions may lead to a different decision being made. For this reason, some scholars have proposed the use of smooth confidence bounds, with which the attraction of agents decreases as their opinion difference increases \cite{deffuant02,deffuant04}. Other generalizations of the confidence bound include introducing heterogeneous tolerance thresholds in a population \cite{weisbuch02,weisbuch05} and considering time-dependent thresholds \cite{weisbuch02}. Additionally, the Deffuant model can be extended naturally to incorporate vector-valued opinions, as this only requires redefining opinion distance \cite{fortunato04}.

Studies of variants of the Deffuant model often compare new results with those of the original model. However, numerical simulations of the original model are usually performed for specific parameter values and networks. Moreover, conclusions are often drawn based on visual inspection and sometimes rely on simplifying assumptions. Furthermore, quantifying the confidence bound and the cautiousness of a population is an open question for many applications. These issues motivate us to take a more systematic approach to the study of the Deffuant model on networks.

We explore the dependence of convergence time and the number of opinion groups at equilibrium on network topology, confidence bound, the number of participating agents, and their willingness to compromise. We conduct regression analyses to model convergence time as a function of the parameters considered and study the qualitative behavior of opinion groups at equilibrium. The networks that we study fall into three categories. The first set of networks --- complete graphs, cycles, prism graphs, square lattices, and complete multipartite graphs --- are synthetic and deterministic. The Deffuant model has been much studied on complete graphs and square lattices due to their simple structures, and we extend this list of simple network structures and compare simulation results on these networks with those on more complex structures. From our simulations on deterministic graphs, we find that network topology and parameter values of the Deffuant model appear to have an intertwined effect on convergence time, with the behavior of convergence time undergoing a transition at a confidence-bound threshold for some network structures. The second set of networks are (synthetic) random graphs, including cycles with random edges, prisms with random edges, and random graphs generated by an Erd\H{o}s--R\'{e}nyi model \cite{gilbert59}. Due to their simplicity, these models are a good starting point for understanding the Deffuant model on random graphs. Our simulations suggest that the behavior of convergence time on random-graph models is similar to that on their counterpart deterministic networks. The third set of networks are empirical and deterministic. In particular, we use two {\sc Facebook100} networks, which are constructed using Facebook ``friendship'' data \cite{traud12}, and which are a type of network in which people have discussions and opinions can change over time. Using all three types of networks, we discuss the number of opinion groups at equilibrium and phenomena such as a confidence-bound threshold for a transition from consensus to multiple-opinion equilibria.

The rest of our paper is organized as follows. First, we introduce relevant definitions from network science, define the Deffuant model in mathematical terms, and present some important known results for the Deffuant model on networks. We then describe our methodology and introduce the networks and the approach that we use for numerical simulations. We then conduct regression analyses on our simulation results to explore the dependence of convergence time on network structure, confidence bound, the number of participating agents, and their cautiousness. We also discuss the phenomena that we observe about the number of opinion groups at equilibrium, and we discuss our results and their implications for sociology. We give further details on our statistical analysis in an appendix.


\section{Background}

In this section, we recall relevant definitions from network science. We then define the Deffuant model, give some intuition about its design, and present some important known results about the Deffuant model on networks.


\subsection{Basic definitions in network science}

A \textit{network} is a set of items (called \textit{nodes}) with connections (called \textit{edges}) between them \cite{newman10}. Many ideas in network science originated in graph theory, and we present some definitions \cite{newman10,west01} that are pertinent to our study. A \textit{graph} \( G \) is a triple consisting of a \textit{node set} \( V(G) \), an \textit{edge set} \( E(G) \), and a relation that associates each edge with two nodes (not necessarily distinct) called its \textit{endpoints}. The simplest type of \textit{network} is a graph. Two nodes are \textit{adjacent}, and are called \textit{neighbors} of each other, if and only if they are endpoints of the same edge. The \textit{degree} of a node is equal to the number of its neighbors. A \textit{regular graph} is a graph in which each node has the same degree. A \textit{random-graph model} is a probability distribution on graphs that has some fixed parameters and generates networks randomly in other respects.


\subsection{The Deffuant model}\label{subsection:bounded-confidence-models}

In the Deffuant model, randomly-selected neighboring agents interact in a pairwise manner and make a compromise toward each other's opinion whenever their opinion difference is below a given threshold. (Otherwise, their opinions do not change.) Consider a population of \( N \) agents, who are connected to each other socially via a network \( G \); and let \( [a,b] \subset \mathbb{R} \) be the opinion space. At time \( t \in \mathbb{N} \), suppose that each agent \( i \) holds a time-dependent opinion \( x_i(t) \in [a,b] \). Given an initial profile \( \vec{x}(0) \in [a,b]^N \), a \textit{confidence bound} \( c \in [0, b-a] \), and a cautiousness parameter that we call the \textit{multiplier} \( m \in (0,0.5] \), the \textit{Deffuant model} is the random process \( (\vec{x}(t))_{t \geq 0} \) defined as follows. At time \( t \), a pair of neighboring agents \( i \) and \( j \neq i \) are selected uniformly at random (i.e., we select an edge uniformly at random) and update their opinions according to the equations
\begin{equation}
	\begin{aligned}
x_i(t+1) &= \left\{
				\begin{array}{ll}
					x_i(t) + m  \Delta_{j,i}(t) \,, & \text{if } \lvert \Delta_{i,j}(t) \rvert < c \,, \\
					x_i(t)\,, & \text{otherwise}\,,
				\end{array}
				\right. \\
x_j(t+1) &= \left\{
				\begin{array}{ll}
					x_j(t) + m \Delta_{i,j}(t)\,, & \text{if } \lvert \Delta_{i,j}(t) \rvert < c \,, \\
					x_j(t)\,, & \text{otherwise}\,,
				\end{array}
				\right.
\end{aligned}
\end{equation}
where $\Delta_{j,i}(t)=x_j(t)-x_i(t)$.

The Deffuant model uses a continuous opinion space, as an individual's stance on a specific matter can vary smoothly from one extreme to another in many real-world scenarios \cite{castellano09}. For instance, a political position (on single dimension) is not typically simply ``left" or``right" but somewhere in between two extremes. The study of opinion-formation processes has traditionally considered an opinion to be a discrete variable, which is a reasonable assumption for some applications. For instance, the classical voter model \cite{clifford73,holley75} considers a binary variable that specifies one's decision in a vote. However, it is important to develop models that incorporate more nuanced opinions.

As in the original paper \cite{deffuant00} that introduced the Deffuant model, most later studies treated the initial opinions as being independent and identically distributed according to the uniform distribution on the opinion space \( [a,b] \). We also adopt this convention, as our goal is to explore the basic version of the Deffuant model in a systematic manner to provide a point of reference for results of the model's variants. Nonuniform initial opinion distributions are considered, for example, in \cite{jacobmeier06}.

The confidence bound \( c \) characterizes a population's tolerance of diverse viewpoints. If the opinion difference between a pair of agents is smaller than this threshold, they reduce their disagreement by making a compromise. Otherwise, the two agents keep their current opinions after they interact (or perhaps are unwilling to discuss the issue at all). In the extreme case of \( c=0 \), no interaction can lead to compromise, and the initial opinion profile is a fixed point. At the other extreme, if \( c=b-a \), any pair of interacting agents will compromise their opinion if they interact with each other

The multiplier \( m \), which also called a \textit{convergence parameter} in some papers \cite{deffuant00,fortunato04,laguna04,weisbuch02}, specifies a population's cautiousness in the modification of judgements. A larger value of \( m \) indicates that individuals are more willing to compromise. In the special case \( m=0.5 \), pairs of interacting agents agree on the mean of their opinions whenever their opinion difference is below the confidence bound. Most past work has examined homogeneous $m$, but it would be interesting to examine the effects of heterogeneous levels of cautiousness. For example, \cite{deffuant04} used a smooth influence function in which agents whose opinions have low uncertainty are more influential than agents whose opinions have high uncertainty, and other types of heterogeneity are also worth exploring.

The Deffuant model, in its original form \cite{deffuant00}, considers the confidence bound and the multiplier to be constant in time and homogeneous across the whole population. In this setting, the mean opinion of two agents is the same before and after their interaction.

Convergence of opinions is generally defined as the appearance of a stable configuration in which no more changes can occur. At equilibrium, the opinion distribution is a superposition of Dirac delta functions in the opinion space \( [a,b] \), such that consecutive spikes are separated by a distance of at least \( c \). In other words, any two agents either hold the same opinion or their viewpoints differ by a distance of at least \( c \). We use the notation \( K \in \mathbb{N} \) to denote the number of opinion groups at equilibrium.


\subsection{The Deffuant model on various networks}

The agents in a Deffuant model are represented by nodes of a network, and a pair of agents on a randomly selected edge can interact with each other. To the best of our knowledge, the Deffuant model has been studied on only a small subset of networks, which includes complete graphs, square lattices, Erd\H{o}s--R\'{e}nyi (ER) random graphs, Watts--Strogatz (WS) random graphs, and Barab\'{a}si--Albert (BA) random graphs \cite{barabasi99}.

The Deffuant model on complete graphs has received considerable attention \cite{deffuant00}. Complete graphs can be used to model small communities, where everyone knows each other, such as high-level political leaders in a country or inhabitants of a village. Complete graphs are also sometimes used as approximations for individual communities in large social networks, as individuals within communities are more closely connected with each other than with outsiders \cite{fortunato16,porter09}. In the homogeneous mixing case, the population's opinions always reach equilibrium \cite{lorenz05a}. It has been shown numerically that a large confidence bound $c$ yields an equilibrium state of consensus, whereas multiple opinion groups can persist for small values of \( c \) \cite{deffuant00,fortunato04b,laguna04,weisbuch02,weisbuch04}. Such results were also obtained in simulations on square lattices, ER random graphs, WS random graphs, and BA random graphs \cite{deffuant00,kozma08a,kozma08b,stauffer03a}. Moreover, numerical simulations on complete graphs imply that one can estimate the number of opinion groups at equilibrium by \( K=1/(2c) \) \cite{deffuant00,weisbuch02,weisbuch04}, and that multiplier \( m \) and the number \( N \) of participating agents do not have a significant effect on \( K \) \cite{deffuant00,weisbuch02}. However, a later study \cite{laguna04} observed that the number of ``major opinion'' groups that include many agents is a function of \( c \), whereas the number of ``minor opinion'' groups (i.e., groups of \textit{minorities}) depends on \( m \).

On square lattices, WS random graphs, and BA random graphs, the Deffuant model includes behavior that differs from the homogeneous mixing case. For instance, simulations on square lattices and BA random graphs suggest that \( K \) depends not only on \( c \), but also on \( N \), when multiple opinion groups persist at equilibrium \cite{deffuant00,stauffer03a}. Simulations on WS random graphs indicate that \( K \) depends on both \( c \) and network structures, and that the presence of disorder (i.e., random ``shortcut'' edges) seems to have only a slight effect on convergence time \( T \) \cite{gandica}.

Existing research on the Deffuant model on ER random graphs has focused mainly on adaptive networks, which evolve along with the game \cite{kozma08a,kozma08b}. For WS random graphs, the study of the model has centered around opinion groups at equilibrium \cite{gandica}.


\section{Methods}\label{section:methods}

For each network structure, we conduct a regression analysis to examine convergence time as a function of confidence bound, the number of participating agents, and the multiplier that measures their cautiousness. We then qualitatively study the behavior of the number of opinion groups at equilibrium, as such an approach is more natural than conducting regression analysis because of the complex nature of opinion-group distributions.

\begin{table*}
\centering
	\begin{tabular}{c  p{10cm}  c}
	\hline
	Network & Definition & Example \\
	\hline
	\( K_n \) & A \textit{complete graph} \( K_n \) has \( n \)  pairwise adjacent nodes \cite{west01}.  &
	\begin{minipage}{.1\textwidth}
		\centering
		\includegraphics[width=\linewidth]{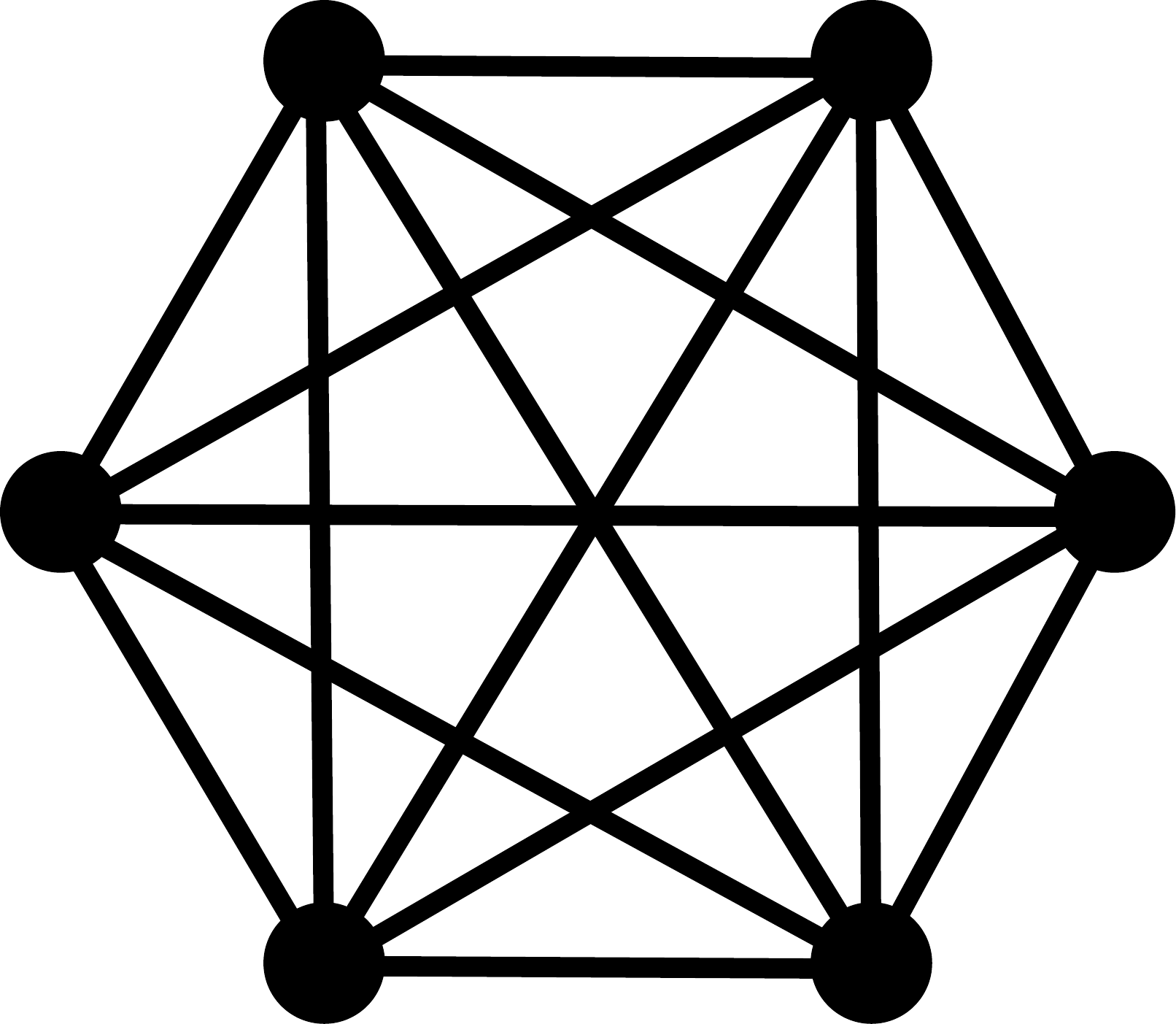}
    \end{minipage} \\
	\hline
	\( C_n \) & For \( n \geq 3 \), a \textit{cycle} \( C_n \) has node set \( \{ v_j \mid j \in \{1, \dots, n \} \} \) and edge set \( \{ v_j v_{j+1} \mid   j \in \{1, \dots, n-1 \} \} \cup \{ v_n v_1 \} \) \cite{west01}. &
	\begin{minipage}{.1\textwidth}
      \includegraphics[width=\linewidth]{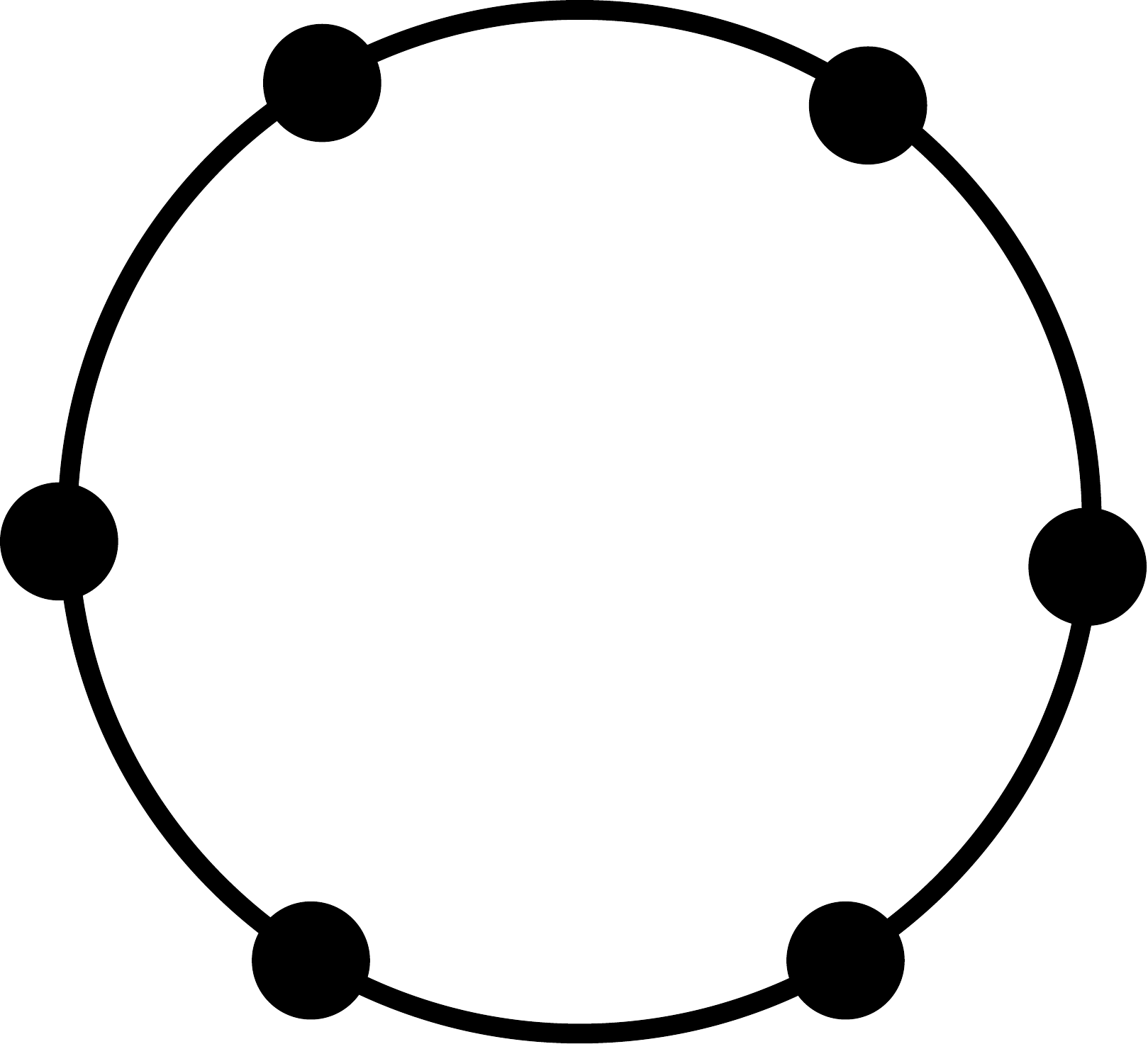}
    \end{minipage} \\
	\hline
	\( Y_n \) & For \( n \geq 3 \), let \( \{ v_j \mid  j \in \{1, \dots, n\} \} \) and \( \{ w_j \mid  j \in \{1, \dots, n\} \} \) be the node sets of two disjoint cycles. The \textit{prism} \( Y_n \) is defined as the graph obtained by joining the two cycles at the set of edges \( \{ v_j w_j \mid   j \in \{1, \dots, n\}\} \) \cite{boesch87}. &
	\begin{minipage}{.1\textwidth}
      \includegraphics[width=\linewidth]{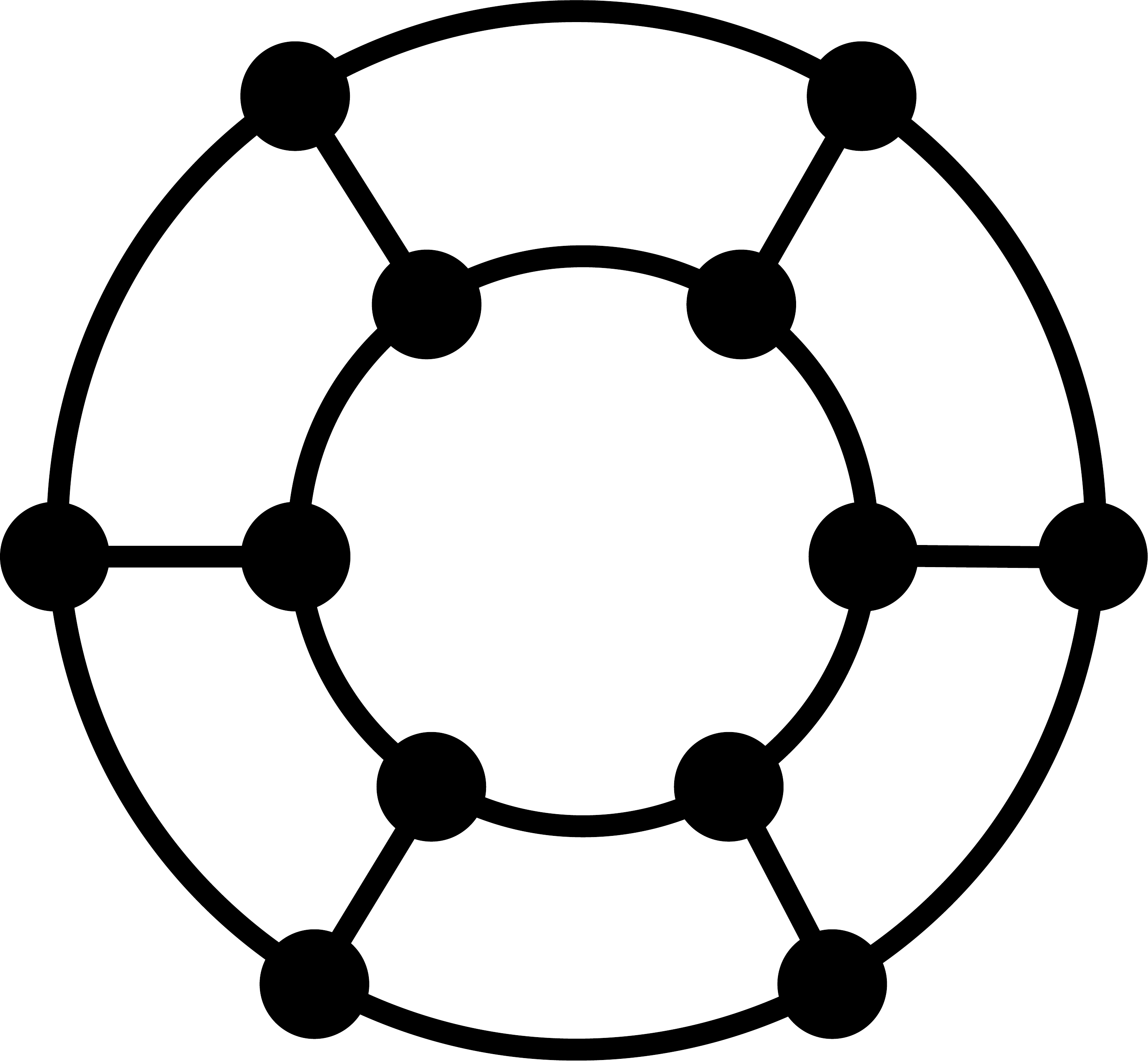}
    \end{minipage} \\
	\hline
	\( S_l \) & For a positive integer \( l \), we define a \textit{square lattice} \( S_l \) of side length \( l \) as the graph with the node set \( \{ (x,y) \mid 0 \leq x,y \leq l, \quad \mathrm{with} \quad x,y \in \mathbb{Z} \} \) and edges \( ((x_1,y_1),(x_2,y_2)) \) such that \( \Vert (x_1-x_2,y_1-y_2) \Vert_2=1 \). &
	\begin{minipage}{.1\textwidth}
      \includegraphics[height=.5in]{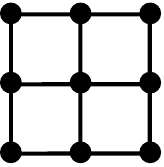}
    \end{minipage} \\
	\hline 
	\( K_{n,r} \) & For an integer \( r \geq 2 \) and positive integers \( n_1, \, n_2, \dots , \, n_r \), a \textit{complete \( r \)-partite graph} \( K_{n_1, n_2, \dots , n_r} \) is a graph whose node set can be partitioned into \( r \) subsets (called \textit{partite sets}) of sizes \( n_1, \, n_2, \dots , \, n_r \), respectively, such that two nodes are adjacent if and only if they are from two distinct subsets. We consider complete \( r \)-partite graphs with equal-sized partite sets and denote such graphs as \( K_{n,r} \), where \( r \) equals the number of partite sets and \( n \) (a multiple of \( r \)) is the size of the node set \cite{chartrand08}. &
	\begin{minipage}{.1\textwidth}
      \includegraphics[height=.5in]{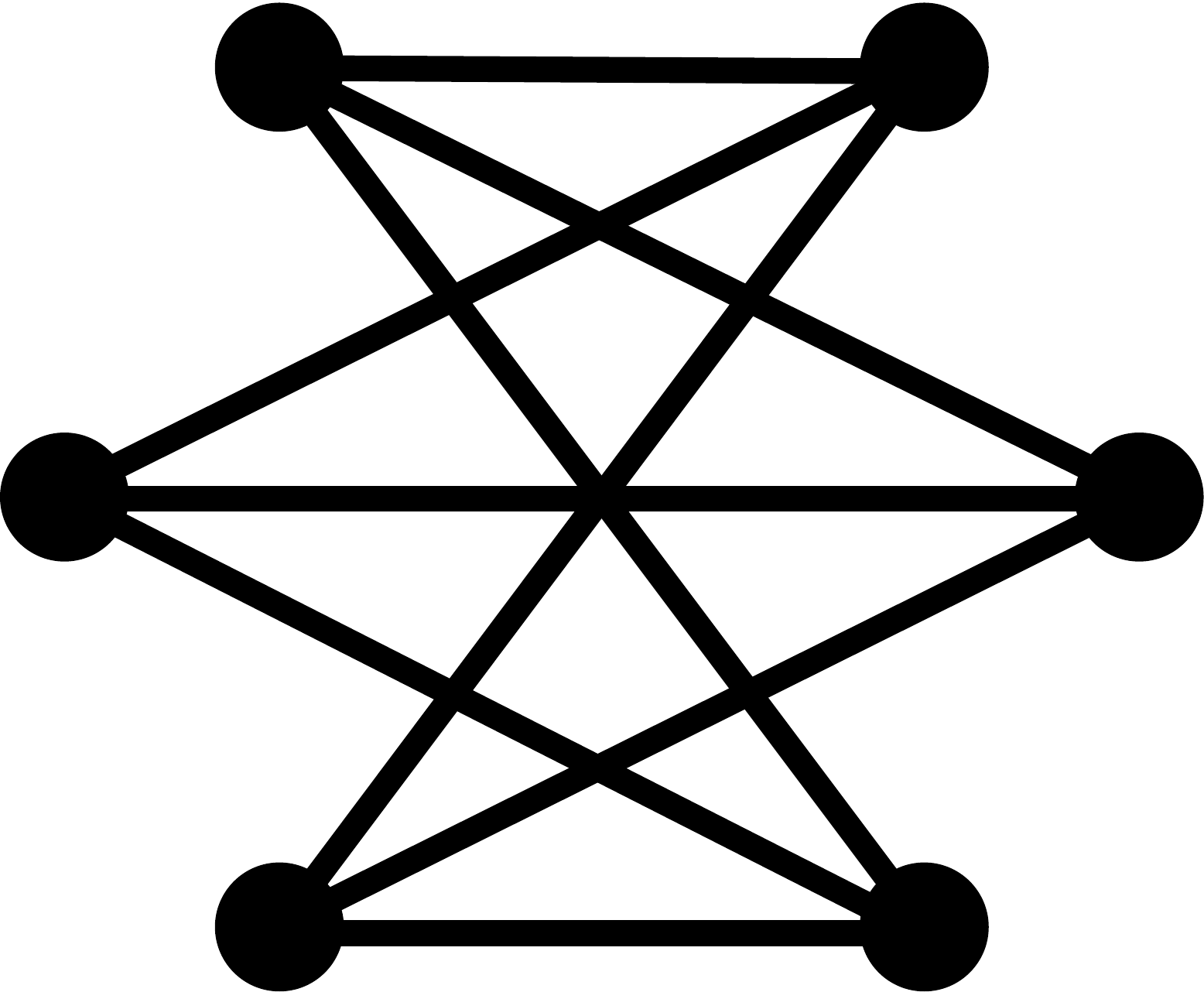}
    \end{minipage} \\
	\hline
	\( C_{n,s} \) & For \( n \geq 3 \) and \( s \in ( 0 , \frac{n-3}{n-1}) \), we define \( C_{n,s} \) as follows: start with \( C_n \) and add edges between non-adjacent nodes uniformly at random until there are \( sn \) extra edges on the cycle \( C_n \). &
	\begin{minipage}{.1\textwidth}
      \includegraphics[height=.6in]{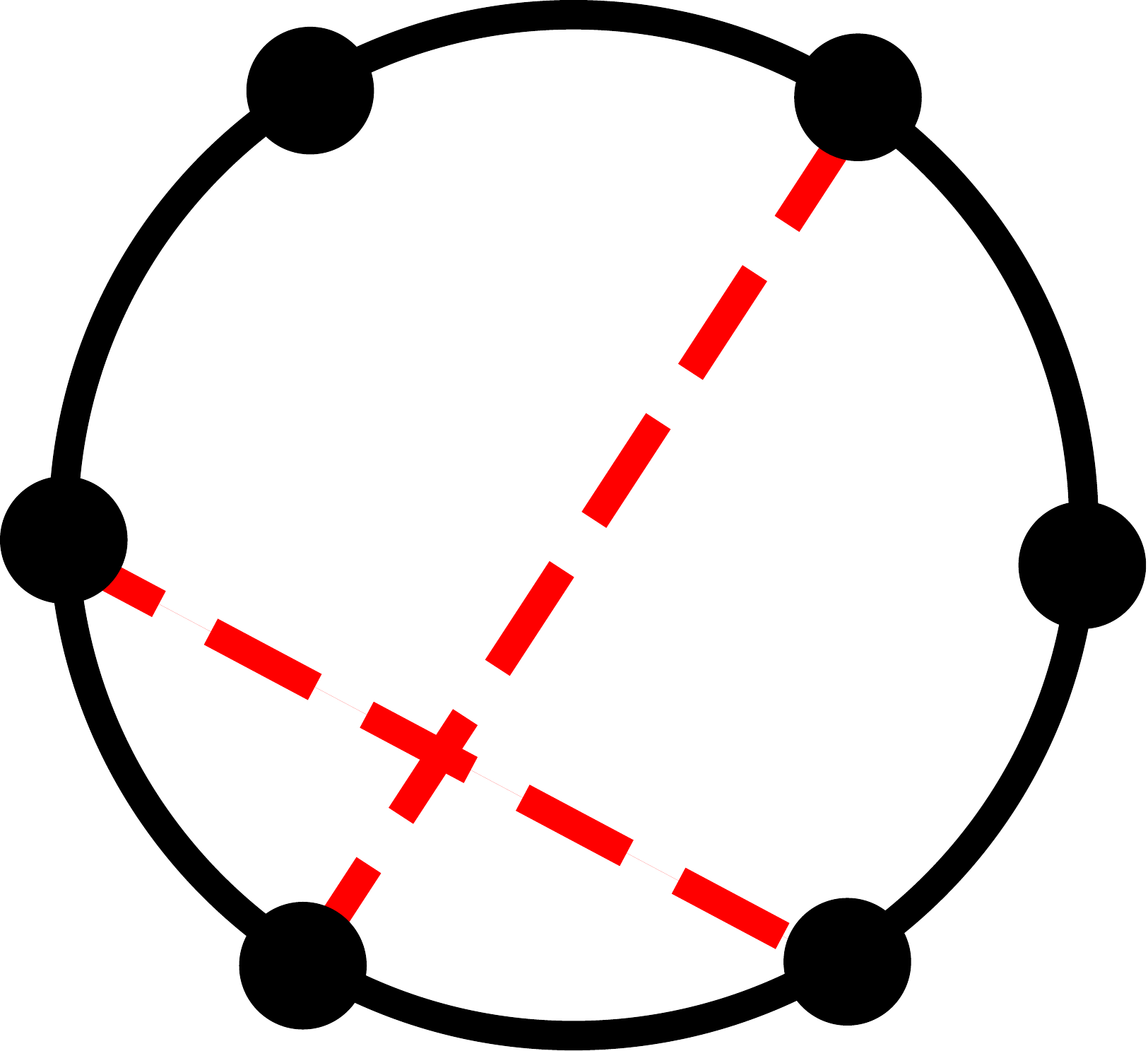}
    \end{minipage} \\
	\hline
	\( Y_{n,s} \) & For \( n \geq 3 \) and \( s \in (   0 , \frac{n-4}{n-1})  \), we define \( Y_{n,s} \) as follows: start with \( Y_n \) and add edges between non-adjacent nodes uniformly at random until there are \( sn \) extra edges on the prism graph \( Y_n \). &
	\begin{minipage}{.1\textwidth}
      \includegraphics[height=.6in]{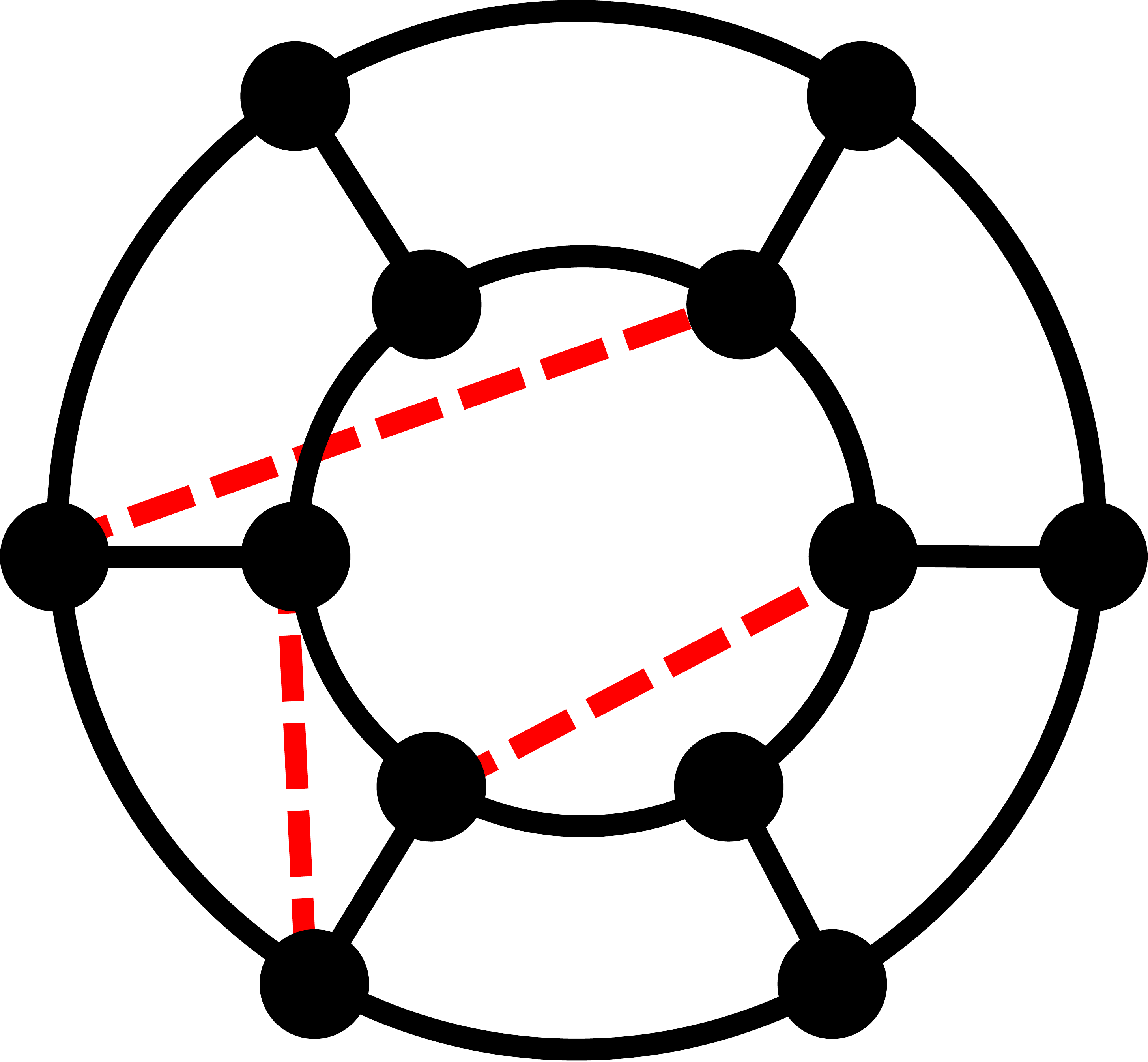}
    \end{minipage} \\
	\hline
	\( G(n,p) \) & For \( n \in \mathbb{N} \) and \( p \in [0,1] \), we generate random graphs from the \textit{Erd\H{o}s--R\'{e}nyi (ER) \( G(n,p) \) model} \cite{gilbert59} as follows: start with \( n \) disconnected nodes and place an edge between each distinct pair with independent probability \( p \). &
	\begin{minipage}{.1\textwidth}
      \includegraphics[height=.6in]{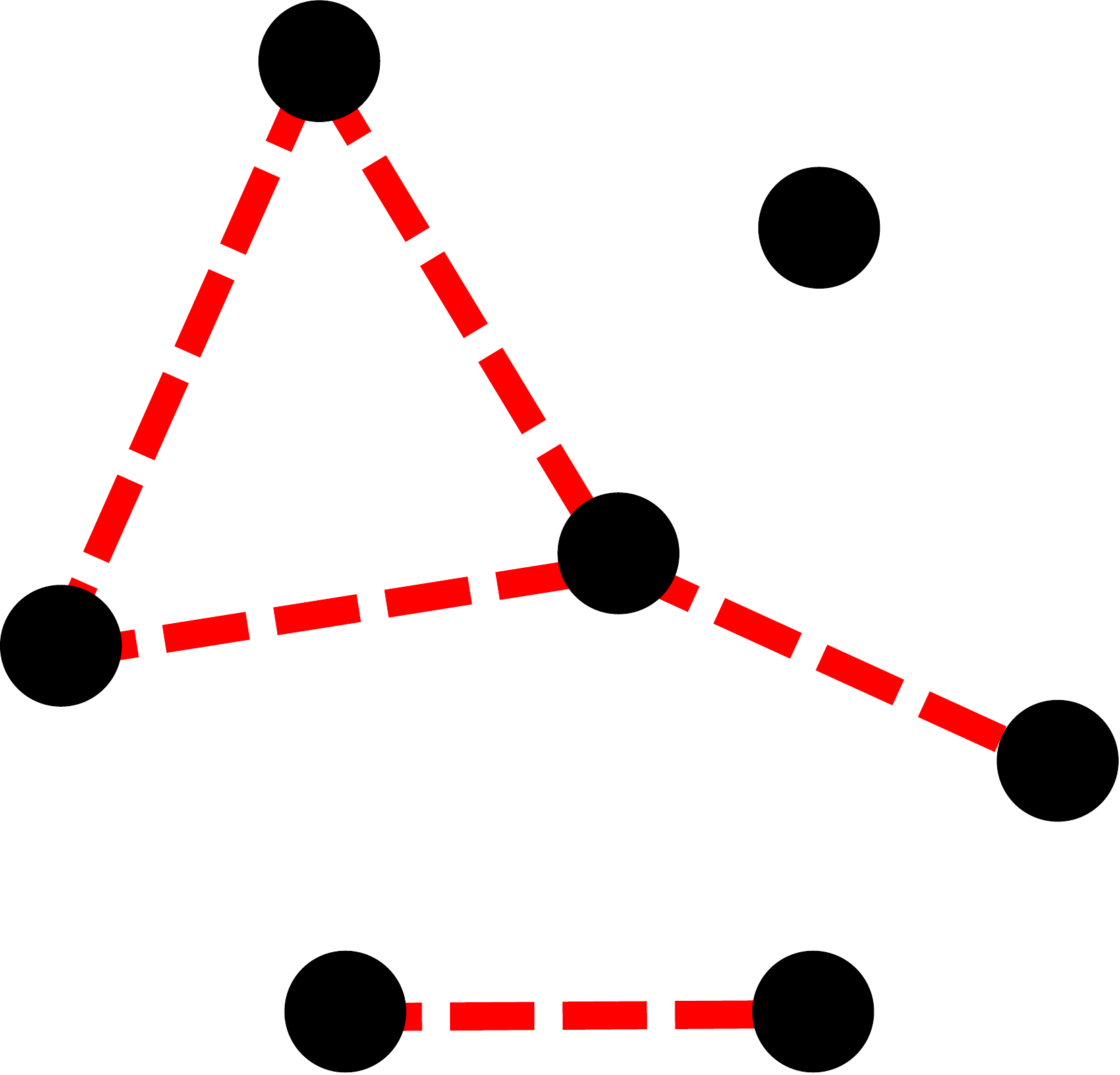}
    \end{minipage} \\
	\hline
	\end{tabular}
\caption{Summary of the definitions of the synthetic networks on which we study the Deffuant model. In each example network, solid black lines denote deterministic edges, and red dashed lines represent edges that are generated randomly.}\label{table:network}
\end{table*}


\subsection{Networks studied}\label{subsection:networks_studied}

We study the Deffuant model on a variety of networks to develop a better understanding of the effect of network structure on convergence time and the number of opinion groups at equilibrium. Some of the networks that we study have deterministic structures, and others are random graphs. In Table~\ref{table:network}, we list the notations, definitions, and examples of these networks. Finally, we conduct numerical simulations using networks that are constructed using Facebook ``friendship'' data \cite{traud12}. The first set of networks that we study are deterministic graphs, including complete graphs (\( K_n \)), cycles (\( C_n \)), prism graphs (\( Y_n \)), square lattices (\( S_l \)), and complete multipartite graphs (\( K_{n,r} \)). These networks have been studied extensively because of their simple structures. Our simulation results on these networks provide references for comparison with conclusions on the variants of the Deffuant model as well on those of the original Deffuant model on more complicated network structures. The second set of networks that we study consists of random graphs, which are cycles with random edges (\( C_{n,s} \)) (which are related to WS small-world networks \cite{watts98,smallworld-scholarpedia}), prism graphs with random edges (\( L_{n,s} \)), and random graphs generated by the Erd\H{o}s--R\'{e}nyi \( G(n,p) \) model. Finally, we investigate the Deffuant model on real social networks constructed using Facebook data.


\subsection{Simulation specifications}

Without loss of generality, we consider the Deffuant model with opinions on the space \( [0,1] \). In other words, we normalize the opinion dynamics so that each agent's opinion lies between \( 0 \) and \( 1 \) at any time step. We also consider the multiplier \( m \in (0,1) \), as opposed with the interval \( (0, 0.5] \) in the original model \cite{deffuant00}. This generalization is useful, as interacting agents can perhaps be convinced to believe in others' opinions more than their own. Moreover, considering \( m \in (0,1) \) reveals interesting phenomena that we will discuss in Section~\ref{sec:numerical_simulations}. A few of the parameter values have specific interpretations.
For example, for \( c=1 \), any pair of interacting agents makes convergent opinion adjustments that correspond to interaction without a confidence bound. For \( m=0.5 \), each pair of interacting agents agrees on their mean opinion whenever their opinion difference is below \( c \). Theoretically, there is no upper bound on the number of agents that one can consider in a population, but running numerical simulations on extremely large populations is computationally intensive. For our simulations, we use a maximum of \( N=1000 \) agents, and one can infer the behavior of the model for larger populations from our regression analysis.

The convergence time \( T \) and the number \( K \) of opinion groups at equilibrium are both difficult to predict, as the initial opinion profile, the pair of agents that interact at each time step, and the particular graphs generated by random graph ensembles are all stochastic. To smooth out these sources of noise, we run \( 10 \) groups of independent simulations for each network in Section~\ref{subsection:networks_studied} and each combination of the values of \( N \), \( c \), and \( m \) that we consider. During one simulation, we first generate a group of \( N \) independent and identically distributed initial opinions from a uniform distribution on \( [0,1] \), and we then simulate the evolution of opinion dynamics according to the Deffuant model.

In principle, equilibrium is reached only at infinitely long times, as the opinion space is continuous and opinions approach each other arbitrarily closely without reaching the same value in finite times unless \( m=0.5 \) \cite{laguna04}. However, the emergence of equilibrium is evident at finite times, as consecutive opinion groups must be separated by a distance of at least \( c \) to avoid merging. Therefore, in practice, we need to set a convergence criterion in our numerical simulations. For our study, we consider an opinion profile to be at equilibrium if consecutive opinion groups are separated by a distance of at least \( c \) and the range of opinions in each group is below \( 0.02 \). Based on some test runs, we also choose a bailout time of \( 10^9 \) iterations for each simulation. If an equilibrium is reached by the bailout time, we record the convergence time (\( T \)) and the number (\( K \)) of opinion groups. Otherwise, we record \( T=3.55 \times 10^9\), a strict upper bound that is higher than all possible convergence times, for the purpose of data visualization. 


\section{Numerical simulations and results}\label{sec:numerical_simulations}

In this section, we study the Deffuant model on various deterministic, randomly generated, and real-world networks by considering different network structures and interaction mechanisms between pairs of agents. For each network structure, we first conduct data exploration and linear regression analysis to model convergence time (\( T \)) as a function of the number (\( N \)) of participating agents,  confidence bound (\( c \)), and multiplier (\( m \)). We then discuss our qualitative observations about the number of opinion groups at equilibrium (\( K \)). Because the process of data exploration and regression analysis is similar, we only give full details in Appendix~\ref{appendix:complete_graphs} for a subset of the parameter space for our simulations on complete graphs.

For our linear regression analysis, we use the method of ordinary least squares, as the estimator is unbiased and consistent if the errors have the same finite variance and are uncorrelated with the explanatory variables \cite{freeman05}. If the errors are also normally distributed, ordinary least squares is also the maximum likelihood estimator \cite{freeman05}. We check these assumptions throughout our model-selection process. For each set of parameters and network structure that we consider, we conduct regression analysis using the mean results of \( 10 \) different simulations. We only use simulation results of networks with \( 100 \) or more agents in order to reduce the stochasticity introduced by the random initial opinion profile and to ensure a sufficient quantity of data for testing the model assumptions.


\subsection{Complete graphs}\label{subsection:complete_graphs}

The simplest form of the Deffuant model allows any pair of agents in a system to interact \cite{deffuant00}. This is equivalent to studying the model on a complete graph. Recall that $N$ denotes the number of nodes in a graph.

\begin{figure*}
	\centering
		\includegraphics[width=.75\textwidth]{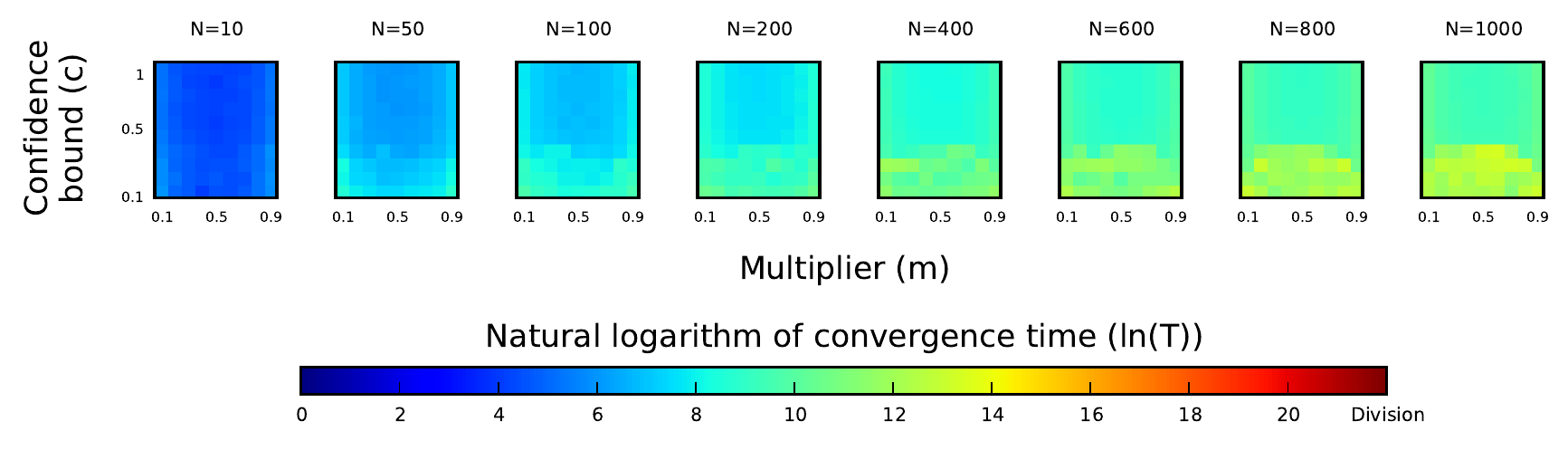}
	\caption{Convergence times for our simulations on $N$-node complete graphs for various $N$. These are representative of the trends that we observe in all simulations. (We generate this and all subsequent figures of this type using the {\scshape matplotlib} library for Python developed by J.~D. Hunter \cite{hunter07}.)}
\label{figure:complete_graphs_panel_plots}
\end{figure*}

In Fig.~\ref{figure:complete_graphs_panel_plots}, we summarize the values of \( \ln(T) \) that we observe in simulations for various $N$, as these are representative of the trends that we observe in all simulations. We present a similar set of plots for all other network structures in the following subsections.

\begin{figure*}
	\centering
		\includegraphics[width=.75\textwidth]{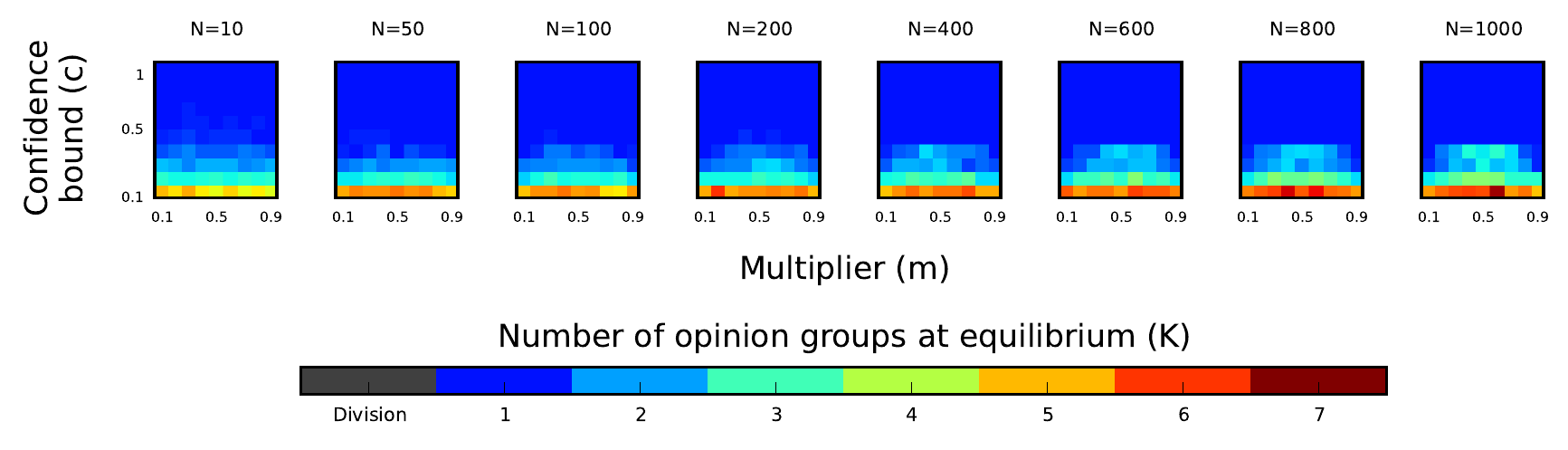}
	\caption{The number of opinion groups that persist at equilibrium in our simulations on complete graphs for various $N$. These are representative of the trends that we observe in all simulations. We use grey color to represent simulations that did not converge by the bailout time (\( 10^9 \) iterations) in this and our subsequent figures of this type.}
\label{figure:complete_graphs_opinion_cliques_panel_plots}
\end{figure*}

Our data exploration suggests that the convergence time has qualitatively different behavior for \( c<0.5 \) and \( c \geq 0.5 \), so we consider different regression models for these two cases. For model selection, we use the Akaike information criterion (AIC) \cite{akaike74} to select the ``best'' subset of predictors, as this method balances the trade-off between the goodness of fit and the complexity of a model. This model selection approach aims to minimize the AIC value, which is defined by
\begin{equation}
	\text{AIC} = 2\left[p-\ln(L)\right]\,,
\end{equation}
where \( p \) is the number of estimated parameters and \( L \) is the maximum value of the likelihood function for the model. The coefficient of determination, \( R^2 \in [0,1]\), is a measure of goodness of fit of a regression model \cite{draper81}. Values of \( R^2 \) that are closer to \( 1 \) indicate better fits. 
For instance, \mbox{\( R^2=0 \)}  implies that the response variable cannot be predicted from the explanatory variables, and \mbox{\( R^2=1 \)} implies that the response variable can be predicted without error from the explanatory variables. 
Let \( \hat{T}_i \) be the predicted value for the observed convergence time \( T_i \) (\( i=1,2, \dots , n\)), and let \( \bar{T} = (\sum_{i=1}^nT_i)/n \). One then calculates 
\begin{equation}
	R^2 = \frac{\sum_{i=1}^n(\hat{T}_i-\bar{T})^2}{\sum_{i=1}^n(T_i-\bar{T})^2}\,.
\label{eqn:R^2}
\end{equation} 
We use the AIC and \( R^2 \) to measure the goodness of fit and the simplicity of our regression models.

For \( c<0.5 \), AIC-based model selection yields 
\begin{equation}
	\ln(\ln(T)) = \beta_0 + \beta_1N + \beta_2N^2 + \beta_3c^2 + \beta_4Nc + \epsilon\,,
\label{model:complete_graph(lower)3}
\end{equation}
and we give our estimates for the coefficients \( \beta_j \) (with \( j=0,1, \dots,4 \)) in Table~\ref{table:complete_graph_summary(lower)3}. (This is part of the regression output given by the software environment \textsc{R} \cite{R08}.) The column for \( t \) values gives the values of the \( t \)-statistic for the hypothesis test with the null hypothesis that the corresponding regression coefficient is \( 0 \). The column for Pr(\( >|t| \)) gives the probability for a test statistic to be at least as extreme as the observed \( t \) value if the null hypothesis were true. A low value of Pr(\( >|t| \)) suggests that it would be rare to obtain a result as extreme as the observed value if the coefficient under consideration were \( 0 \), and hence we should keep the corresponding term in our model. For Eq.~\eqref{model:complete_graph(lower)3}, the values of the AIC and \( R^2 \) are \( -2037.1 \) and \( 0.8246 \), respectively.

\begin{table}
	\small
	\centering
	\begin{tabular}{ccccc}
	\hline
$\beta_n$	 & {Estimate} & {Std. Error} & {\( t \) value} & {Pr(\( >|t| \))} \\ 
	\hline
	\( \beta_0 \) & \( \phantom{-}2.139 \) & \( 1.380 \times 10^{-2} \) & \( \phantom{-}1.550 \times 10^2 \) & \( < 2 \times 10^{-16} \) \\ 
	\( \beta_1 \) & \( \phantom{-}7.124 \times 10^{-4} \) & \( 5.255 \times 10^{-5} \) & \( \phantom{-}1.356 \times 10 \) & \( < 2 \times 10^{-16} \) \\ 
	\( \beta_2 \) & \( -3.763 \times 10^{-7} \) & \( 4.178 \times 10^{-8} \) & \( -9.006 \) & \( < 2 \times 10^{-16} \) \\ 
	\( \beta_3 \) & \( -9.922 \times 10^{-1} \) & \( 1.076 \times 10^{-1} \) & \( -9.220 \) & \( < 2 \times 10^{-16} \) \\ 
	\( \beta_4 \) & \( \phantom{-}3.696 \times 10^{-4} \) & \( 8.983 \times 10^{-5} \) & \( \phantom{-}4.114 \) & \( 4.850 \times 10^{-5} \) \\ 
	\hline
	\end{tabular}
\caption{Estimates of regression coefficients for Eq.~\eqref{model:complete_graph(lower)3}. We present estimates for regression coefficients accurate to four significant figures in this and the following regression analyses unless we state otherwise.}\label{table:complete_graph_summary(lower)3}
\end{table}

For \( c \geq 0.5 \), regression analysis suggests the model
\begin{equation}
	\ln(T) = \beta_0 + \beta_1\ln(N) + \beta_2(c-1)^2 + \beta_3(m-0.5)^2 + \epsilon\,,
\label{model:complete_graph(upper)3}
\end{equation}
where we list our coefficient estimates in Table~\ref{table:complete_graph_summary(upper)3}. For Eq.~\eqref{model:complete_graph(upper)3}, the values of the AIC and \( R^2 \) are \( -3240.9 \) and \( 0.9964 \), respectively.

\begin{table}
	\small
	\centering
	\begin{tabular}{ccccc}
	\hline
	$\beta_n$ & {Estimate} &{Std. Error} & {\( t \) value} & {Pr(\( >|t| \))} \\ 
	\hline
	\( \beta_0 \) & \( 1.865 \)  & \( 1.916 \times 10^{-2} \) & \( 9.734 \times 10 \) & \( < 2 \times 10^{-16} \) \\ 
	\( \beta_1 \) & \( 1.062 \) & \( 3.067 \times 10^{-3} \) & \( 3.463 \times 10^2 \) & \( < 2 \times 10^{-16} \) \\ 
	\( \beta_2 \) & \( 4.530 \times 10^{-1} \) & \( 2.398 \times 10^{-2} \) & \( 1.889 \times 10 \) & \( < 2 \times 10^{-16} \) \\ 
	\( \beta_3 \) & \( 6.262 \) & \( 3.646 \times 10^{-2} \) & \( 1.718 \times 10^2 \) & \( < 2 \times 10^{-16} \) \\ 
	\hline
	\end{tabular}
\caption{Estimates of regression coefficients for Eq.~\eqref{model:complete_graph(upper)3}.}
\label{table:complete_graph_summary(upper)3}
\end{table}

The different forms of Eqs.~\eqref{model:complete_graph(lower)3} and \eqref{model:complete_graph(upper)3} confirm our conjecture based on data exploration that \( T \) undergoes a transition at \( c=0.5 \). More precisely, the regression results suggest that the behavior of \( T \) differs for \( c \leq 0.4 \) and \( c \geq 0.5 \). To determine a more precise transition point for \( c \), one should conduct numerical simulations using \( c \in (0.4, 0.5) \). For \( c<0.5 \), the multiplier \( m \) has no statistically significant impact on \( T \). Moreover, \( T \) increases with \( N \) for \( N < (\beta_1 + \beta_4c)/(2\beta_2) \), and it decreases with \( c<0.5 \). For \( c \geq 0.5 \), the effects of \( N \), \( c \), and \( m \) on \( T \) seem to be independent (or at least predominantly independent) of each other. In particular, \( T \) increases with \( N \) roughly linearly. We also observe that \( T \) increases with \( (c-1)^2 \) exponentially and has a minimum at \( c=1 \), which corresponds to interactions without a confidence bound. In other words, for fixed \( N \) and \( m \), the convergence time on complete graphs is minimal when any pair of interacting agents makes a convergent compromise. Furthermore, \( T \) increases with \( (m-0.5)^2 \) exponentially and has a minimum at \( m=0.5 \). This corresponds to the case in which each pair of interacting agents agrees at their mean opinion whenever their opinion difference is below the confidence bound.

For each combination of \( N \), \( c \), and \( m \), we average the number \( K \) of opinion groups at equilibrium if and only if at least \( 60\% \) of simulations reach equilibrium within the bailout time. Otherwise, we state that we observe a ``division'' of opinion for the associated parameter combination. We also use the same standard to determine the number of opinions at equilibrium in our subsequent numerical experiments.

In Fig.~\ref{figure:complete_graphs_opinion_cliques_panel_plots}, we summarize the number of opinion groups that persist at equilibrium in our simulations on complete graphs. We observe that \( K \) depends on \( N \) only when the confidence bound is \( c<0.5 \), with the most dramatic changes occurring in the region of \( c=0.1 \). For \( c \geq 0.5 \), consensus is reached consistently. For \( c \in [0.1, 0.4] \), we observe that \( K \) generally increases with \( N \). For \( c \in [0.2 , 0.4] \), we obtain \( K \in [1 , 4] \). At \( c=0.1 \), we observe that \( K \geq 5 \) for \( N \geq 200 \). Additionally, for \( c<0.5 \) and \( N \geq 600 \), we observe that \( K \) is generally larger for \( m \) closer to \( 0.5 \). This is reasonable because, as \( m \to 0.5 \), agents tend to agree on the mean of their opinions, which reduces the length of time for opinions to stabilize, so more opinion groups tend to persist at equilibrium.


\subsection{Cycles}\label{subsection:cycle}

\begin{figure*}
	\centering
		\includegraphics[width=.75\textwidth]{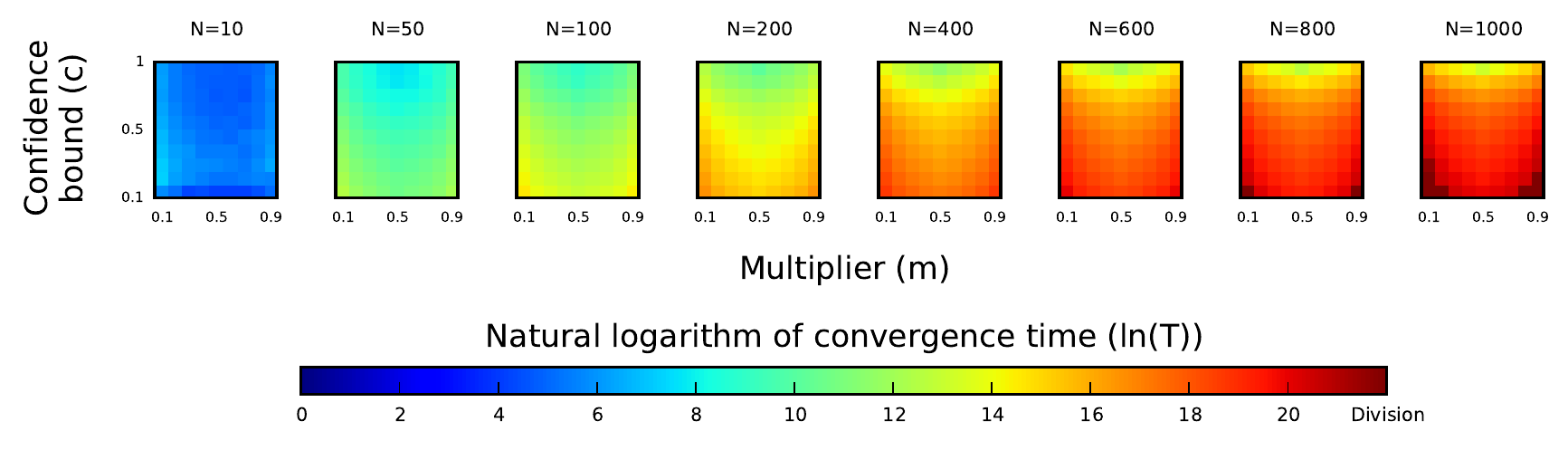}
	\caption{Convergence time for simulations on $N$-node cycles for various $N$.}
\label{figure:cycle_panel_plots}
\end{figure*}

\begin{figure*}
	\centering
		\includegraphics[width=.75\textwidth]{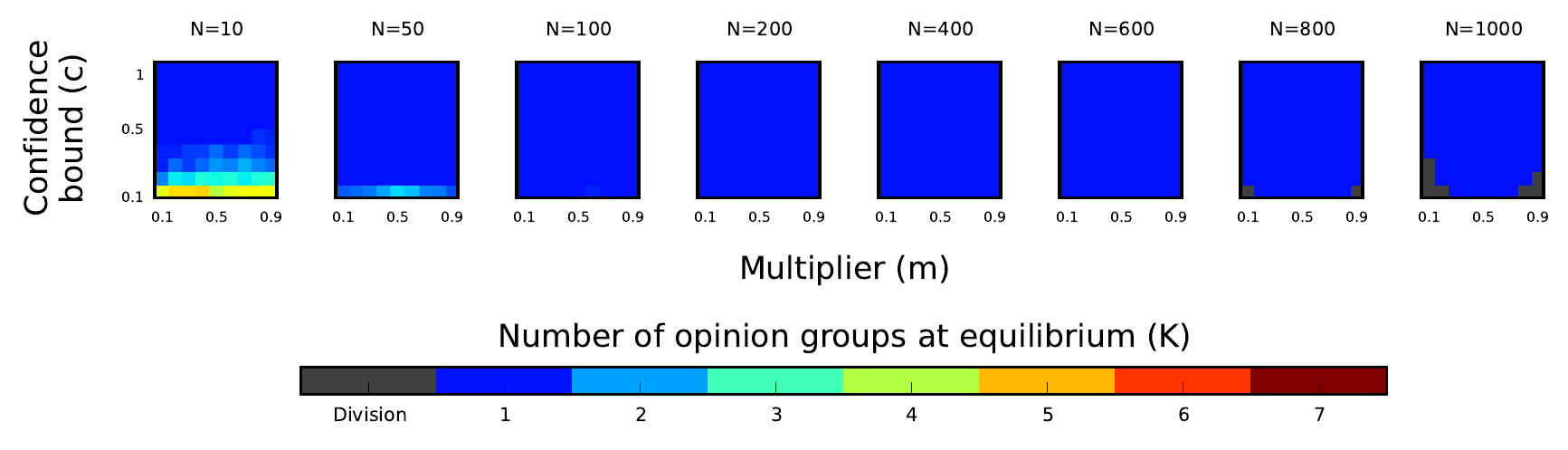}
	\caption{The number of opinion groups that persist at equilibrium in our simulations on cycles for various $N$.}
\label{figure:cycle_opinion_cliques_panel_plots}
\end{figure*}

In this subsection, we explore the behavior of convergence time and the number of opinion groups at equilibrium by simulating the Deffuant model on $N$-node cycles. We will compare these simulation results to ones on cycles with additional, randomly-placed ``shortcut'' edges in Section~\ref{subsection:cycle_random_edges}.

In Fig.~\ref{figure:cycle_panel_plots}, we summarize the values of \( \ln(T) \) that we observe in our simulations on cycles. Our simulations suggest that \( \ln(T) \) changes rapidly with \( m \) when \( c \) is close to \( 1 \). We speculate that a singularity arises at \( c=1 \) and \( m=0.5 \) as \( N \to \infty \). Our linear regression models cannot capture singular points, so we exclude data points that correspond to \( c \geq 0.7 \) from our regression analysis for cycles. Our regression analysis gives the model
\begin{equation}
	\ln(T) = \beta_0 + \beta_1\ln(N) + \beta_2c + \beta_3c^2 + \beta_4(m-0.5)^2 + \beta_5Nm + \epsilon\,,
\label{model:cycle}
\end{equation}
where we list our coefficient estimates in Table~\ref{table:cycle_summary}. For Eq.~\eqref{model:cycle}, we obtain \( R^2 \approx 0.9991 \) and \( \text{AIC} \approx -3257.6 \).

\begin{table}
	\small
	\centering
	\begin{tabular}{ccccc}
	\hline
	 & {Estimate} & {Std. Error} & {\( t \) value} & {Pr(\( >|t| \))} \\ 
	\hline
	  \( \beta_0 \) & \( -6.313 \times 10^{-1} \) & \( 3.054 \times 10^{-2} \) & \( -2.067 \times 10 \) & \( < 2 \times 10^{-16} \) \\
	  \( \beta_1 \) & \( \phantom{-}3.018 \) & \( 5.142 \times 10^{-3} \) & \( \phantom{-}5.870 \times 10^2 \) & \( < 2 \times 10^{-16} \) \\
	  \( \beta_2 \) & \( -2.630 \) & \( 6.357 \times 10^{-2} \) & \( -4.137 \times 10 \) & \( < 2 \times 10^{-16} \) \\
	  \( \beta_3 \) & \( -1.624 \) & \( 7.708 \times 10^{-2} \) & \( -2.107 \times 10 \) & \( < 2 \times 10^{-16} \) \\
	  \( \beta_4 \) & \( \phantom{-}9.371 \) & \( 4.669 \times 10^{-2} \) & \( \phantom{-}2.007 \times 10^2 \) & \( < 2 \times 10^{-16} \) \\
	  \( \beta_5 \) & \( -7.642 \times 10^{-5} \) & \( 1.713 \times 10^{-5} \) & \( -4.461 \) & \( 9.770 \times 10^{-6} \) \\
   \hline
   \end{tabular}
\caption{Estimates of regression coefficients for Eq.~\eqref{model:cycle}.}\label{table:cycle_summary}
\end{table}

In contrast to complete graphs, our simulations on cycles indicate that the dependence of \( T \) on \( N \), \( c \), and \( m \) does not undergo a transition with respect to \( c \). We observe that \( T \) decreases with \( c \in (0,1] \), and, as \( c \) gets closer to \( 1 \), the value of \( \ln(T) \) changes with \( m \) more rapidly as \( N \) increases. Moreover, for \( N<-\beta_4/\beta_5 \), the convergence time \( T \) obtains a global minimum at \( m =0.5-(\beta_5N)/(2\beta_4) \in (0.5,1) \) if \( N \) and \( c \) are held constant. Furthermore, \( T \) increases with \( N \) for \( N<-\beta_1/(\beta_5m) \). Additionally, the effects of \( N \) and \( m \) on \( T \) appear to be weakly coupled.

In Fig.~\ref{figure:cycle_opinion_cliques_panel_plots}, we summarize the number of opinion groups that arise in our simulations on cycles. A consensus is reached for \( N \in [100 , 700] \). Athough some of our simulations for \( N \in [800 , 1000] \) do not converge by the bailout time, we conjecture that all simulations on cycles with large values of \( N \) will eventually converge, independent of the values of \( c \) and \( m \), if the Deffuant dynamics are continued for sufficiently many iterations. A consensus is reached when \( c \geq 0.5 \) for \( N=10 \) and when \( c \geq 0.2 \) for \( N=50 \). This observation is reasonable as, with fewer agents adjacent to each other on a cycle, their initial opinions are more disperse, which compels them to form more groups. Similar to complete graphs, we observe that more opinion groups tend to emerge in the final state as \( m \to 0.5 \) if multiple opinion groups persist at equilibrium.


\subsection{Prism graphs}\label{subsection:cylindrical_lattice}

In this subsection, we explore the behavior of convergence time and the number of opinion groups at equilibrium by simulating the Deffuant model on prism graphs. Prism graphs are a special type of \textit{generalized Petersen graph} \cite{coxeter50}. We will compare our simulation results on prism graphs to those on prisms with additional random edges in Section~\ref{subsec:prism_random_edges}.

\begin{figure*}
	\centering
		\includegraphics[width=.75\textwidth]{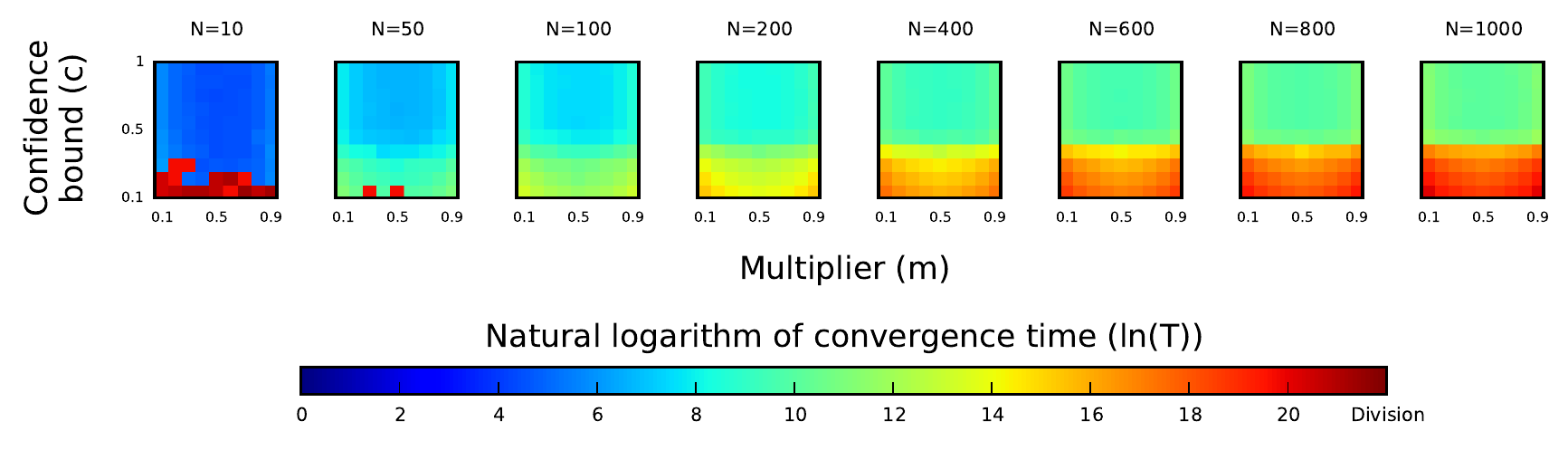}
	\caption{Convergence times for our simulations on $N$-node prism graphs for various $N$.}
\label{figure:annular_lattice_panel_plots}
\end{figure*}

\begin{figure*}
	\centering
		\includegraphics[width=.75\textwidth]{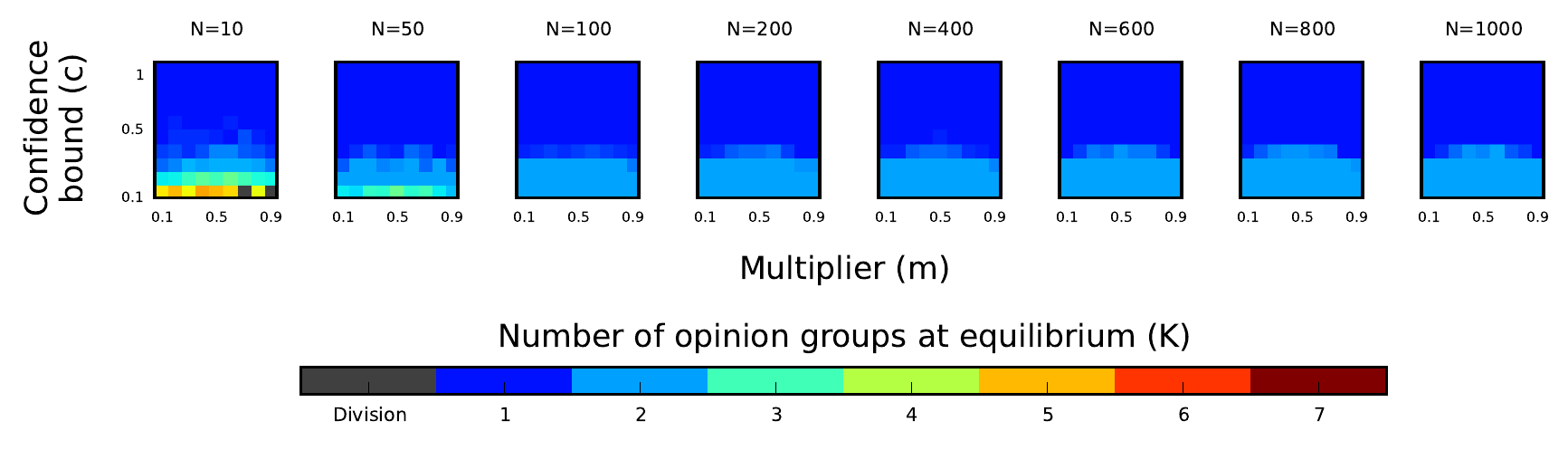}
	\caption{The number of opinion groups that persist at equilibrium in our simulations on prism graphs for various $N$.}
\label{figure:annular_lattice_opinion_cliques_panel_plots}
\end{figure*}

\begin{table}
	\small
	\centering
	\begin{tabular}{ccccc}
	\hline
	 & {Estimate} & {Std. Error} & {\( t \) value} & {Pr(\( >|t| \))} \\ 
	\hline
	  \( \beta_0 \) & \( \phantom{-}1.062 \times 10^2 \) & \( 2.803 \) & \( \phantom{-}3.789 \times 10 \) & \( < 2 \times 10^{-16} \) \\
	  \( \beta_1 \)  & \( \phantom{-}4.319 \times 10^{-1} \) & \( 6.206 \times 10{-3} \) & \( \phantom{-}6.960 \times 10 \) & \( < 2 \times 10^{-16} \) \\
	  \( \beta_2 \) & \( -1.790 \times 10^{-4} \) & \( 4.822 \times 10{-6} \) & \( -3.712 \times 10 \) & \( < 2 \times 10^{-16} \) \\
	  \( \beta_3 \) & \(  \phantom{-}7.759 \times 10\) & \( 1.830 \times 10 \) & \( \phantom{-}4.239 \) & \( 2.890 \times 10^{-5} \) \\
	  \( \beta_4 \) & \( -5.946 \times 10^2 \) & \( 3.400 \times 10 \) & \( -1.749 \times 10 \) & \( < 2 \times 10^{-16} \) \\
	  \( \beta_5 \) & \( \phantom{-}2.839 \times 10^2 \) & \( 5.924 \) & \( \phantom{-}4.792 \times 10 \) & \( < 2 \times 10^{-16} \) \\
	  \( \beta_6 \) & \(  -1.332 \times 10^{-1}\) & \( 1.083 \times 10{-2} \) & \( -1.230 \times 10 \) & \( < 2 \times 10^{-16} \) \\
   \hline
\end{tabular}
\caption{Estimates of regression coefficients for Eq.~\eqref{model:annular_lattice(lower)}.}\label{table:annular_lattice_summary(lower)}
\end{table}

\begin{table}
	\small
	\centering
	\begin{tabular}{ccccc}
	\hline
	 & {Estimate} & {Std. Error} & {\( t \) value} & {Pr(\( >|t| \))} \\ 
	\hline
	  \( \beta_0 \) & \( \phantom{-}2.263 \) & \( 2.634 \times 10^{-2} \) & \( \phantom{-}8.592 \times 10 \) & \( < 2 \times 10^{-16} \) \\
	  \( \beta_1 \) & \( \phantom{-}2.072 \times 10^{-1} \) & \( 3.160 \times 10^{-3} \) & \( \phantom{-}6.556 \times 10 \) & \( < 2 \times 10^{-16} \) \\
	  \( \beta_2 \) & \( -1.212 \) & \( 4.976 \times 10^{-2} \) & \( -2.436 \times 10 \) & \( < 2 \times 10^{-16} \) \\
	  \( \beta_3 \) & \( \phantom{-}7.507 \times 10^{-1} \) & \( 3.273 \times 10^{-2} \) & \( \phantom{-}2.294 \times 10 \) & \( < 2 \times 10^{-16} \) \\
	  \( \beta_4 \) & \( \phantom{-}1.064 \) & \( 1.395 \times 10^{-2} \) & \( \phantom{-}7.624 \times 10 \) & \( < 2 \times 10^{-16} \) \\
	  \( \beta_5 \) & \( -6.056 \times 10^{-5} \) & \( 9.865 \times 10^{-6} \) & \( -6.138 \) & \( 1.650 \times 10^{-9} \) \\
   \hline
\end{tabular}
\caption{Estimates of regression coefficients for Eq.~\eqref{model:annular_lattice(upper)}.}\label{table:annular_lattice_summary(upper)}
\end{table}

In Fig.~\ref{figure:annular_lattice_panel_plots}, we summarize the values of \( \ln(T) \) that we observe in our simulations on prism graphs. Similar to our computations for complete graphs in Section~\ref{subsection:complete_graphs}, scatter plots of \( \ln(T) \) versus \( N \), \( c \), and \( m \), exhibit qualitatively distinct behavior for \( c<0.5 \) and \( c \geq 0.5 \). We thus conduct separate regression analyses for \( c<0.5 \) and \( c \geq 0.5 \).
For \(c < 0.5\), regression analysis suggests the model
\begin{equation}\begin{aligned}
	\ln(T)^2 & = \beta_0 + \beta_1N + \beta_2N^2 + \beta_3c + \beta_4c^2 + \beta_5(m-0.5)^2 \\
& \qquad + \beta_6Nc + \epsilon\,,
\label{model:annular_lattice(lower)}
\end{aligned}\end{equation}
where we list our coefficient estimates in Table~\ref{table:annular_lattice_summary(lower)}. For Eq.~\eqref{model:annular_lattice(lower)}, we obtain \( R^2 \approx 0.9919 \) and \( \text{AIC} \approx 1301.9 \).
For \(c \geq 0.5\), regression analysis suggests the model
\begin{equation}\begin{aligned}
	\sqrt{\ln(T)} & = \beta_0 + \beta_1\ln(N) + \beta_2c + \beta_3c^2 + \beta_4(m-0.5)^2 \\
& \qquad + \beta_5Nc + \epsilon\,,
\label{model:annular_lattice(upper)}
\end{aligned}\end{equation}
where we list our coefficient estimates in Table~\ref{table:annular_lattice_summary(upper)}. For Eq.~\eqref{model:annular_lattice(upper)}, we obtain \( R^2 \approx 0.9845 \) and \( \text{AIC} \approx -4219.1 \).

Similar to the case of complete graphs, the different forms of Eqs.~\eqref{model:annular_lattice(lower)} and \eqref{model:annular_lattice(upper)} confirm our conjecture based on data exploration that \( T \) undergoes a transition at \( c=0.5 \). According to Eqs.~\eqref{model:annular_lattice(lower)} and \eqref{model:annular_lattice(upper)}, \( T \) increases with \( N < -(\beta_1+\beta_6c)/(2\beta_2) \) for \( c<0.5 \), and \( T \) increases with \( N<-\beta_1/(\beta_5c) \) for \( c \geq 0.5 \). The effects of \( N \), \( c \), and \( m \) on \( T \) are coupled to each other for prism graphs. Additionally, \( T \) increases with respect to \( N \) more rapidly for \( c<0.5 \) than for \( c \geq 0.5 \). For \( c<0.5 \) and \( N \geq -\beta_3/\beta_6 \), the convergence time \( T \) decreases with \( c \in (0,1] \) if \( N \) and \( m \) are held constant. For \( c<0.5 \) and \( N < -\beta_3/\beta_6 \), however, \( T \) obtains a global maximum at \( c = -(\beta_3 + \beta_6N)/(2\beta_4) \in (0,0.5) \) if \( N \) and \( m \) are held constant. For \( c \geq 0.5 \) and \( N \geq -(2\beta_3+\beta_2)/\beta_5 \), the convergence time \( T \) decreases with \( c \in (0,1] \) if \( N \) and \( m \) are held constant. For \( c \geq 0.5 \) and \( N < -(2\beta_3+\beta_2)/\beta_5 \), however, \( T \) obtains a global minimum at \( c=-(\beta_2+\beta_5N)/(2\beta_3) \in (0.5,1) \) if \( N \) and \( m \) are held constant. With fixed values of \( N \) and \( c \), the convergence time \( T \) obtains a global minimum at \( m=0.5 \).

In Fig.~\ref{figure:annular_lattice_opinion_cliques_panel_plots}, we summarize the number of opinion groups that persist in our simulations on prism graphs. For \( c \geq 0.5 \), a consensus is reached for all simulations on prism graphs. For \( c<0.5 \), the equilibrium state is mostly polarized into \( 2 \) distinct opinion groups if \( N \geq 100 \) and can sometimes have more than \( 2 \) opinion groups for \( N \in \{10, 50\} \). Similar to our simulations on cycles in Section~\ref{subsection:cycle}, we observe that large discrepancies in the initial opinion distribution hinder the agents from agreeing with each other through their interactions on a prism graph.


\subsection{Square lattices}\label{subsection:square_lattice}

Apart from complete graphs, square lattices are the most common deterministic networks on which the Deffuant model has been studied previously \cite{weisbuch02}.

\begin{figure*}
	\centering
		\includegraphics[width=.75\textwidth]{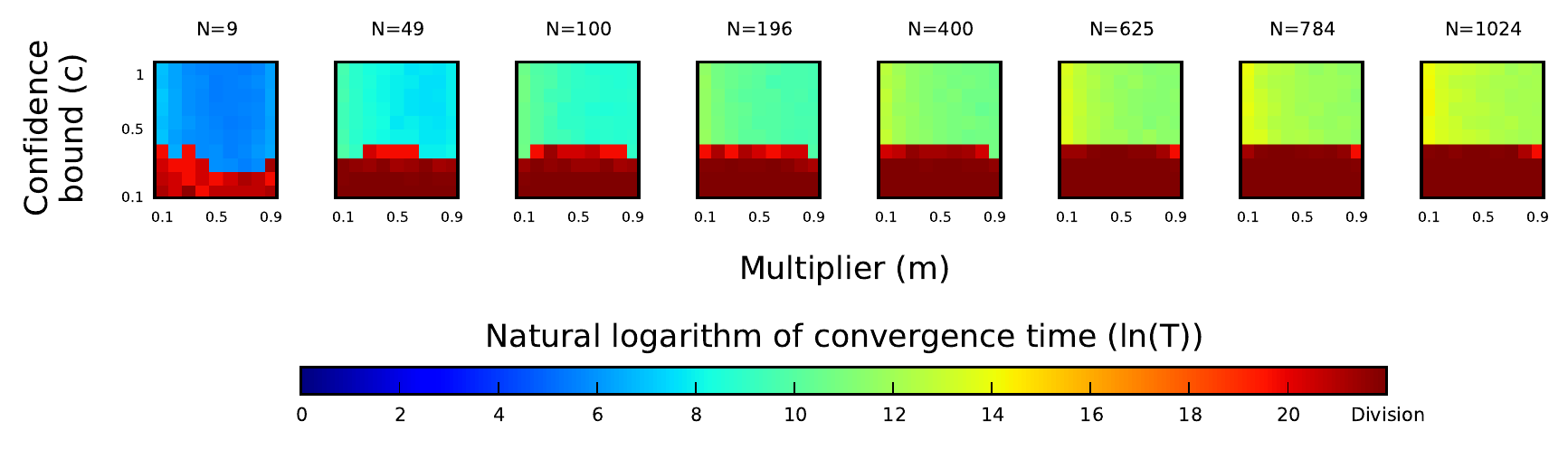}
	\caption{Convergence times for our simulations on various square lattices.}
\label{figure:square_lattice_panel_plots}
\end{figure*}

\begin{figure*}
	\centering
		\includegraphics[width=.75\textwidth]{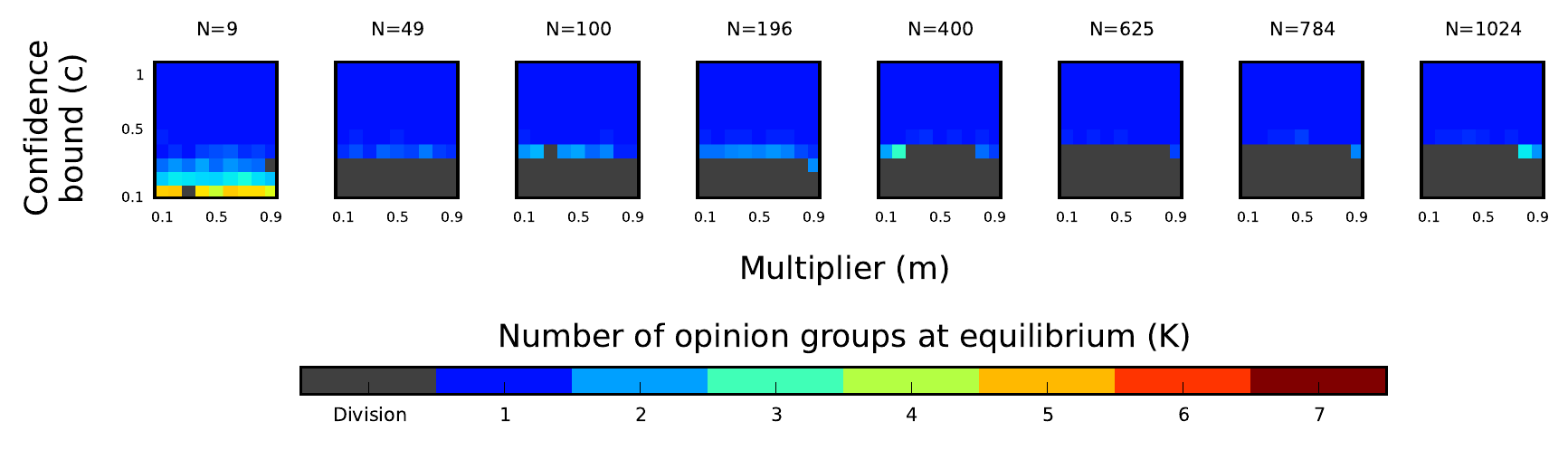}
	\caption{Summary of the number of opinion groups that persist at equilibrium in our simulations on square lattices.}
\label{figure:square_lattice_opinion_cliques_panel_plots}
\end{figure*}

In Fig.~\ref{figure:square_lattice_panel_plots}, we summarize the values of \( \ln(T) \) that we observe in our simulations on square lattices. For \( c<0.5 \), most of the simulations do not converge by the bailout time, so we conduct regression analysis for \( c \geq 0.5 \).
Our regression analysis suggests for \( c \geq 0.5 \) that
\begin{equation}\begin{aligned}
	\ln(T)^{1/4} & = \beta_0 + \beta_1\ln(N) + \beta_2N + \beta_3N^2 + \beta_4c + \beta_5c^2 \\
& \qquad + \beta_6m + \beta_7m^2 + \beta_8Nm + \epsilon\,,
\label{model:square_lattice}
\end{aligned}\end{equation}
where we list our coefficient estimates in Table~\ref{table:square_lattice_summary}. For Eq.~\eqref{model:square_lattice}, we obtain \( R^2 \approx 0.9908 \) and \( \text{AIC} \approx -5684.4 \).

Similar to our observations for prism graphs, the effects of \( N \), \( c \), and \( m \) on \( T \) are coupled to each other for square lattices. According to Eq.~\eqref{model:square_lattice}, \( T \) increases with \( N \). If \( N \geq -\beta_6/\beta_8 \), the convergence time \( T \) increases with \( m\in (0,1) \). Otherwise, \( T \) obtains a global minimum at \( m=-(\beta_6+\beta_8N)/(2\beta_7) \in (0,1) \) if \( N \) and \( c \) are held constant. Moreover, \( T \) has a minimum at \( c=-\beta_4/(2\beta_5) \in (0.5,1) \) if \( N \) and \( m \) are held constant.

\begin{table}
	\small
	\centering
	\begin{tabular}{ccccc}
	\hline
	 & {Estimate} & {Std. Error} & {\( t \) value} & {Pr(\( >|t| \))} \\ 
	\hline
	  \( \beta_0 \) & \( \phantom{-}1.505 \) & \( 1.194\times 10^{-2} \) & \( \phantom{-}1.261\times 10^2 \) & \( <2\times 10^{-16} \) \\
	  \( \beta_1 \)  & \( \phantom{-}7.361 \times 10^{-2} \) & \( 2.550\times 10^{-3} \) & \( \phantom{-}2.887\times 10 \) & \( <2\times 10^{-16} \) \\
	  \( \beta_2 \) & \( -4.931\times 10^{-5} \) & \( 1.358\times 10^{-5} \) & \( -3.630 \) & \( 3.110\times 10^{-4} \) \\
	  \( \beta_3 \) & \( \phantom{-}1.336\times 10^{-8} \) & \( 7.154\times 10^{-9} \) & \( \phantom{-}1.868 \) & \( 6.227\times 10^{-2} \) \\
	  \( \beta_4 \) & \( -5.129\times 10^{-2} \) & \( 1.288\times 10^{-2} \) & \( -3.982 \) & \( 7.790\times10^{-5} \) \\
	  \( \beta_5 \) & \( \phantom{-}3.154\times 10^{-2} \) & \( 8.543\times 10^{-3} \) & \( \phantom{-}3.692 \) & \( 2.460\times10^{-4} \) \\
	  \( \beta_6 \) & \( -2.430\times 10^{-1} \) & \( 4.093\times 10^{-3} \) & \( -5.938\times 10 \) & \( <2\times 10^{-16} \) \\
	  \( \beta_7 \) & \( \phantom{-}1.396\times10^{-1} \) & \( 3.654\times 10^{-3} \) & \( \phantom{-}3.821\times 10 \) & \( <2\times 10^{-16} \) \\
	  \( \beta_8 \) & \( \phantom{-}1.544\times10^{-5} \) & \( 2.862\times 10^{-6} \) & \( \phantom{-}5.395 \) & \( 1.030\times 10^{-7} \) \\
   \hline
\end{tabular}
\caption{Estimates of regression coefficients for Eq.~\eqref{model:square_lattice}.}\label{table:square_lattice_summary}
\end{table}

In Fig.~\ref{figure:square_lattice_opinion_cliques_panel_plots}, we summarize the number of opinion groups that persist at equilibrium in our simulations on square lattices. As with our results on prism graphs, a consensus is reached for all simulations on square lattices for \( c \geq 0.5 \).


\subsection{Complete multipartite graphs}\label{subsection:complete_multipartite_graph}

In this subsection, we consider complete multipartite graphs \( K_{n,r} \) with \( r=n \). We use the values \( r=2,5,10 \), and we note that one construe a complete graph \( K_n \) (see Section \ref{subsection:complete_graphs}) as a complete multipartite graph \( K_{n,r} \) with \( r=n \). By varying the value of \( r \), we explore the effect of network density (i.e., the ratio of the number of edges to the maximum possible number of edges \cite{west01}) on the behavior of the Deffuant model.

\begin{figure*}
	\centering
		\includegraphics[width=.75\textwidth]{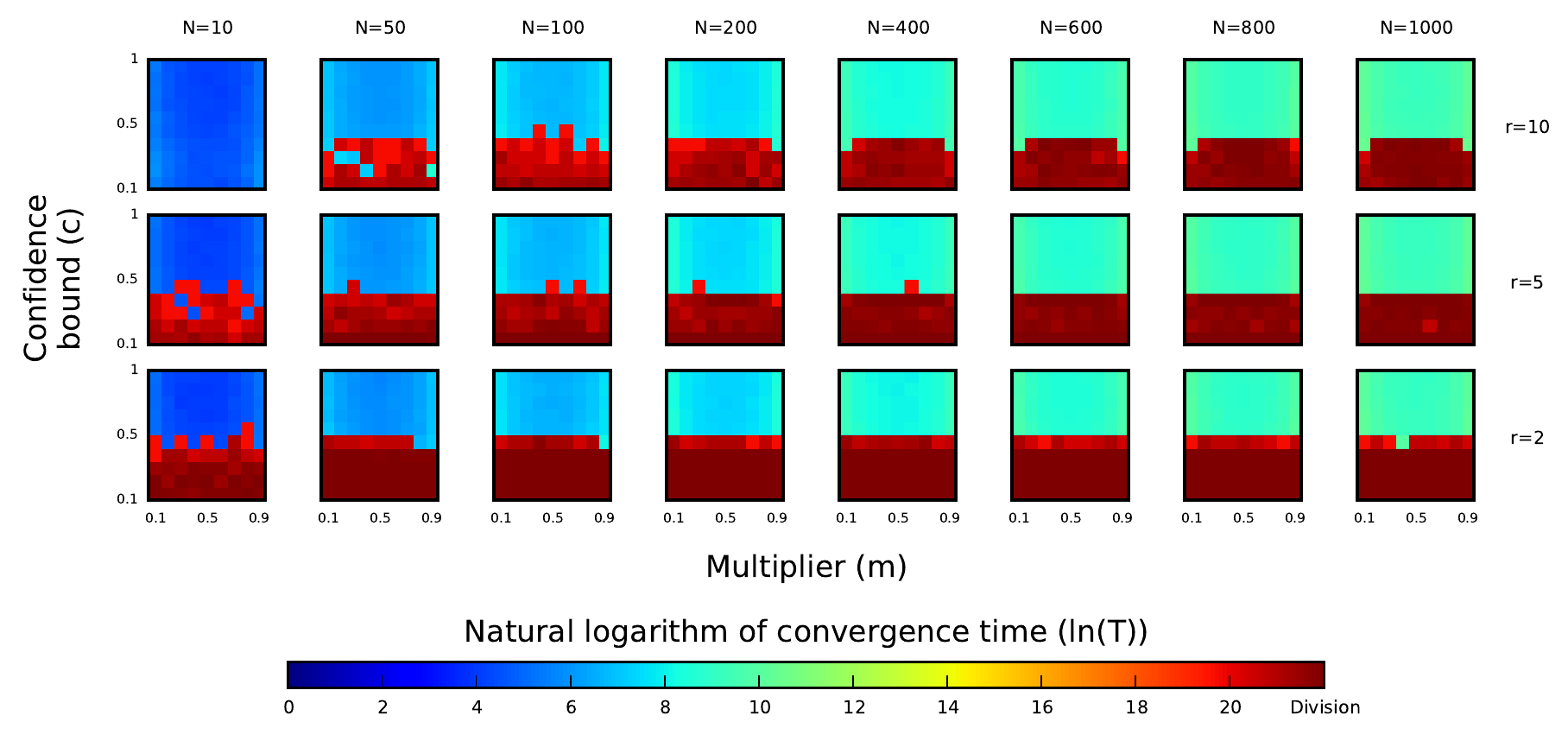}
	\caption{Convergence times for our simulations on $N$-node complete \( r \)-partite graphs (with \( r=2,5,10 \)) for various $N$.
	}
\label{figure:complete_multipartite_graph_panel_plots}
\end{figure*}

\begin{figure*}
	\centering
		\includegraphics[width=.75\textwidth]{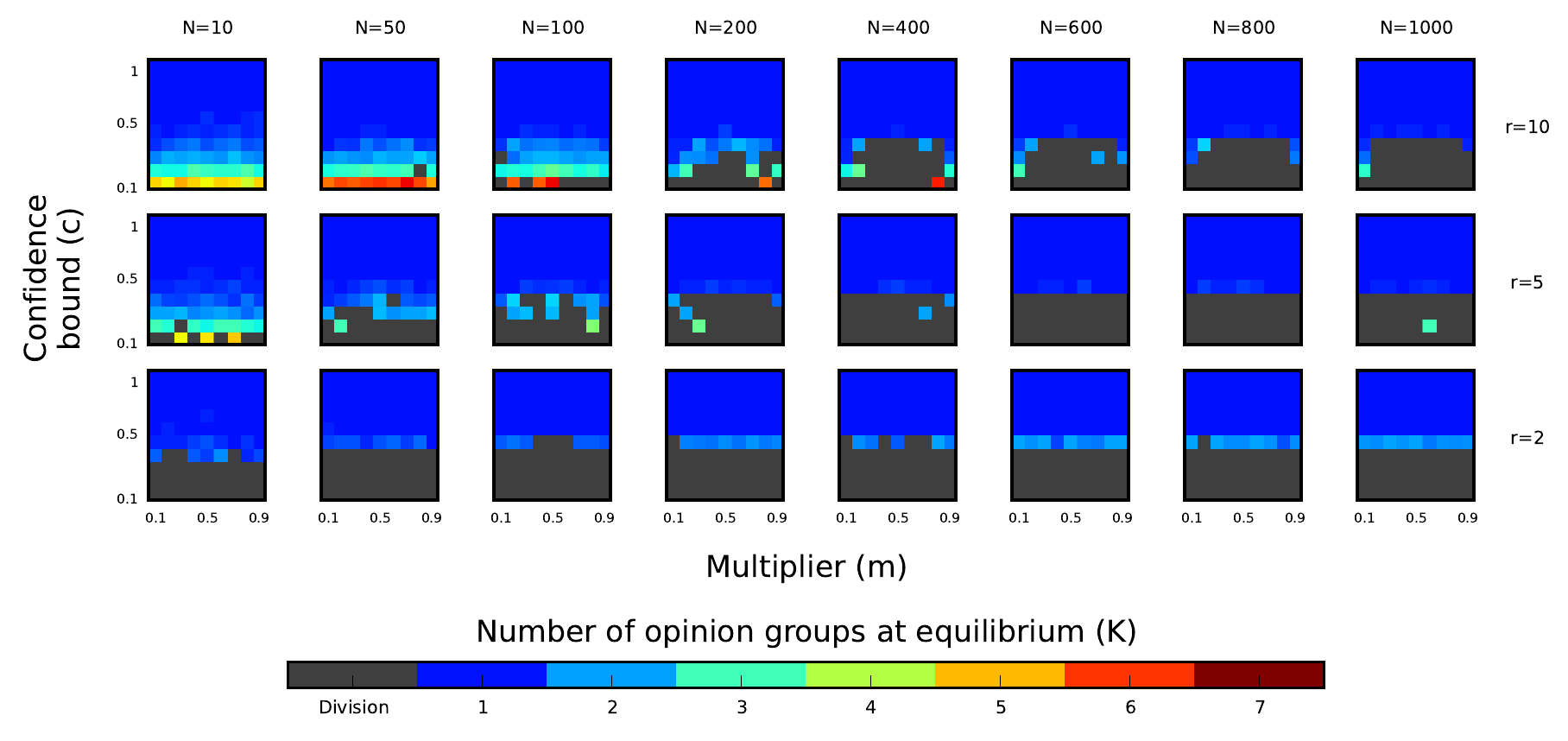}
	\caption{The number of opinion groups that persist at equilibrium in our simulations on complete \( r \)-partite graphs (with \( r=2,5,10 \)) for various $N$.}
\label{figure:complete_multipartite_graph_opinion_cliques_panel_plots}
\end{figure*}

In Fig.~\ref{figure:complete_multipartite_graph_panel_plots}, we summarize the values of \( \ln(T) \) that we observe in our simulations on complete \( r \)-partite graphs (with \( r=2,5,10 \)) and number \( N\) of nodes with values \(10 \), \( 50 \), \( 100 \), \( 200\), \( 400 \), \( 600\), \( 800 \), and \( 1000 \). For \( c<0.6 \), most of our simulations do not converge by the bailout time (\( 10^9 \) iterations), so we conduct regression analysis for \( c \geq 0.6 \).
For \( c \geq 0.6 \), our regression analysis suggests the model given by Eq.~\eqref{model:complete_graph(upper)3}, which has the same form as the regression model of complete graphs when \( c\geq 0.5 \) but has different coefficient values (see Table~\ref{table:complete_multipartite_graph_coefficients}). Note that one can construe a complete graph of size \( N \) as a complete \( N \)-partite graph.

\begin{table}
	\small
	\centering
	\begin{tabular}{ccccccc}
	\hline
	{\( r \)} & {\( \beta_0 \)} & {\( \beta_1 \)} & {\( \beta_2 \)} & {\( \beta_3 \)} & {AIC} & {\( R^2 \)} \\ 
	\hline
	\( 2 \) & \( 1.489 \) & \( 1.093 \) & \( 4.843\times10^{-1} \) & \( 6.553 \)  & \( -2675.68 \) & \( 0.9966 \) \\
	\( 5 \) & \( 1.672 \) & \( 1.074 \) & \( 3.252\times10^{-1} \) & \( 6.415 \)  & \( -2726.16 \) & \( 0.9969 \) \\
	\( 10 \) & \( 1.763 \) & \( 1.068 \) & \( 3.330\times10^{-1} \) & \( 6.305 \)  & \( -2709.59 \) & \( 0.9965 \) \\
	   \hline
	\end{tabular}
\caption{Estimates of regression coefficients, AIC values, and coefficients of determination (\( R^2 \)) for Eq.~\eqref{model:complete_graph(upper)3} using our simulation results on complete \( r \)-partite graphs with \( r=2,5,10 \).}\label{table:complete_multipartite_graph_coefficients}
\end{table}

The regression model in Eq.~\eqref{model:complete_graph(upper)3} suggests that the behavior of the convergence time on a general multipartite graph is similar to that on a complete graph. As the number \( r \) of partite sets increases, the growth rate of \( T \) with respect to \( N \) decreases slightly if \( c \) and \( m \) are held constant. In other words, as a complete multipartite graph becomes more densely connected, adding agents to a network increases the convergence time of the Deffuant model at a slower rate if all other conditions remain the same. Additionally, \( T \) increases with \( (m-0.5)^2 \) more slowly as \( r \) increases.

In Fig.~\ref{figure:complete_multipartite_graph_opinion_cliques_panel_plots}, we summarize the number of opinion groups that persist at equilibrium in our simulations on complete \( r \)-partite graphs (with \( r=2,5,10 \)). For \( r \in \{5,10\} \), consensus is reached for all \( c \geq 0.5 \). We obtain consensus in all of our simulations on bipartite graphs with \( c \geq 0.6 \), whereas some simulations fail to converge by the bailout time for \( c=0.5 \).


\subsection{Cycles with random edges}\label{subsection:cycle_random_edges}

We consider random graphs generated by the ensemble \( C_{N,s} \) (see Table~\ref{table:network}) for \( s=0.1 \), \(s =  0.2 \), and \( s = 0.3 \). Cycles with additional, random ``shortcut'' edges are related to Watts--Strogatz small-world networks \cite{watts98,newman99,smallworld-scholarpedia} (see also earlier work by Bollob\'as and Chung \cite{boll88}), except that nodes initially have degree \( 2\), which yields (for cycles that are not too small) a clustering coefficient of $0$ for each node before random edges are added.

\begin{figure*}
	\centering
		\includegraphics[width=.75\textwidth]{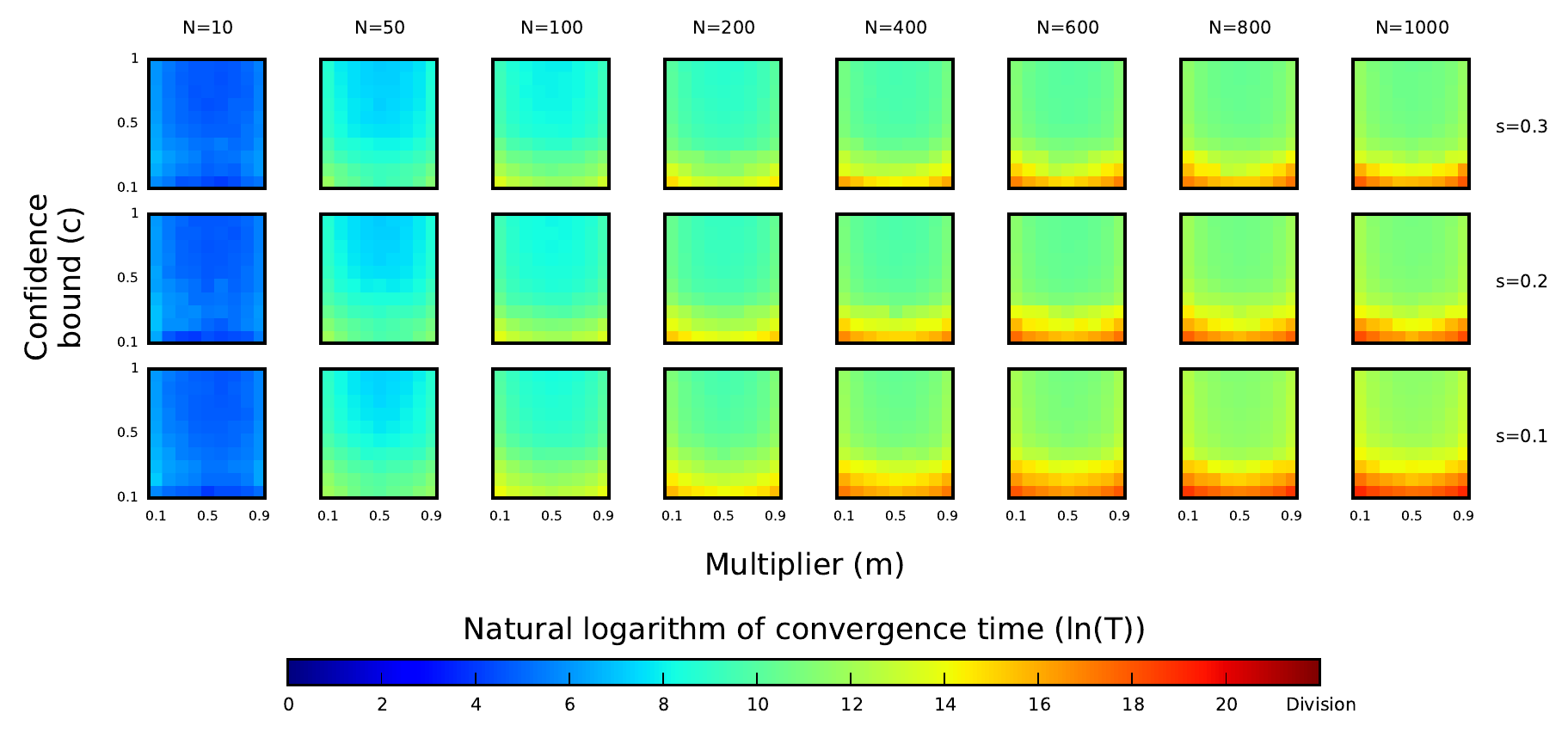}
	\caption{Convergence times for our simulations on \( C_{N,s} \) for \( s=0.1 \), \( s = 0.2 \), and \( s = 0.3 \) for various values of $N$.}
\label{figure:cycle_random_edges_panel_plots}
\end{figure*}

\begin{figure*}
	\centering
		\includegraphics[width=.75\textwidth]{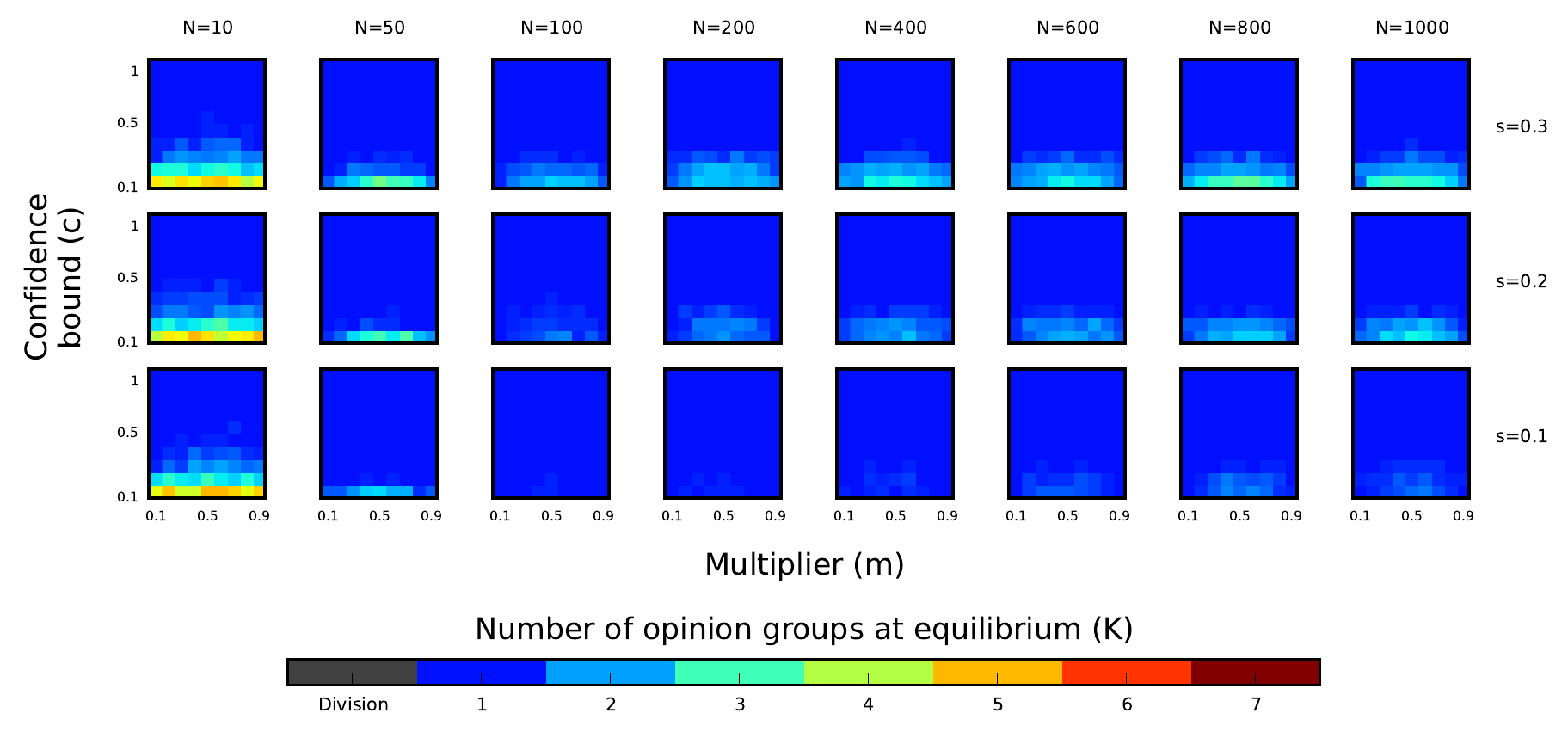}
	\caption{The number of opinion groups that persisted at equilibrium in simulations on \( C_{N,s} \) given \( s=0.1 \), \( s = 0.2 \), and \( s = 0.3 \) for various values of $N$.}
\label{figure:cycle_random_edges_opinion_cliques_panel_plots}
\end{figure*}

In Fig.~\ref{figure:cycle_random_edges_panel_plots}, we summarize the values of \( \ln(T) \) that we observe in our simulations on \( C_{N,s} \) for \( s=0.1 \), \( s = 0.2 \), and s = \( 0.3 \).
Regression analysis suggests the model
\begin{equation}\begin{aligned}
	\ln(T)^{\alpha} & = \beta_0 + \beta_1\ln(N) + \beta_2N + \beta_3N^2 + \beta_4c + \beta_5c^2 \\
& \qquad + \beta_6(m-0.5)^2 + \beta_7Nc + \epsilon\,,
\label{model:cycle_random_edges}
\end{aligned}\end{equation}
where the power-transformation parameter is \( \alpha=-1/3 \), \( \alpha = -2/3 \), and \( \alpha = -5/6 \) for \( s=0.1 \), \( s = 0.2 \), and \( s = 0.3 \), respectively. For \( s=0.1 \), the \( Nc \) term is statistically insignificant, and we thus drop it. In Table~\ref{table:cycle_random_edges_coefficients}, we summarize our coefficient estimates for Eq.~\eqref{model:cycle_random_edges}. For \( s=0.1 \), we obtain \( \text{AIC} \approx -10378.2 \) and \( R^2 \approx 0.9853 \); for \( s=0.2 \), we obtain \( \text{AIC} \approx -10443.3 \) and \( R^2 \approx 0.9829 \); and for \( s=0.3 \), we obtain \( \text{AIC} \approx -10719.2 \) and \( R^2 \approx 0.9816 \).

\begin{table*}
	\small
	\centering
	\begin{tabular}{ccccccccc}
	\hline
	{\( s \)} & {\( \beta_0 \)} & {\( \beta_1 \)} & {\( \beta_2 \)} & {\( \beta_3 \)} & {\( \beta_4 \)} & {\( \beta_5 \)} & {\( \beta_6 \)} & {\( \beta_7 \)} \\ 
	\hline
	\( 0.1 \) & \( 5.485 \times 10^{-1} \) & \( -3.140 \times 10^{-2} \) & \( 6.760 \times 10^{-5} \) & \( -2.908 \times 10^{-8} \)  & \( 1.820 \times 10^{-1} \) & \( -1.101 \times 10^{-1} \) & \( -9.576 \times 10^{-2} \) & N/A \\
	\( 0.2 \) & \( 2.763 \times 10^{-1} \) & \( -2.351 \times 10^{-2} \) & \( 4.343 \times 10^{-5} \) & \( -1.653 \times 10^{-8} \)  & \( 1.779 \times 10^{-1} \) & \( -1.084 \times 10^{-1} \) & \( -8.901 \times 10^{-2} \) & \( -1.008 \times 10^{-5} \) \\
	\( 0.3 \) & \( 2.031 \times 10^{-1} \) & \( -1.977 \times 10^{-2} \) & \( 3.548 \times 10^{-5} \) & \( -1.090 \times 10^{-8} \)  & \( 1.452 \times 10^{-1} \) & \( -8.829 \times 10^{-2} \) & \( -7.934 \times 10^{-2} \) & \( -1.093 \times 10^{-5} \) \\
	   \hline
	\end{tabular}
\caption{Estimates of regression coefficients for Eq.~\eqref{model:cycle_random_edges}. For \( s=0.1 \), the \( Nc \) term is statistically insignificant, so we drop it.}\label{table:cycle_random_edges_coefficients}
\end{table*}

Our data exploration and regression analysis suggest that \( T \) does not experience a transition with respect to \( c \). According to Eq.~\eqref{model:cycle_random_edges}, \( T \) increases with \( N \) for \( s=0.1 \), \( s = 0.2 \), and \( s = 0.3 \). For \( s=0.1 \), the convergence time \( T \) obtains a global minimum at \( c=-\beta_4/(2\beta_5) \in (0,1) \) if \( N \) and \( m \) are held constant. For \( s\in \{0.2,0.3\} \), if \( N \geq -\beta_4/\beta_7 \), the convergence time \( T \) increases with \( c \in (0,1] \). If \( N<-\beta_4/\beta_7 \), then \( T \) obtains a global minimum at \( c=-(\beta_4+\beta_7N)/(2\beta_5) \) if \( N \) and \( m \) are held constant. Finally, \( T \) obtains a global minimum at \( m=0.5 \) if \( N \) and \( c \) are held constant.

Similar to our observations for prism graphs and square lattices, the effects of \( N \), \( c \), and \( m \) on \( T \) are coupled to each other for cycles with random edges, in contrast to what we observed using our regression model of cycles (see Eq.~\eqref{model:cycle}), which has only one weak coupling term \( Nm \). Adding random shortcut edges to cycles significantly decreases the convergence time. Additionally, \( T \) increases much more slowly with \( N \) on \( C_{N,s} \) than it does for cycles.

In Fig.~\ref{figure:cycle_random_edges_opinion_cliques_panel_plots}, we summarize the number of opinion groups that persist at equilibrium in our simulations on cycles with random edges. With only a small proportion (i.e., \( s=0.1 \)) of random edges, the number \( K \) of opinion groups at equilibrium is roughly the same as what we observed in our simulations on cycles (see Fig.~\ref{figure:cycle_opinion_cliques_panel_plots}). However, as more random edges are added, multiple opinion groups start to emerge at equilibrium for \( c\leq 0.3 \). We conjecture that, as the proportion of random edges increases, the behavior of \( K \) is more similar to the case of complete graphs than that of cycles.


\subsection{Prism graphs with random edges}\label{subsec:prism_random_edges}

We consider random graphs generated by the ensemble \( Y_{N,s} \) (see Table~\ref{table:network}) for \( s=0.1 \), \( s = 0.2 \), and \( s = 0.3 \). We study the effect of random edges on the behavior of the Deffuant model by comparing our simulation results with the ones that we obtained for prism graphs in Section~\ref{subsection:cylindrical_lattice}.

\begin{figure*}
	\centering
		\includegraphics[width=.75\textwidth]{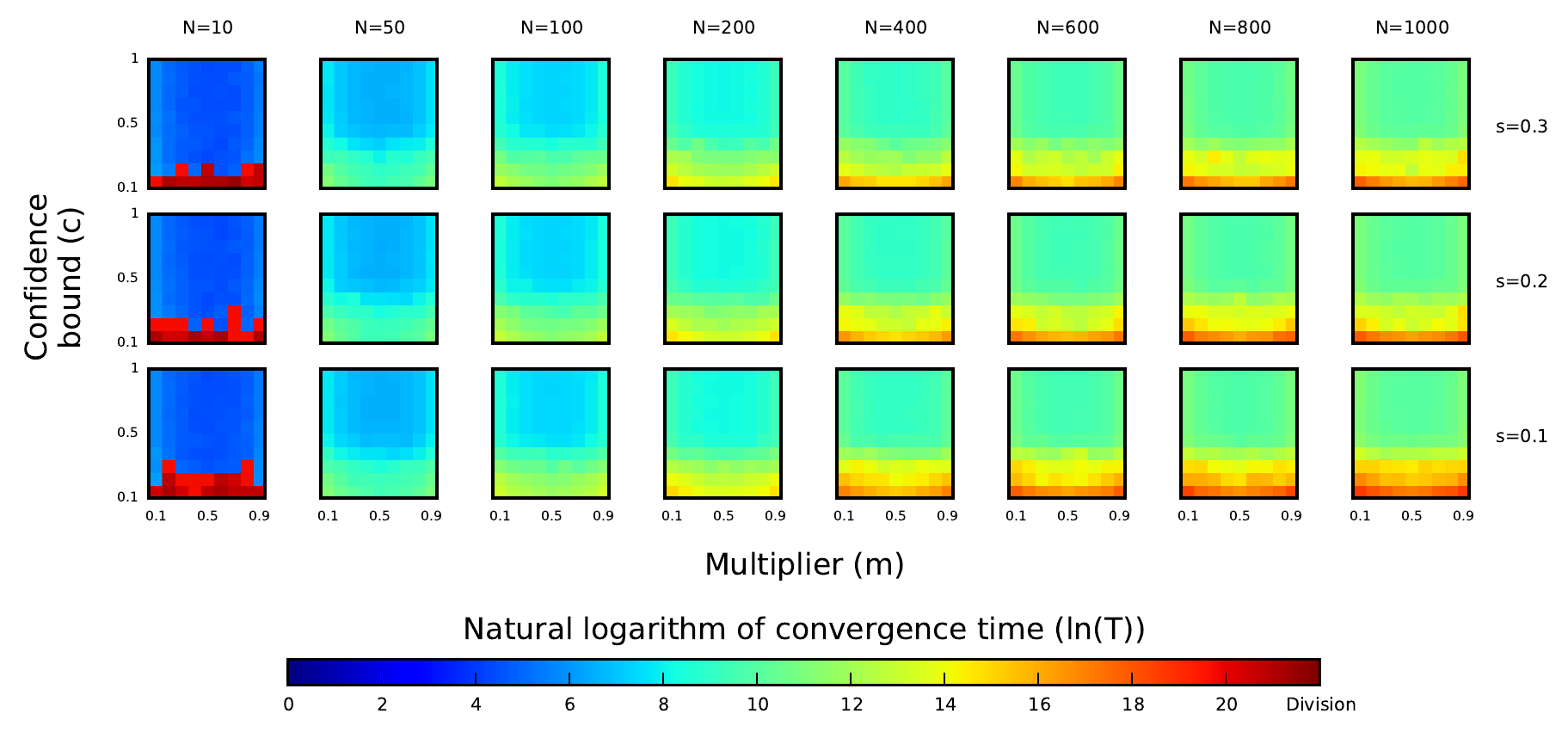}
	\caption{Convergence times for our simulations on \( Y_{N,s} \) for \( s=0.1 \), \( s = 0.2 \), and \( s = 0.3 \) for various values of $N$.}
\label{figure:annular_lattice_random_edges_panel_plots}
\end{figure*}

\begin{figure*}
	\centering
		\includegraphics[width=.75\textwidth]{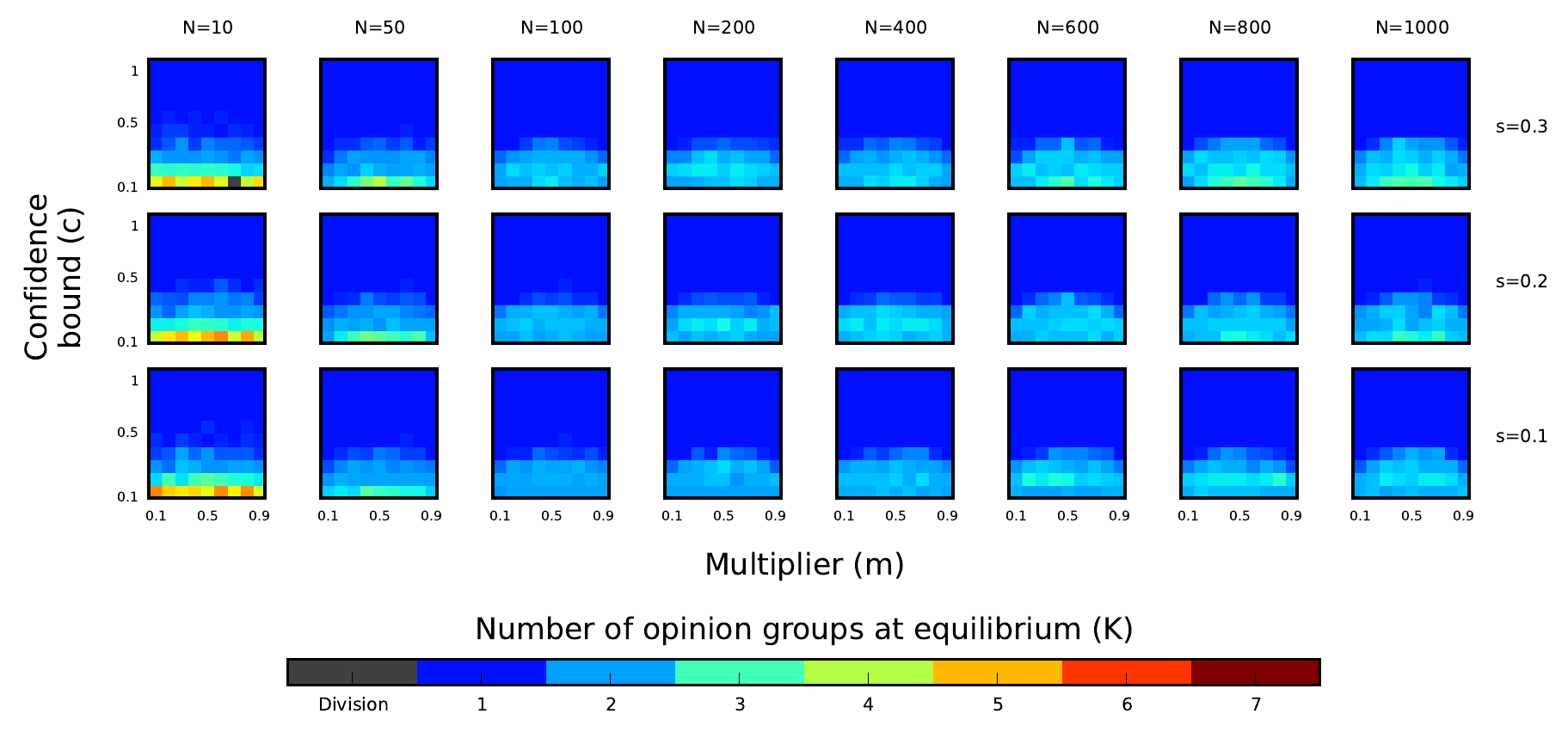}
	\caption{The number of opinion groups that persist at equilibrium in our simulations on \( Y_{N,s} \) for \( s=0.1 \), \( s = 0.2 \), and \( s = 0.3 \) for various values of $N$.}
\label{figure:annular_lattice_random_edges_opinion_cliques_panel_plots}
\end{figure*}

In Fig.~\ref{figure:annular_lattice_random_edges_panel_plots}, we summarize the values of \( \ln(T) \) that we observe in our simulations on \( Y_{N,s} \) for \( s=0.1 \), \( s = 0.2 \), and \( s = 0.3 \). Similar to our results for prism graphs in Section~\ref{subsection:cylindrical_lattice}, we observe qualitatively distinct behavior of the convergence time for \( c<0.5 \) and \( c\geq 0.5 \) for the Deffuant model on \( Y_{N,s} \). Therefore, we conduct separate regression analyses for these two cases.
For \( c<0.5 \), our regression analysis suggests the model
\begin{equation}\begin{aligned}
	\ln(T) & = \beta_0 + \beta_1N + \beta_2N^2 + \beta_3c + \beta_4c^2 \\
& \qquad + \beta_5(m-0.5)^2 + \beta_6Nc + \epsilon\,,
\label{model:annular_lattice_random_edges(lower)}
\end{aligned}\end{equation}
where we list our coefficient estimates in Table~\ref{table:annular_lattice_random_edges_coefficients(lower)}. For \( s=0.1 \), we obtain \( \text{AIC} \approx -592.1 \) and \( R^2 \approx 0.9613 \); for \( s=0.2 \), we obtain \( \text{AIC} \approx -427.99 \) and \( R^2 \approx 0.9283 \); and for \( s=0.3 \), we obtain \( \text{AIC} \approx -482.58 \) and \( R^2 \approx 0.9336 \).
For \( c \geq 0.5 \), our regression analysis suggests the model
\begin{equation}
	\ln(T) = \beta_0 + \beta_1\ln(N) + \beta_2c + \beta_3c^2 + \beta_4(m-0.5)^2 + \beta_5Nc + \epsilon\,,
\label{model:annular_lattice_random_edges(upper)}
\end{equation}
where we list our coefficient estimates in Table~\ref{table:annular_lattice_random_edges_coefficients(upper)}. For \( s=0.1 \), we obtain \( \text{AIC} \approx -2596.67 \) and \( R^2 \approx 0.9914 \); for \( s=0.2 \), we obtain \( \text{AIC} \approx -2693.1 \) and \( R^2 \approx 0.9922 \); and for \( s=0.3 \), we obtain \( \text{AIC} \approx -2912.41 \) and \( R^2 \approx 0.9947 \).

\begin{table*}
	\small
	\centering
	\begin{tabular}{cccccccc}
	\hline
	{\( s \)} & {\( \beta_0 \)} & {\( \beta_1 \)} & {\( \beta_2 \)} & {\( \beta_3 \)} & {\( \beta_4 \)} & {\( \beta_5 \)} & {\( \beta_6 \)} \\ 
	\hline
	\( 0.1 \) & \( 1.225 \times 10 \) & \( 1.240 \times 10^{-2} \) & \( -6.048 \times 10^{-6} \) & \( -6.132 \) & \( -1.087 \times 10 \) & \( 6.183 \) & \( -5.739 \times 10^{-3} \) \\
	\( 0.2 \) & \( 1.309 \times 10 \) & \( 1.067 \times 10^{-2} \) & \( -5.535 \times 10^{-6} \) & \( -1.699 \times 10 \) & \( \phantom{-}1.088 \times 10 \) & \( 6.964 \) & \( -4.199 \times 10^{-3} \) \\
	\( 0.3 \) & \( 1.353 \times 10 \) & \( 8.670 \times 10^{-3} \) & \( -4.073 \times 10^{-6} \) & \( -2.060 \times 10 \) & \( \phantom{-}1.764 \times 10 \) & \( 6.879 \) & \( -3.324 \times 10^{-3} \) \\
	   \hline
	\end{tabular}
\caption{Estimates of regression coefficients for Eq.~\eqref{model:annular_lattice_random_edges(lower)}.}\label{table:annular_lattice_random_edges_coefficients(lower)}
\end{table*}

\begin{table}
	\small
	\centering
	\begin{tabular}{ccccccc}
	\hline
	{\( s \)} & {\( \beta_0 \)} & {\( \beta_1 \)} & {\( \beta_2 \)} & {\( \beta_3 \)} & {\( \beta_4 \)} & {\( \beta_5 \)} \\ 
	\hline
	\( 0.1 \) & \( 3.528 \) & \( 1.200 \) & \( -4.263 \) & \( 2.628 \) & \( 6.720 \) & \( -2.061 \times 10^{-4} \) \\
	\( 0.2 \) & \( 3.300 \) & \( 1.161 \) & \( -3.242 \) & \( 2.002 \) & \( 6.634 \) & \( -1.808 \times 10^{-4} \) \\
	\( 0.3 \) & \( 3.193 \) & \( 1.116 \) & \( -2.436 \) & \( 1.465 \) & \( 6.671 \) & \( -5.580 \times 10^{-5} \) \\
	\hline
	\end{tabular}
\caption{Estimates of regression coefficients for Eq.~\eqref{model:annular_lattice_random_edges(upper)}.}\label{table:annular_lattice_random_edges_coefficients(upper)}
\end{table}

The different forms of Eqs.~\eqref{model:annular_lattice_random_edges(lower)} and \eqref{model:annular_lattice_random_edges(upper)} support our conjecture based on our data exploration that \( T \) undergoes a transition at \( c=0.5 \). If \( c \) and \( m \) are held constant, the convergence time \( T \) obtains a maximum at \( N=-(\beta_1+\beta_6c)/(2\beta_2) \) for \( c<0.5 \) and at \( N=-\beta_1/(\beta_5c) \) for \( c \geq 0.5 \). Moreover, \( T \) increases with \( (m-0.5)^2 \) exponentially and has a minimum at \( m=0.5 \). If \( N \) and \( m \) are held constant, \( T \) decreases with \( c<0.5 \). For \( c \geq 0.5 \), if \( N \geq -(\beta_2+2\beta_3)/\beta_5 \), the convergence time \( T \) decreases with \( c \); otherwise, \( T \) has a minimum at \( c=-(\beta_2+\beta_5N)/(2\beta_3) \in (0.5,1] \) for \( N \) and \( m \) are held constant. In contrast to the coupling effects of \( N \), \( c \), and \( m \) that we observed in our simulations on prism graphs, simulations on \( Y_{N,s} \) suggest that only a weak coupling term \( Nc \) exists in the regression model (see Eqs.~\eqref{model:annular_lattice_random_edges(lower)} and \eqref{model:annular_lattice_random_edges(upper)}). Additionally, adding random edges to prism graphs decreases \( T \) more significantly for \( c<0.5 \) than for \( c \geq 0.5 \).

In Fig.~\ref{figure:annular_lattice_random_edges_opinion_cliques_panel_plots}, we summarize the number of opinion groups that arise in our simulations on prism graphs with randomly-generated extra edges. As we observed for prism graphs, consensus is always reached on prism graphs with random edges for \( c \geq 0.5 \). However, for \( c<0.5 \) and \( N \geq 50 \), we observe \( K \geq 2 \), in contrast to \( K \approx 2 \) for prism graphs. Therefore, when a population's confidence bound is small, adding random edges to prism graphs is more favorable to expediting the process of opinions dividing into distinct groups than reaching agreement among the population.


\subsection{Erd\H{o}s--R\'{e}nyi networks}\label{subsection:erdos-renyi_model}

We now consider random graphs generated by the Erd\H{o}s--R\'{e}nyi \( G(N,p) \) model, where \( p \in [0,1] \) is an independent probability for there to be an edge between a pair of nodes. Erd\H{o}s--R\'{e}nyi graphs are one of the best-studied models of network science, and they have been used in previous studies of the Deffuant model on networks \cite{alaali08,fortunato04b,kozma08a,kozma08b}. However, existing research on the Deffuant model on ER random graphs has focused primarily on adaptive networks that evolve along with the game \cite{kozma08a,kozma08b}. In our simulations, we consider the ER \( G(N,p) \) model for \( p = 0.1, \, 0.2, \, \dots , \, 0.9 \). Complete graphs are a special case of the ER \( G(N,p) \) model, as one obtains a complete graph for the parameter value \( p=1 \).

\begin{figure*}
	\centering
		\includegraphics[width=.75\textwidth]{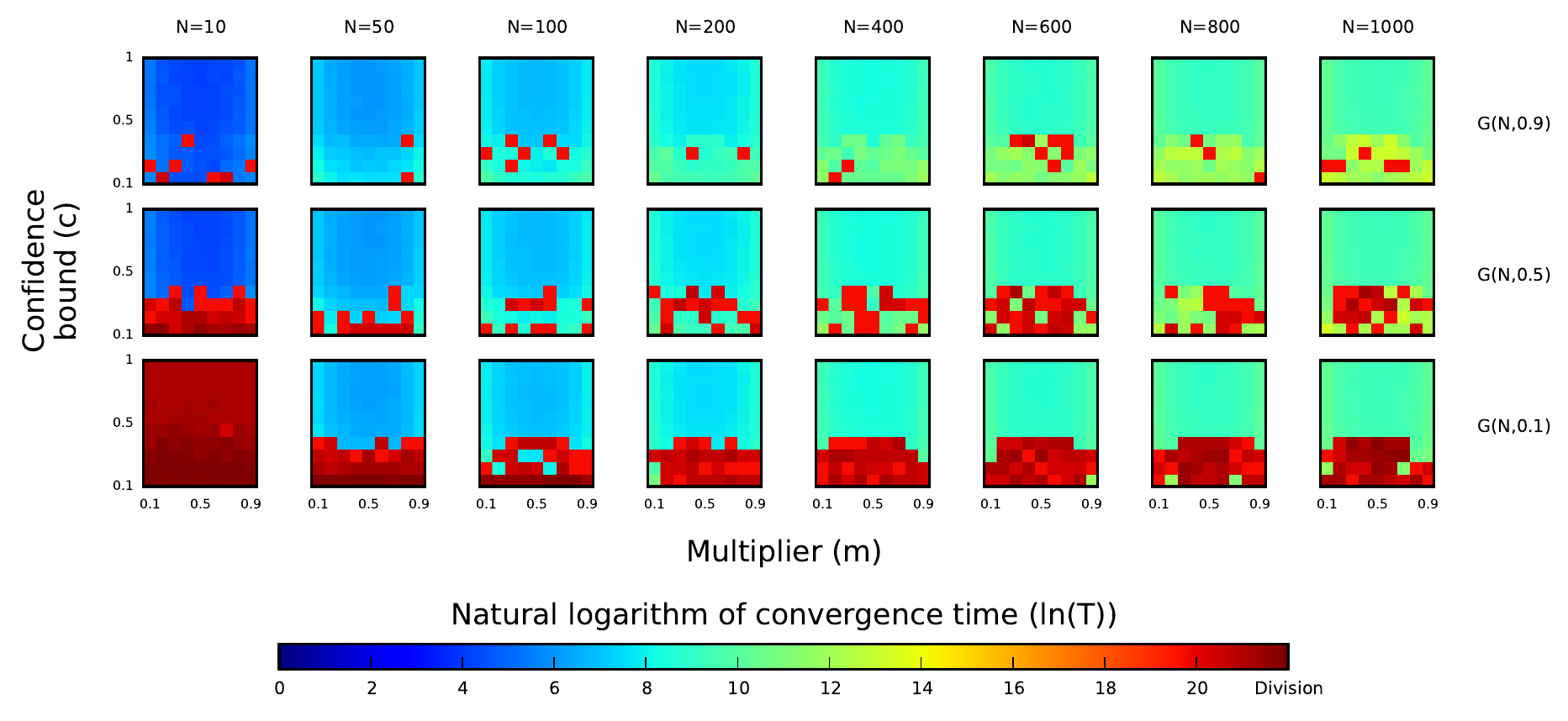}
	\caption{Convergence times for our simulations on random graphs generated by the Erd\H{o}s--R\'{e}nyi \( G(N,p) \) model. We conduct simulations for \( p = 0.1, \, 0.2, \, \dots , \, 0.9 \), and we present a subset of our plots to illustrate the observed trends.}
\label{figure:er_model_interaction_plots}
\end{figure*}

\begin{figure*}
	\centering
		\includegraphics[width=.75\textwidth]{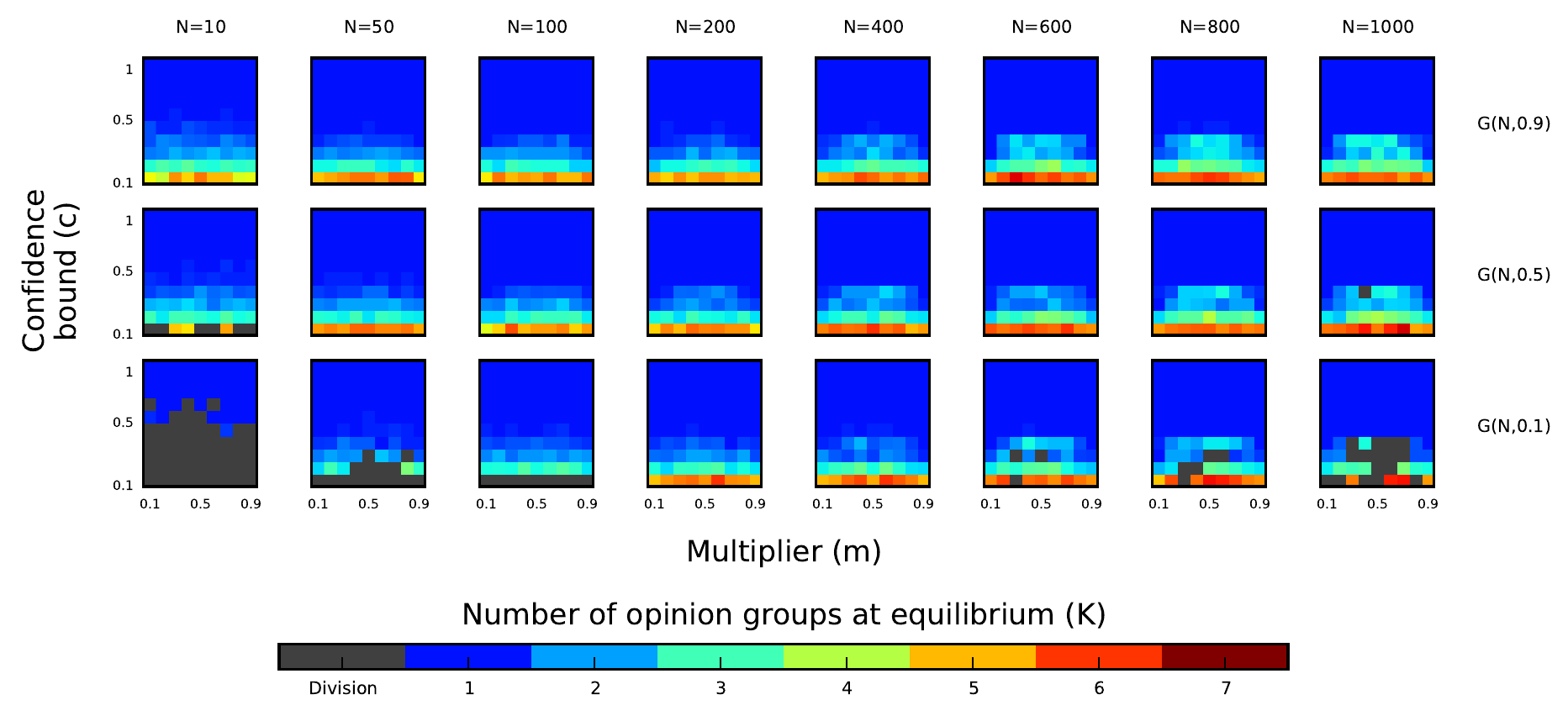}
	\caption{The number of opinion groups that persist at equilibrium in our simulations on random graphs generated by the Erd\H{o}s--R\'{e}nyi \( G(N,p) \) model. We conduct simulations for \( p = 0.1, \, 0.2, \, \dots , \, 0.9 \), and we present a subset of our plots to illustrate the observed trends.}
\label{figure:er_model_opinion_cliques_panel_plots}
\end{figure*}

In Fig.~\ref{figure:er_model_interaction_plots}, we show a subset of the values of \( \ln(T) \) that we obtain in our simulations. These values are representative of the observed trends in all of our simulations. As in our simulations on complete graphs, we observe qualitatively distinct behavior for \( T \) for \( c<0.5 \) and \( c \geq 0.5 \). Therefore, we conduct separate regression analyses for these two cases.
For \( c < 0.5 \), regression analysis suggests the model
\begin{equation}
	\ln(\ln(T)) = \beta_0 + \beta_1N + \beta_2N^2 + \beta_3c^2  + \beta_4(m-0.5)^2 + \beta_5Nc + \epsilon\,,
\label{model:er_model(lower)4}
\end{equation}
where we list estimates for the coefficients in Table~\ref{table:er_model_lower_coefficients}. Random graphs generated by the ER \( G(N,p) \) model are a source of stochasticity for the opinion dynamics. It is thus not surprising that we observe a larger number of outliers for our ER simulations than for complete graphs. Let \( q \in [0,1] \) be the proportion of data points that we identify as outliers and thus exclude from our regression analysis.\footnote{We construe a data point as outlier if $T > 600,000$. Each data point is a mean over 10 simulations, and we chose this value so that all 10 simulations converge by the bailout time of $3.55 \times 10^9$.}
For \( c < 0.5 \), we conduct a regression analysis for the \( G(N,p) \) graphs for \( p=0.7 \), \( p = 0.8 \), and \( p = 0.9 \), as \( q>0.15 \) for smaller values of \( p \) (and this would undermine the reliability of the regression analysis). For \( p=0.7 \), we obtain \( q \approx 0.1133 \), \( \text{AIC} \approx -1515.5 \), and \( R^2 \approx 0.8360 \); for \( p=0.8 \), we obtain \( q \approx 0.0767 \), \( \text{AIC} \approx -1658.1 \), and \( R^2 \approx 0.8114 \); and for \( p=0.9 \), we obtain \( q=0.05 \), \( \text{AIC} \approx -1774.1 \), and \( R^2 \approx 0.7984 \).
For \( c \geq 0.5 \), our regression analysis suggests the model in Eq.~\eqref{model:complete_graph(upper)3} for each value of \( p \) that we consider. For each \( p \), Table~\ref{table:er_model_upper_coefficients} summarizes our estimates for the coefficients \( \beta_j \) (\( j=0,1, 2,3 \)), together with the corresponding values of AIC and \( R^2 \).

\begin{table*}
	\small
	\centering
	\begin{tabular}{ccccccc}
	\hline
	{\( p \)} & {\( \beta_0 \)} & {\( \beta_1 \)} & {\( \beta_2 \)} & {\( \beta_3 \)} & {\( \beta_4 \)} & {\( \beta_5 \)} \\ 
	\hline
	\( 0.7 \) & \( 2.098 \) & \( 8.744 \times 10^{-4} \) & \( -4.854 \times 10^{-7} \) & \( -8.606 \times 10^{-1} \) & \( 2.294 \times 10^{-1} \) & N/A \\
	\( 0.8 \) & \( 2.111 \) & \( 8.328 \times 10^{-4} \) & \( -4.349 \times 10^{-7} \) & \( -7.874 \times 10^{-1} \) & \( 1.255 \times 10^{-1} \) & N/A \\
	\( 0.9 \) & \( 2.117 \) & \( 7.901 \times 10^{-4} \) & \( -4.327 \times 10^{-7} \) & \( -8.926 \times 10^{-1} \) & \( 1.200 \times 10^{-1} \) & \( 2.323 \times 10^{-4} \) \\
	\hline
	\end{tabular}
\caption{Estimates of regression coefficients for Eq.~\eqref{model:er_model(lower)4}.}\label{table:er_model_lower_coefficients}
\end{table*}

\begin{table}
	\small
	\centering
	\begin{tabular}{ccccccc}
	\hline
	{\( p \)} & {\( \beta_0 \)} & {\( \beta_1 \)} & {\( \beta_2 \)} & {\( \beta_3 \)} & {AIC} & {\( R^2 \)} \\ 
	\hline
	\( 0.1 \) & \( 1.953 \) & \( 1.050 \) & \( 4.412 \times 10^{-1} \) & \( 6.362 \) & \( -3156.1 \) & \( 0.9958 \) \\ 
	\( 0.2 \) & \( 1.931 \) & \( 1.053 \) & \( 4.500 \times 10^{-1} \) & \( 6.290 \) & \( -3194.2 \) & \( 0.9961 \) \\ 
	\( 0.3 \) & \( 1.918 \) & \( 1.055 \) & \( 4.512 \times 10^{-1} \) & \( 6.275 \) & \( -3215.1 \) & \( 0.9962 \) \\ 
	\( 0.4 \) & \( 1.886 \) & \( 1.060 \) & \( 4.453 \times 10^{-1} \) & \( 6.270 \) & \( -3233.7 \) & \( 0.9963 \) \\ 
	\( 0.5 \) & \( 1.827 \) & \( 1.068 \) & \( 4.548 \times 10^{-1} \) & \( 6.284 \) & \( -3209.2 \) & \( 0.9963 \) \\ 
	\( 0.6 \) & \( 1.870 \) & \( 1.062 \) & \( 4.499 \times 10^{-1} \) & \( 6.255 \) & \( -3233.6 \) & \( 0.9964 \) \\
	\( 0.7 \) & \( 1.851 \) & \( 1.065 \) & \( 4.470 \times 10^{-1} \) & \( 6.242 \) & \( -3213.7 \) & \( 0.9963 \) \\
	\( 0.8 \) & \( 1.873 \) & \( 1.061 \) & \( 4.555 \times 10^{-1} \) & \( 6.289 \) & \( -3267.4 \) & \( 0.9966 \) \\
	\( 0.9 \) & \( 1.838 \) & \( 1.067 \) & \( 4.676 \times 10^{-1} \) & \( 6.261 \) & \( -3251.8 \) & \( 0.9965 \) \\
	\( 1 \) & \( 1.865 \) & \( 1.062 \) & \( 4.530 \times 10^{-1} \) & \( 6.262 \) & \( -3240.9 \) & \( 0.9964 \) \\
	\hline
	\end{tabular}
\caption{Estimates of regression coefficients, AIC values, and coefficients of determination (\( R^2 \)) for Eq.~\eqref{model:complete_graph(upper)3} from our simulation results on ER random graphs. For comparison, we also include the coefficients for the complete graphs (which arises from the ER model with connection probability $p = 1$) that we studied in Section~\ref{subsection:complete_graphs}.}\label{table:er_model_upper_coefficients}
\end{table}

The different forms of Eqs.~\eqref{model:complete_graph(upper)3} and \eqref{model:er_model(lower)4} support our conjecture from our data exploration that \( T \) undergoes a transition at \( c=0.5 \). For \( c<0.5 \), the convergence time \( T \) increases with \( N < -\beta_1/(2\beta_2) \) for \( p \in \{ 0.7,0.8 \} \) and with \( N < -(\beta_1+\beta_5c)/(2\beta_2) \) for \( p=0.9 \). For \( p \in \{0.7,0.8\} \), the convergence time \( T \) decreases with \( c<0.5 \). For \( p =0.9 \), if \( N \geq -\beta_3/\beta_5 \), then \( T \) increases with \( c<0.5 \); otherwise, \( T \) obtains a maximum at \( c=-\beta_5N/(2\beta_3) \in (0,0.5) \). For \( c<0.5 \), Eq.~\eqref{model:er_model(lower)4} suggests that \( \ln(T) \) is proportional to \( \exp \left( (m-0.5)^2 \right) \), which contrasts with our regression model for complete graphs (see Eq.~\eqref{model:complete_graph(lower)3}), which do no exhibit a statistically significant influence of \( m \) on \( T \). For \( c \geq 0.5 \), our regression model for the ER \( G(N,p) \) model is the same as what we obtained for complete graphs, so the behavior of \( T \) with respect to \( N \), \( c \), and \( m \) is similar in that parameter regime. For large values of \( p \), the estimated coefficients are very close to those for complete graphs. This suggests that it is probably accurate to use a mean-field approximation to study convergence time on the Erd\H{o}s--R\'{e}nyi \( G(N,p) \) model if \( p \) is close to \( 1 \).

In Fig.~\ref{figure:er_model_opinion_cliques_panel_plots}, we summarize the number of opinion groups that arise in our simulations of the Deffuant model on random graphs generated by the ER \( G(N,p) \) model. When the connection probability \( p \) is close to \( 1 \), the behavior of \( K \) is similar to what we observed for complete graphs. As \( p \to 0 \), the major qualitative difference is that opinions sometimes fail to converge within the bailout time for small values of \( c \).


\subsection{Facebook friendship networks}\label{subsection:fb_networks}

We now simulate the Deffuant model on two Facebook ``friendship'' networks \cite{traud12} --- one of Swarthmore College and the other of the California Institute of Technology (Caltech) --- from one day in autumn 2005. We consider the largest connected component (LCC) of each network. For the Swarthmore network, the LCC has \( 1657 \) nodes and \( 61049 \) edges. The LCC of the Caltech network has \( 762 \) nodes and \( 16651 \) edges.

\begin{figure}
	\centering
		\includegraphics[width=.45\textwidth]{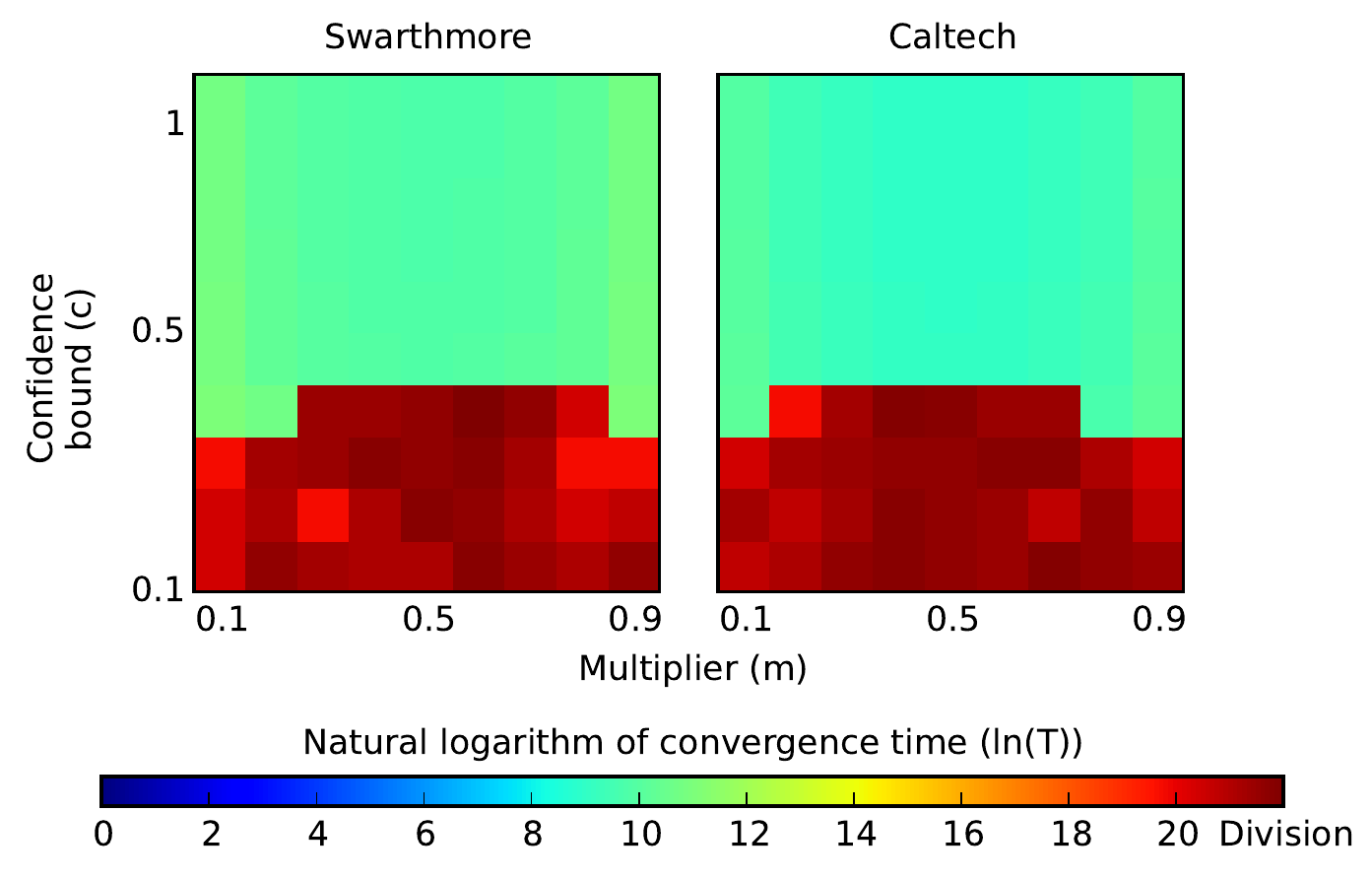}
    \caption{Convergence times for our simulations on the largest connected components of the Swarthmore and Caltech Facebook networks from the {\sc Facebook100} data set \cite{traud12}.}
\label{figure:fb_convergence_time}
\end{figure}

In Fig.~\ref{figure:fb_convergence_time}, we summarize the values of \( \ln(T) \) that we observe in simulations. For \( c<0.5 \), most of the simulations on both networks fail to converge by the bailout time, so we consider only the results of \( c \geq 0.5 \) in our regression analysis. For the Swarthmore network, we obtain a regression model of 
\begin{equation}
	T^{-\frac{7}{8}} = \beta_0 + \beta_1c + \beta_2c^2 + \beta_3(m-0.5)^2 + \epsilon\,,
\label{model:fb_swarthmore}
\end{equation}
where we list our estimates for the coefficients in Table~\ref{table:fb_swarthmore}. For Eq.~\eqref{model:fb_swarthmore}, we obtain \( \text{AIC} \approx -1279.02 \) and \( R^2 \approx 0.9987 \). For the Caltech network, we obtain a regression model of
\begin{equation}
	T^{-\frac{2}{3}} = \beta_1c + \beta_2c^2 + \beta_3(m-0.5)^2 + \epsilon\,,
\label{model:fb_caltech}
\end{equation}
where we list our estimates for the coefficients in Table~\ref{table:fb_caltech}. For Eq.~\eqref{model:fb_caltech}, we obtain \( \text{AIC} \approx -1001.8 \) and \( R^2 \approx 0.9981 \).

\begin{figure}
	\centering
		\includegraphics[width=.45\textwidth]{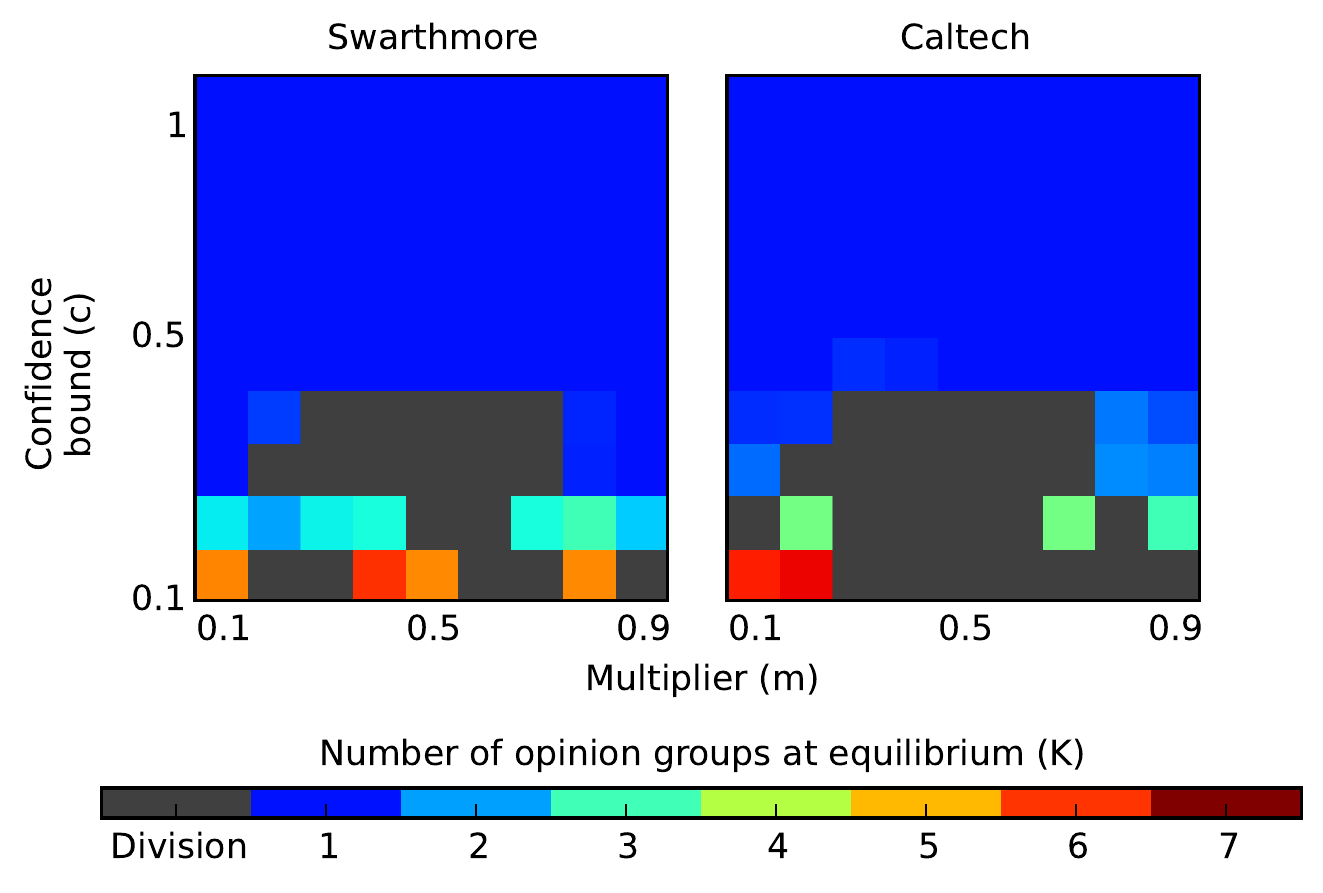}
    \caption{The number of opinion groups that persist at equilibrium in our simulations on the largest connected components of the Swarthmore and Caltech Facebook networks.}
\label{figure:fb_opinion_cliques}
\end{figure}

\begin{table}[h]
	\small
	\centering
	\begin{tabular}{ccccc}
	\hline
	 & {Estimate} & {Std. Error} & {\( t \) value} & {Pr(\( >|t| \))} \\ 
	\hline
	\( \beta_0 \) & \( \phantom{-}1.338 \times 10^{-4} \) & \( 5.718 \times 10^{-6} \) & \( \phantom{-}2.340 \times 10 \) & \( < 2 \times 10^{-16} \) \\ 
	\( \beta_1 \) & \( \phantom{-}1.419 \times 10^{-4} \) & \( 1.533 \times 10^{-5} \) & \( \phantom{-}9.258 \) & \( 6.81 \times 10^{-12} \) \\ 
	\( \beta_2 \) & \( -7.964 \times 10^{-5} \) & \( 9.963 \times 10^{-6} \) & \( -7.993 \) & \( 4.13 \times 10^{-10} \) \\ 
	\( \beta_3 \) & \( -7.208 \times 10^{-4} \) & \( 3.989 \times 10^{-6} \) & \( -1.807 \times 10^2 \) & \( < 2 \times 10^{-16} \) \\ 
	\hline
	\end{tabular}
\caption{Estimates of regression coefficients for Eq.~\eqref{model:fb_swarthmore}.}
\label{table:fb_swarthmore}
\end{table}

\begin{table}[h]
	\small
	\centering
	\begin{tabular}{ccccc}
	\hline
	 & {Estimate} & {Std. Error} & {\( t \) value} & {Pr(\( >|t| \))} \\ 
	\hline
	\( \beta_1 \) & \( \phantom{-}6.662 \times 10^{-3} \) & \( 9.431 \times 10^{-5} \) & \( \phantom{-}7.063 \times 10 \) & \( < 2 \times 10^{-16} \) \\ 
	\( \beta_2 \) & \( -4.201 \times 10^{-3} \) & \( 1.056 \times 10^{-4} \) & \( -3.979 \times 10 \) & \( < 2 \times 10^{-16} \) \\ 
	\( \beta_3 \) & \( -7.608 \times 10^{-3} \) & \( 2.117 \times 10^{-4} \) & \( -3.593 \times 10 \) & \( < 2 \times 10^{-16} \) \\ 
	\hline
	\end{tabular}
\caption{Estimates of regression coefficients for Eq.~\eqref{model:fb_caltech}.}
\label{table:fb_caltech}
\end{table}

For both networks, the variables \( c \) and \( m \) have an intertwined effect on \( T \). Moreover, if \( m \) is held constant, the convergence time \( T \) has a minimum at \( c=-\beta_1/(2\beta_2) \in (0.5,1) \). If \( c \) is held constant, \( T \) increases with \( (m-0.5)^2 \). The convergence time for both of the networks has qualitatively similar behavior as what we observed for cycles with random edges (see Section~\ref{subsection:cycle_random_edges}) of corresponding sizes for \( s=0.1 \) and \( 0.2 \). This empirical observation suggests that simulating the Deffuant model on random graphs generated by these and similar networks (e.g., WS networks) may yield some useful insights about the convergence time for the Deffuant model on social networks.

In Fig.~\ref{figure:fb_opinion_cliques}, we summarize the number of opinion groups that persist at equilibrium in our simulations on the LCCs of the Swarthmore and the Caltech Facebook networks. In both networks, consensus occurs for all \( c \geq 0.5 \). For \( c<0.5 \), at least half of the simulations fail to converge within the bailout time, but those that converge suggest that \( K \) increases as \( c \) approaches \( 0 \). In contrast, our simulations of Deffuant dynamics on cycles with random edges reached equilibrium within bailout time (see Section~\ref{subsection:cycle_random_edges}). 


\section{Conclusions and Discussion}

We studied the Deffuant model on several types of deterministic and random networks. For each of these networks, we systematically examined the number of groups of different opinions in these networks and the convergence time to reach equilibrium as a function of the number (\( N \)) of agents that participate in the opinion dynamics, the population's confidence bound (\( c \)), and their cautiousness (which we measure using the multiplier \( m \)). For the convergence time to equilibrium, we used both numerical simulations and regression analyses to obtain qualitative and quantitative insights. For the number $K$ of opinion groups at equilibrium, we used our numerical simultions to examine the qualitative behavior of different types of networks.

We obtained many insights from our systematic computations. Studying the effect of network structure on dynamical processes (such as opinion models) is a difficult problem, and we were able to achieve several interesting insights about the intertwined effect of network topology and the parameter values of the Deffuant model on the convergence time \( T \). For example, our regression analysis suggests that the convergence time \( T \) undergoes a transition at a critical value of the confidence bound (\( c=0.5 \)) on complete graphs and prism graphs but not on cycles. We also illustrated that the interplay among the effects of \( N \), \( c \), and \( m \) on the convergence time \( T \) of the Deffuant model can be rather different qualitatively for different families of networks. For instance, the effects of the three parameters on \( T \) are independent on complete graphs and complete multipartite graphs for large values of \( c \) (these are mean-field situations), whereas \( N \) and \( m \) are weakly coupled for cycles, and all three parameters are coupled for prism graphs and square lattices. 

Our results also shed further light on educated guesses and other claims that have appeared in the literature. We examined quantitatively how convergence time $T$ increases with the number $N$ of agents in a network, and we thereby obtained several insights for different network topologies. For example, although \cite{laguna04} speculated that \( T \) is proportional to \( N \), our regression results indicate that the linear relationship need not hold and that it depends on the underlying network topology. Additionally, several papers have concluded based on numerical simulations for a few values of the multiplier \( m \) that consensus occurs for the Deffuant model for several networks (e.g., ER networks, WS networks, and BA networks) when the confidence bound is large, whereas multiple opinion groups persist at equilibrium for low confidence bounds \cite{fortunato04b,laguna04,weisbuch02,weisbuch04}. However, different transition thresholds (e.g., \( 0.25 \), \( 0.3\), and \( 0.5 \)) have been proposed for the confidence bound. In the synthetic networks that we study (except for cycles and cycles with random edges), our simulation results suggest that a transition threshold of \( c \in [0.4,0.5] \) is most likely for large populations. For \( c \geq 0.5 \), consensus occurs on all of our families of deterministic networks (both synthetic and empirical) except for bipartite graphs. For \( c<0.5 \), more opinion groups persist at equilibrium as node degree increases for our simulations on $k$-regular graphs given by cycles (for which the degree is $k = 2$ for each node), prism graphs ($k = 3$ for each node), and complete graphs (of progressively larger size, starting from $N = 10$ nodes, and hence of progressively larger degree for each node).
 This is possibly because agents who have a higher node degree in a regular graph have more neighbors with ``competing'' opinions, which gives the agents less time to make up their minds, so more opinion groups remain at equilibrium. Additionally, it was proposed in \cite{weisbuch02,weisbuch04} based on numerical simulations that one can approximate the number of opinion groups at equilibrium by \( K= \lfloor 1/(2c) \rfloor \) for large population. Our simulations show that this statement is not true in general. For instance, for simulations on prism graphs, \( K=2 \) for \( c \geq 0.3 \) when \( N \) is large.

Our simulations suggest that the equilibrium number \( K \) of opinion groups is similar for random graph models and appropriate counterpart deterministic networks (at least for the network families that we study). For example, adding a small number of random edges per node (i.e., the number of random edges divided by total number of nodes is small) to cycles and prism graphs does not have an obvious impact on \( K \). We conjecture that \( K \) approaches the situation for complete graphs as the proportion of random edges on cycles and prism graphs increases. For the Erd\H{o}s--R\'{e}nyi \( G(N,p) \) model, we observed that the behavior of \( K \) is similar to that on complete graphs when the edge generation probability \( p \) is close to \( 1 \). This suggests that it would be useful to study the Deffuant model on ER graphs using a mean-field approximation, especially as useful results have been obtained for other dynamical processes in this way \cite{porter2016}.

Our results provide insight into the convergence of opinion dynamics into stable groups of different opinions and on how long it takes to achieve such groups in differently-structured populations. For instance, when it is desirable to achieve a consensus among many individuals (especially in a potentially contentious situation), one may try to obtain agreement as quickly as possible, and it is useful to obtain a better understanding of which network structures can best achieve such useful outcomes. It is also noteworthy that one topic in early studies of bounded-confidence models such as the Deffuant model was to examine how extremism can take hold in a population \cite{deffuant02,amblard04,deffuant04}, and (perhaps especially given recent events) it seems useful to revisit such applications of these models. In developing models further for such applications, it will be important to incorporate recent insights, such as those in \cite{friedkin2016}.

Our systematic approach for studying the Deffuant model on various network structures is also applicable to other bounded-confidence models and models of opinion dynamics more generally. For example, the Hegselmann--Krause model was invented and subsequently attracted much research about the same time as the Deffuant model. It would be interesting to study the HK model using a systemic approach that is similar to the one in the present paper. One can also generalize bounded-confidence models to incorporate population heterogeneity, such as by drawing cautiousness parameters from a distribution (analogous to what is done in threshold models of social influence \cite{porter2016,watts2002}) rather using the same constant for all individuals, as openness to compromise is different for different people. Our regression approach should also be useful more generally for studying dynamical processes on networks, including more general structures such as multilayer networks \cite{kivela14}, temporal networks \cite{holme12}, and adaptive networks \cite{thilo-adaptive}.


\appendix


\section{Statistical Analysis}\label{appendix:complete_graphs}

In this appendix, we illustrate our statistical analysis in detail. For concreteness, we discuss our analysis in the context of the Deffuant model on complete graphs. We performed the same procedure for all of our regression analyses.

\begin{figure*}
	\centering
		\includegraphics[width=.85\textwidth]{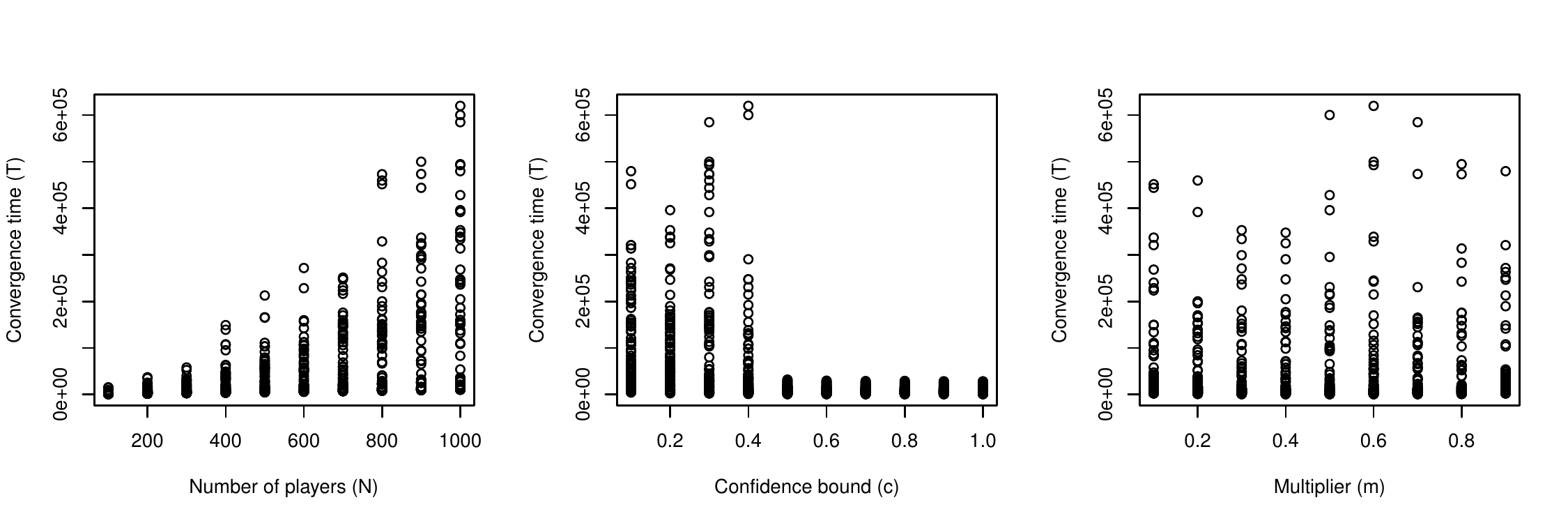}
	\caption{Scatter plots of convergence time (\( T \)) on complete graphs versus the number of agents (\( N \)), the confidence bound (\( c \)), and the multiplier (\( m \)). (We drew this figure using the software environment {\scshape R} \cite{R08}.)}
\label{figure:complete_graphs_scatter_plots(full)}
\end{figure*}

\begin{figure*}
	\centering
		\includegraphics[width=.85\textwidth]{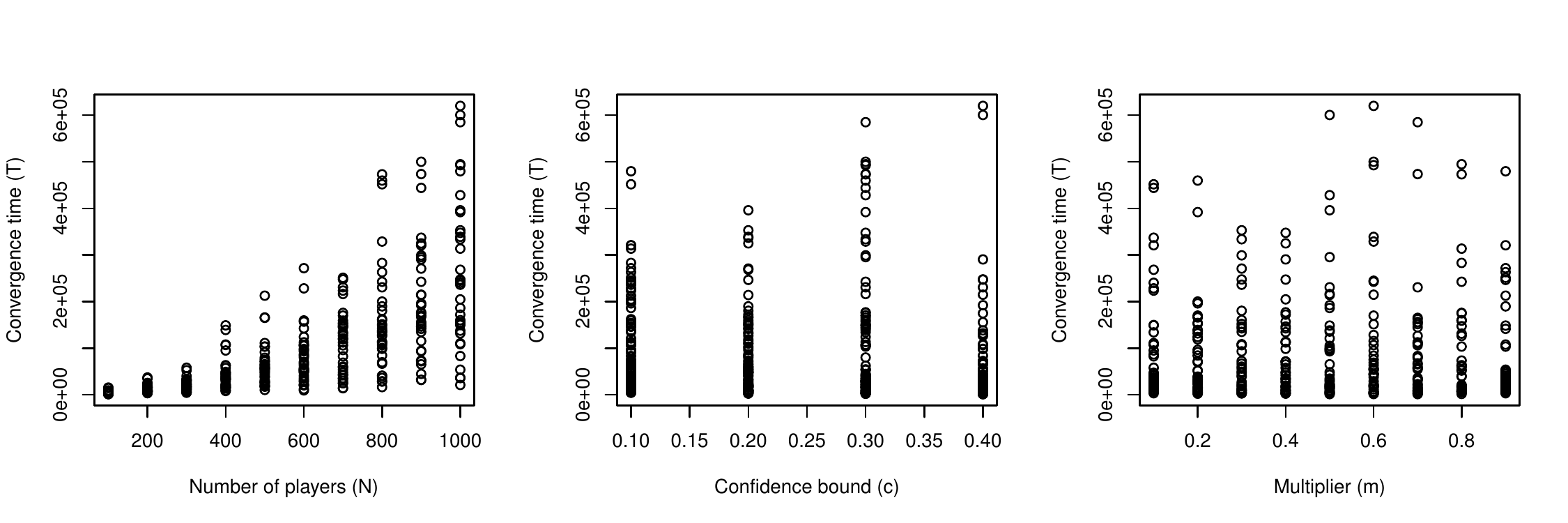}\\
		\includegraphics[width=.85\textwidth]{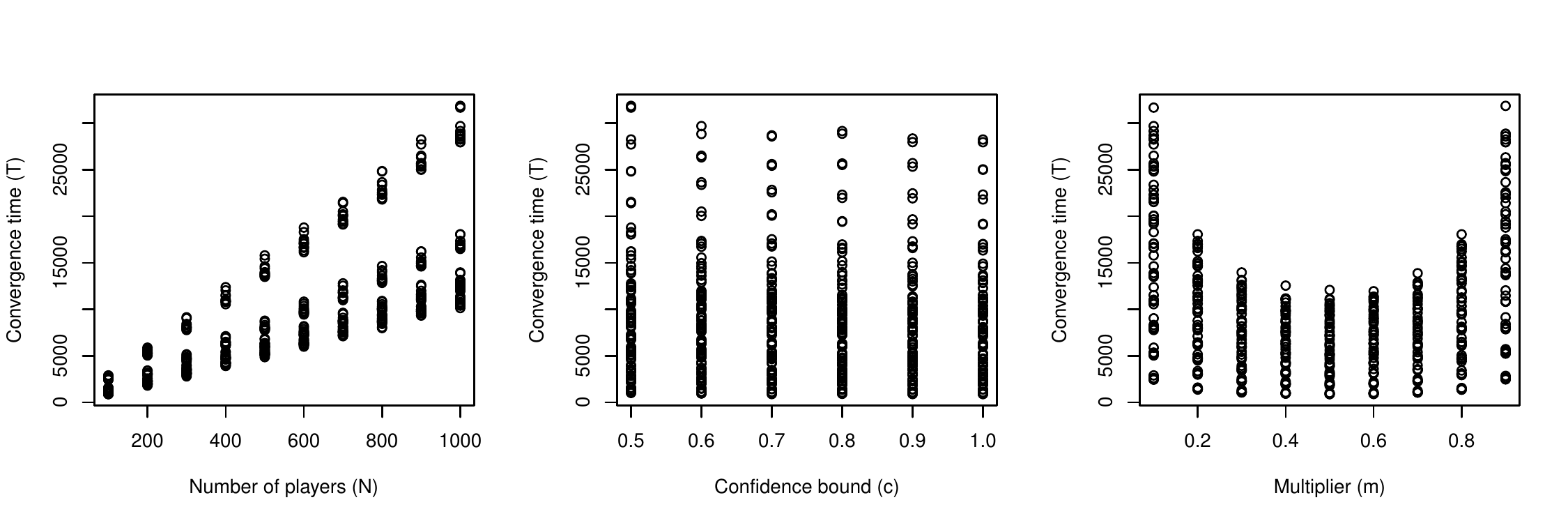}
	\caption{Scatter plots of \( T \) versus \( N \), \( c \), and \( m \) using simulation results on complete graphs with confidence bound (top) \( c<0.5 \) and (bottom) \( c \geq 0.5 \). (We drew this figure using the software environment {\scshape R} \cite{R08}.)}
\label{figure:complete_graphs_scatter_plots(lower)}
\end{figure*}

The scatter plots in Fig.~\ref{figure:complete_graphs_scatter_plots(full)} suggest that the convergence time (\( T \)) depends on the number of participating agents (\( N \)), the population's confidence bound (\( c \)), and possibly on their cautiousness (which we measure using the multiplier \( m \)). In particular, the relationship between \( T \) and \( c \) seems to undergo a transition at a critical value \( c=0.5 \), below which we observe a larger variation in \( T \). In Fig.~\ref{figure:complete_graphs_scatter_plots(lower)}, we show separate scatter plots for \( c<0.5 \) and \( c \geq 0.5 \) to illustrate the qualitatively distinct behavior in the two regimes. 

First, let's consoder the case \( c<0.5 \). We start by fitting a normal linear model
\begin{equation}\begin{aligned}
	T & = \beta_0 + \beta_1N + \beta_2N^2 + \beta_3c + \beta_4c^2 + \beta_5m + \beta_6m^2 \\
& \qquad + \beta_7mc + \beta_8Nc+ \beta_9Nm + \epsilon\,,
\label{model:complete_graph(lower)0}
\end{aligned}\end{equation}
where \( \beta_j \) (with \( j=0,1 \dots , 9 \)) are coefficients to be estimated and we assume that \( \epsilon \) is an independent and normally distributed error with mean $0$ and constant variance for every observation. To account for the curvature observed in Fig.~\ref{figure:complete_graphs_scatter_plots(lower)}, we include explanatory variables up to the second order in the model in Eq.~\eqref{model:complete_graph(lower)0}. We will subsequently drop statistically insignificant variables in a model-reduction procedure.

\begin{figure}
    \centerline{
    	(a) \hspace{-3mm}\includegraphics[width=.5\linewidth]{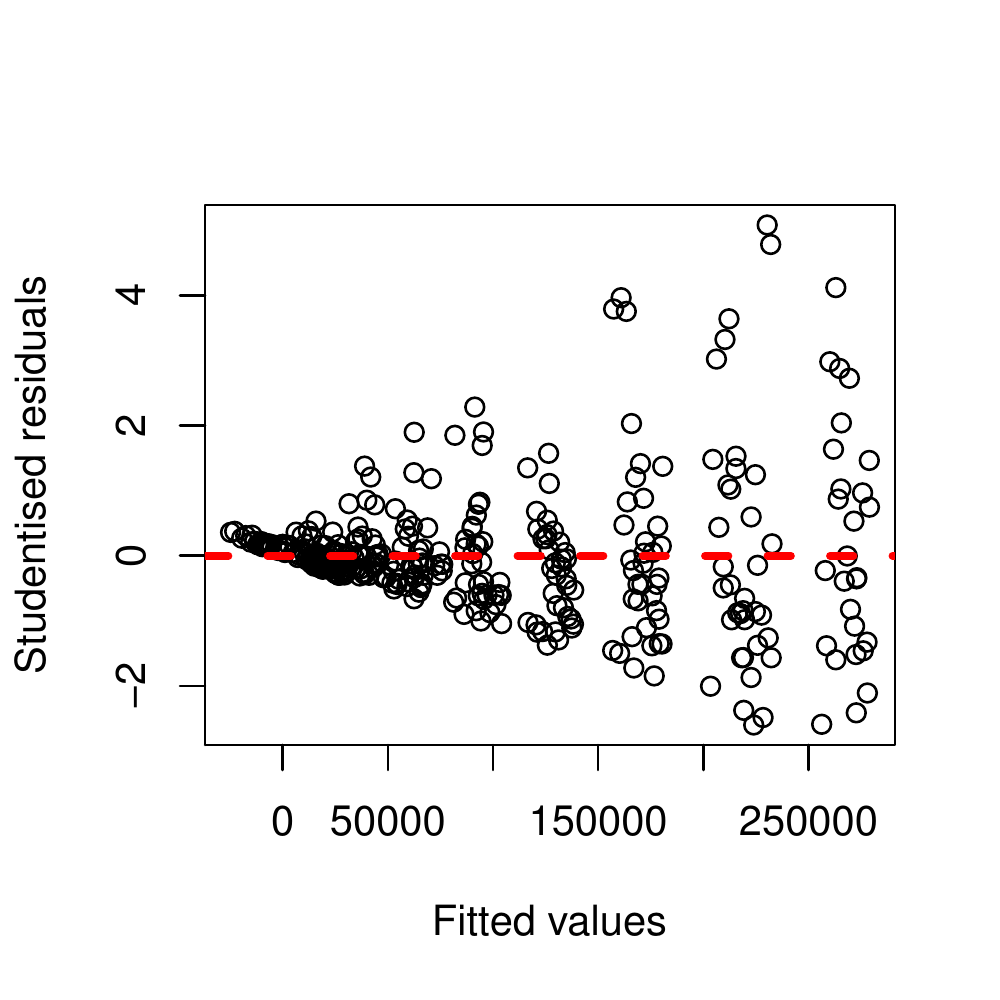}
    	\label{fig:complete_graph_residuals(lower)0}
    	(b) \hspace{-3mm}\includegraphics[width=.5\linewidth]{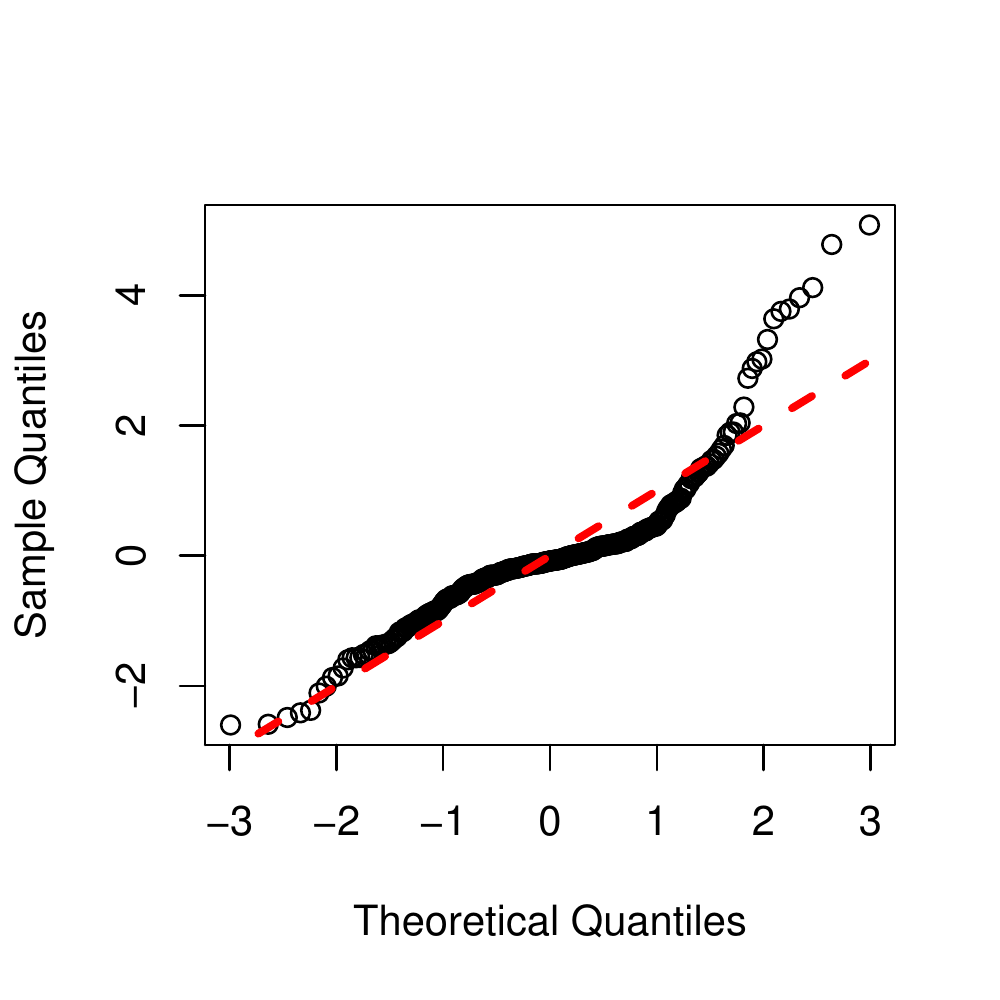}
    	\label{fig:complete_graph_qqplot(lower)0}
    	}
    \caption{(a) Studentized residuals versus fitted values and (b) normal Q--Q plot of studentized residuals for Eq.~\eqref{model:complete_graph(lower)0} using our simulation results on complete graphs with \( c < 0.5 \). In panel (a), the red dashed reference line is the horizontal line through the origin. Ideally, variance should be constant in the vertical direction, and the scatter should be symmetric vertically about \( 0 \). In panel (b), the red dashed reference line is the \( 45^{\circ} \) line through the origin. Data generated from a normal distribution should closely follow the red dashed line. \label{fig22}}
\end{figure}

Before proceeding with model selection, we check the validity of our model assumptions. In Fig.~\ref{fig22}, we check the assumption that the errors have $0$ mean and constant variance by plotting studentized residuals versus the response values predicted by Eq.~\eqref{model:complete_graph(lower)0}. Ideally, variance should be constant in the vertical direction, and the scatter should be symmetric vertically about \( 0 \). However, Fig.~\ref{fig22} indicates that the variance is not constant, as the points follow a clear wedge-shaped pattern, with the vertical spread of the points increasing with the fitted values. In Fig.~\ref{fig22}, we check the assumption of normality by plotting the sample quantiles versus the quantiles of a normal distribution. Data generated from a normal distribution should closely follow the \( 45^{\circ} \) line through the origin, but this is contradicted by the Q--Q plot in Fig.~\ref{fig22}. Therefore, the diagnostics show the necessity of stabilizing the variance and thereby making the data more like a normal distribution.

The one-parameter Box--Cox method \cite{box64} is a popular way to determine a transformation on strictly positive responses \cite{osbourne10}. A Box--Cox transformation maps \( T \) to \( T^{(\lambda)} \), where the family of transformations indexed by \( \lambda\in \mathbb{R} \) is defined by
\begin{equation}
T^{(\lambda)}=
	\begin{cases}
		\frac{T^{\lambda} -1}{\lambda} \,, & \text{if } \lambda \neq 0\,, \\
		\ln(T) \,, & \text{if } \lambda = 0\,.
	\end{cases}
\end{equation}

\begin{figure}
	\centering
		\includegraphics[width=.4\textwidth]{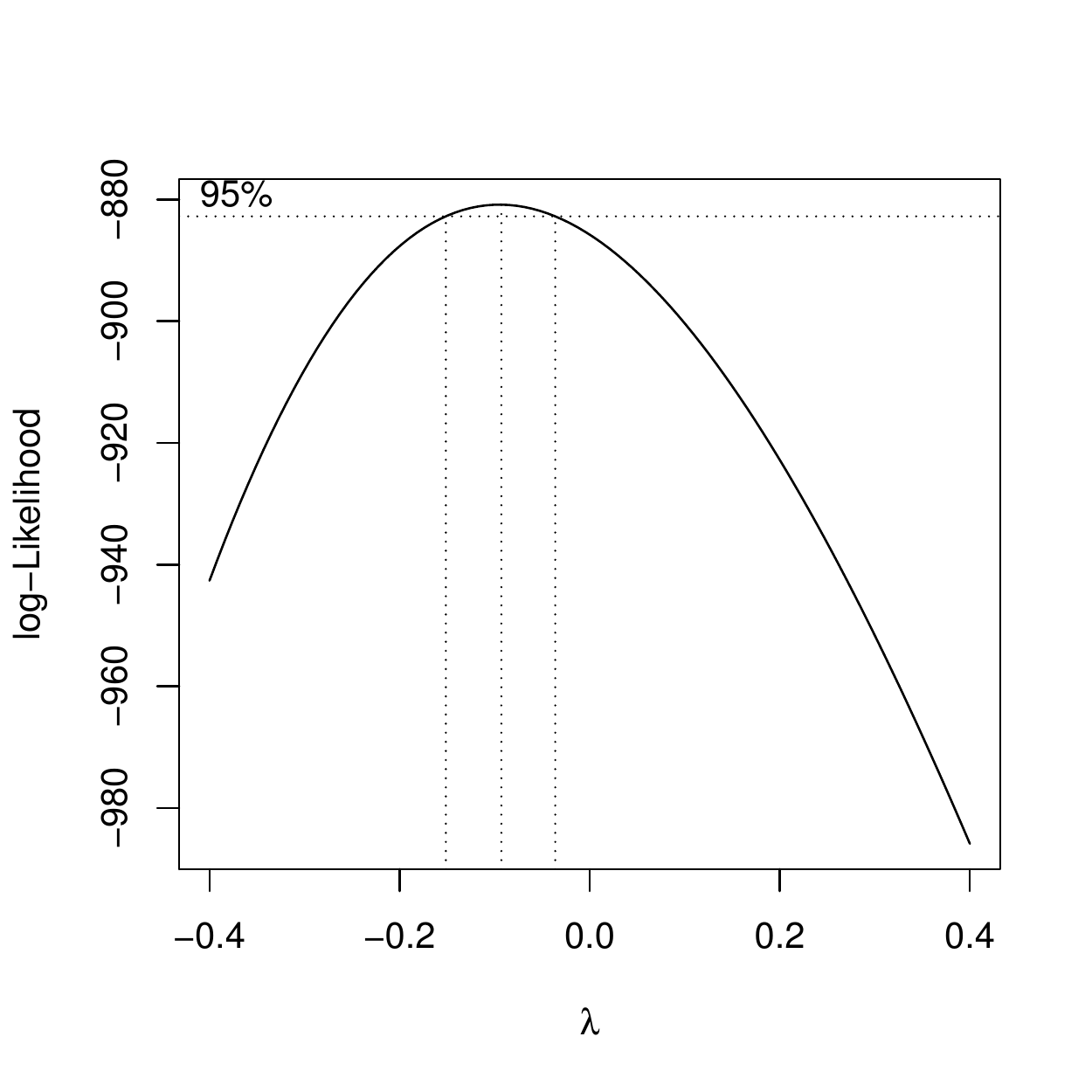}
	\caption{Profile log-likelihood plot for the parameter \( \lambda \) of the Box--Cox transformation.}
\label{figure:complete_graph_box-cox(lower)0}
\end{figure}

In Fig.~\ref{figure:complete_graph_box-cox(lower)0}, we show that the confidence interval for \( \lambda \) at the \( 95\% \) confidence level is roughly \( [-0.2,0] \). We choose to set \( \lambda=0 \), as this corresponds to taking a natural logarithm. The diagnostics of the new model suggest another log transformation, leading to the model
\begin{equation}\begin{aligned}
	\ln(\ln(T)) & = \beta_0 + \beta_1N + \beta_2N^2 + \beta_3c + \beta_4c^2 + \beta_5m \\
& \qquad + \beta_6m^2 + \beta_7mc + \beta_8Nc+ \beta_9Nm + \epsilon\,,
\label{model:complete_graph(lower)2}
\end{aligned}\end{equation}
where we assume that \( \epsilon \) is an independent and normally-distributed error with mean $0$ and constant variance for every observation. The variance for \( \epsilon \) is not necessarily the same for Eq.~\eqref{model:complete_graph(lower)0} and \eqref{model:complete_graph(lower)2}. However, we use the same notation for \( \epsilon \), with the understanding that it is of course allowed to be different for different models.

\begin{figure}
    \centerline{
    	 (a) \hspace{-3mm}\includegraphics[width=.5\linewidth]{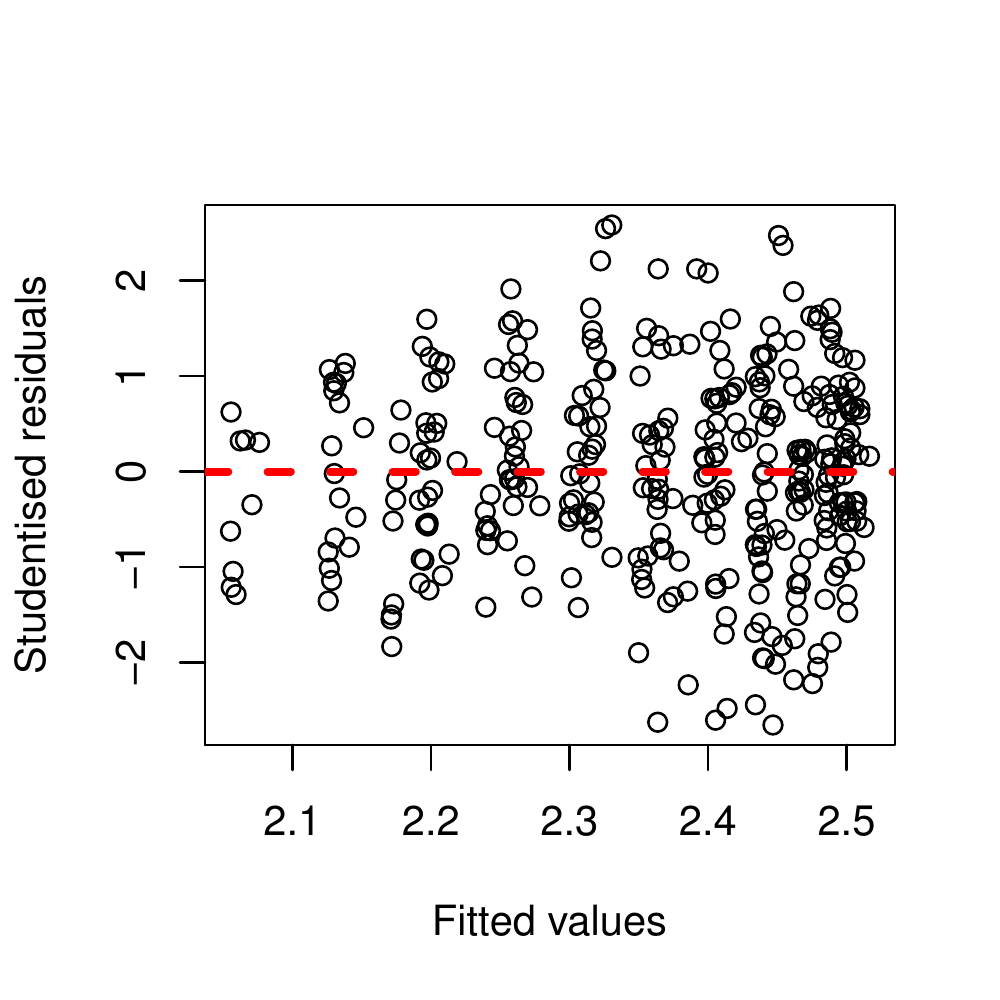}
    	\label{fig:complete_graph_residuals(lower)2}
    	 (b) \hspace{-3mm}\includegraphics[width=.5\linewidth]{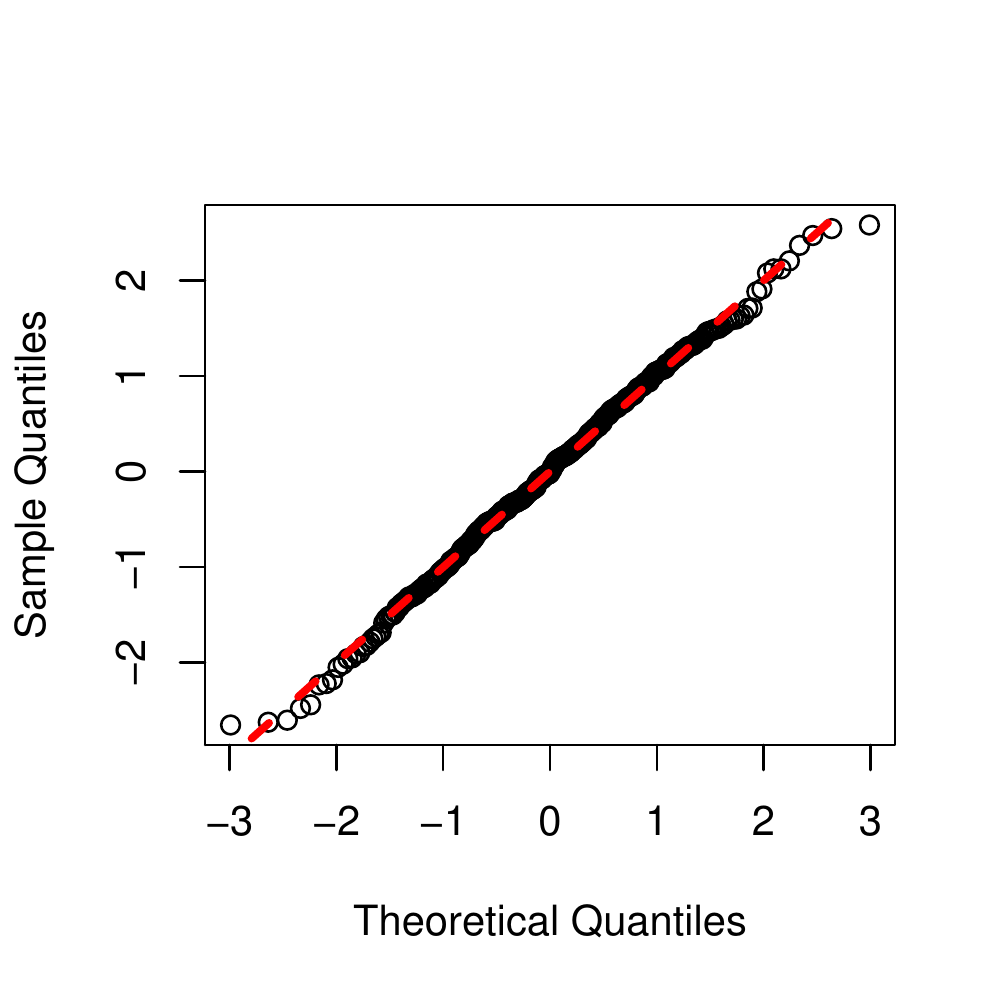}
    	\label{fig:complete_graph_qqplot(lower)2}
    }
    \caption{(a) Studentized residuals versus fitted values and (b) normal Q--Q plot of studentized residuals for Eq.~\eqref{model:complete_graph(lower)2} using simulation results on complete graphs with \( c < 0.5 \). The red dashed reference lines in panels (a) and (b) are, respectively, the horizontal line through the origin and the \( 45^{\circ} \) line through the origin. These diagnostic plots show that our model assumptions of normally distributed errors with mean $0$ and constant variance are reasonable for Eq.~\eqref{model:complete_graph(lower)2}. \label{fig24}}
\end{figure}

This time, Fig.~\ref{fig24} shows approximately constant variance in the vertical direction, and the scatter is roughly symmetric vertically about \( 0 \). There are no studentized residuals outside the \( [-3,3] \) range, revealing no serious outliers. In Fig.~\ref{fig24}, the points closely follow the \( 45^{\circ} \) line through the origin. Therefore, our model assumptions appear to be reasonable for Eq.~\eqref{model:complete_graph(lower)2}.

\begin{table}
	\centering
	\begin{tabular}{lcc}
	\hline
	{Model} & {AIC} & {\( R^2 \)} \\ 
	\hline
	Eq.~\eqref{model:complete_graph(lower)0} & \( \phantom{-}8146.1 \) & \( 0.5272 \) \\ 
	Eq.~\eqref{model:complete_graph(lower)2} & \( -2040.4 \) & \( 0.8164 \) \\ 
	Eq.~\eqref{model:complete_graph(lower)3} & \( -2037.1 \) & \( 0.8246 \) \\ 
	 \hline
	\end{tabular}
\caption{Values of AIC and \( R^2 \) of regression models that we consider for our simulations on complete graphs with \( c<0.5 \). They are accurate to \( 5 \) and \( 4 \) significant figures, respectively.}\label{table:complete_graph_compare_models(lower)}
\end{table}

\begin{table}
	\small
	\centering
	\begin{tabular}{ccccc}
	\hline
	 & {Estimate} & {Std. Error} & {\( t \) value} & {Pr(\( >|t| \))} \\ 
	\hline
	\( \beta_0 \) & \( \phantom{-}4.024 \)  & \( 5.038 \times 10^{-2} \) & \( \phantom{-}7.988 \times 10 \) & \( < 2 \times 10^{-16} \) \\ 
	  \( \beta_1 \) & \( \phantom{-}1.062 \) & \( 3.039 \times 10^{-3} \) & \( \phantom{-}3.495 \times 10^2 \) & \( < 2 \times 10^{-16} \) \\ 
	  \( \beta_2 \) & \( -1.316 \) & \( 1.277 \times 10^{-1} \) & \( -1.031 \times 10 \) & \( < 2 \times 10^{-16} \) \\ 
	  \( \beta_3 \) & \( \phantom{-}7.346 \times 10^{-1} \) & \( 8.472 \times 10^{-2} \) & \( \phantom{-}8.671 \) & \( < 2 \times 10^{-16} \) \\ 
	  \( \beta_4 \) & \( -6.261 \) & \( 3.704 \times 10^{-2} \) & \( -1.690 \times 10^2 \) & \( < 2 \times 10^{-16} \) \\ 
	  \( \beta_5 \) & \( \phantom{-}6.262 \) & \( 3.612 \times 10^{-2} \) & \( \phantom{-}1.733 \times 10^2 \) & \( < 2 \times 10^{-16} \) \\ 
	   \hline
	\end{tabular}
\caption{Estimates of regression parameters for Eq.~\eqref{model:complete_graph(upper)2}.}\label{table:complete_graph_summary(upper)2}
\end{table}

\begin{figure}
	\centering
		\includegraphics[width=.3\textwidth]{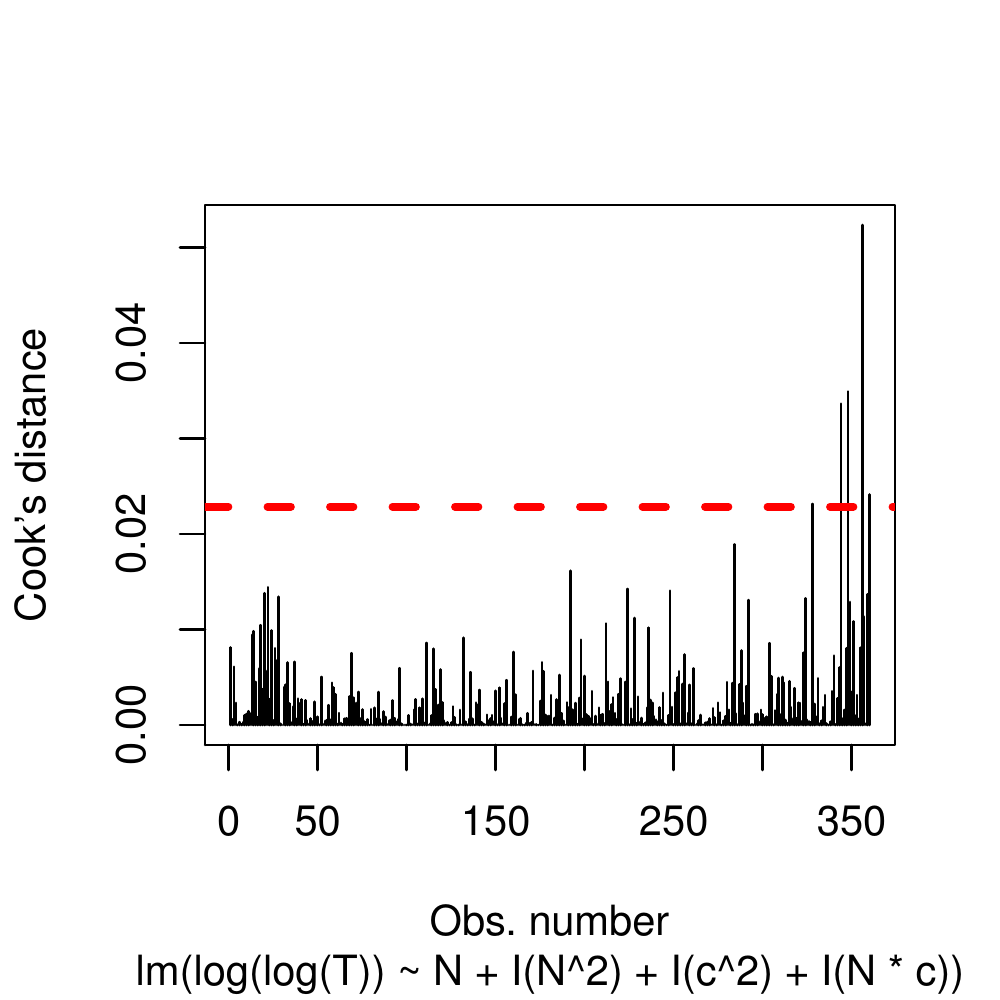}
	\caption{Cook's distances for the regression model defined by Eq.~\eqref{model:complete_graph(lower)3}. The red dashed line is a horizontal line through \mbox{\( 8/(\tilde{n}-2\tilde{p}) \)}, where \( \tilde{n} \) is the number of observations and \( \tilde{p} \) is the number of fitting parameters. This line gives the threshold for detecting highly influential observations that are particularly worth checking for validity.
	}
\label{figure:complete_graph_cooks_distance(lower)3}
\end{figure}

It is also important to minimize the number of regression terms in our models. AIC-based model selection drops the first-order term of \( c \) and all terms that include \( m \) to yield Eq.~\eqref{model:complete_graph(upper)3}. The diagnostic graphs of Eq.~\eqref{model:complete_graph(lower)3} are similar to those in Fig.~\ref{fig24} and are therefore acceptable.

Cook's distance \cite{cook77} measures the influence of a data point in a least-squares regression analysis. A commonly used threshold for detecting highly influential observations is \mbox{\( 8/(\tilde{n}-2\tilde{p}) \)}, where \( \tilde{n} \) is the number of observations and \( \tilde{p} \) is the number of fitting parameters. Fig.~\ref{figure:complete_graph_cooks_distance(lower)3} reveals \( 3 \) influential observations. We remove these \( 3 \) points and give the resulting estimates (accurate to \( 4 \) significant figures) for the coefficients \( \beta_j \) (with \( j=0,1, \dots , 4 \)) of Eq.~\eqref{model:complete_graph(lower)3} in Table~\ref{table:complete_graph_summary(lower)3}.

In Table~\ref{table:complete_graph_compare_models(lower)}, we summarize the values of AIC and \( R^2 \) for the regression models that we consider for simulations on complete graphs with \( c<0.5 \). The substantial increase in \( R^2 \) and decrease in AIC indicate that our final model (see Eq.~\eqref{model:complete_graph(lower)3}) has a much better goodness-of-fit and a considerably simpler form than our original model (see Eq.~\eqref{model:complete_graph(lower)0}).

For \( c\geq 0.5 \), we go through a similar model-selection process for \( c<0.5 \) and thereby obtain
\begin{equation}
	\ln(T) = \beta_0 + \beta_1\ln(N) + \beta_2c + \beta_3c^2 + \beta_4m + \beta_5m^2 + \epsilon\,.
\label{model:complete_graph(upper)2}
\end{equation}
We include an \( \ln(N) \) term in the full model (see Eq.~\eqref{model:complete_graph(upper)2}) to account for the linear dependence of \( T \) on \( N \) that Fig.~\ref{figure:complete_graphs_scatter_plots(lower)} suggests. AIC-based model selection indicates the statistical significance of the \( \ln(N) \) term. For Eq.~\eqref{model:complete_graph(upper)2}, we obtain \( \text{AIC} \approx -3248.9 \) and \( R^2 \approx 0.9965 \). In Table~\ref{table:complete_graph_summary(upper)2}, we give the estimates for the coefficients \( \beta_j \) (with \( j=0,1, \dots , 5 \)) of Eq.~\eqref{model:complete_graph(upper)2}.

Table~\ref{table:complete_graph_summary(upper)2} suggests combining \( m \) and \( m^2 \) into a single term \( (m-0.5)^2 \), and it also suggests combining \( c \) and \( c^2 \) into \( (c-1)^2 \). The model with the combined terms has \( \text{AIC} \approx -3240.9 \) and \( R^2 \approx 0.9964 \), which are very close to those of Eq.~\eqref{model:complete_graph(upper)2} but have two fewer coefficients to estimate. Therefore, we update our model for \( c \geq 0.5 \) to obtain the simpler model in Eq.~\eqref{model:complete_graph(upper)3}.


\section*{Acknowledgements}

We thank Mariano Beguerisse D\'iaz for helpful comments.






\end{document}